\documentclass[11pt]{article}
\usepackage{mathrsfs}
\usepackage{amsfonts}
\usepackage{amsfonts}
\usepackage{mathrsfs}
\usepackage{amsmath}
\usepackage{amssymb}

\numberwithin{equation}{section}

\author{ Yu Hou$^1$, \   Engui Fan$^1$\footnote{Corresponding
author and  e-mail address:
      faneg@fudan.edu.cn} \  and \  Zhijun Qiao$^2$  }
\date{   \small{ $^1$ School of Mathematical Sciences, Institute of Mathematics \\
 and Key Laboratory of Mathematics for Nonlinear Science, \\ Fudan
University, Shanghai 200433, P.R. China  \\
$^2$ Department of Mathematics, University of Texas-Pan American,\\
Edinburg, TX 78539, U.S.A }}
\title{\bf \Large{The algebro-geometric solutions for the modified Camassa-Holm
hierarchy} }
\begin{document}
\maketitle
\begin{abstract}
This paper is dedicated to provide theta function
representation of
algebro-geometric
solutions and
related crucial quantities for the modified Camassa-Holm (MCH) hierarchy through 
studying a
algebro-geometric initial value problem.
Our main tools
include the polynomial recursive formalism to derive the MCH
hierarchy, the hyperelliptic curve with finite number of genus,
the Baker-Akhiezer functions, the meromorphic function,
the Dubrovin-type equations for auxiliary divisors, and the
associated trace formulas. With the help of these tools, the
explicit
representations of the Baker-Ahhiezer
functions, the meromorphic function, and the algebro-geometric
solutions  are obtained for the entire MCH hierarchy.
\end{abstract}

\section{Introduction}

  Algebro-geometric solution,
 an important feature
 of integrable system, is a kind of explicit
 solutions closely related to the inverse spectral theory \cite{1},\cite{4},\cite{6}-\cite{8}.
 As a degenerated case of the algebro-geometric solution, the
 multi-soliton solution and periodic solution in elliptic function type may be
 obtained \cite{4},\cite{5},\cite{21}.
 A systematic approach, proposed by Gesztesy and Holden to construct algebro-geometric solutions for integrable equations,   has been extended to
 the whole (1+1) dimensional integrable hierarchy,
 such as the AKNS hierarchy, the Camassa-Holm (CH) hierarchy
etc. \cite{9}-\cite{12}. Recently, we investigated the Gerdjikov-Ivanov
 hierarchy and the Degasperis-Procesi hierarchy 
 and obtained their algebro-geometric
 solutions \cite{15},\cite{16}.

 The CH equation
  \begin{equation}\label{1.1}
    u_t-u_{xxt}+2\gamma u_x +3uu_x-2u_xu_{xx}-uu_{xxx}=0
  \end{equation}
 where $\gamma$ is a real constant, was 
 derived
 from the two-dimensional Euler equations in a search for integrable
 shallow water equations, with the $u(x,t)$ is the
 fluid velocity in the $x$ direction and $\gamma$ is a constant
 related to the critical shallow water wave speed \cite{2}. The CH equation describes
 the propagation of two-dimensional shallow water waves over a flat
 bed, also the propagation of axially symmetric waves in
 hyperelastic rods \cite{20},\cite{25}. The CH and the Degasperis-Procesi (DP)
 equations  are only two integrable members with $b=2$
 and $b=3$ from the following family
     \begin{equation}\label{1.2}
      u_t-u_{txx}+ \gamma b u_x +(b+1)uu_x=bu_xu_{xx}+uu_{xxx},
    \end{equation}
 where $b$ is a constant.

 Due to its many remarkable integrable properties, the CH equation has extensively been
 studied in the last twenty years. The bi-Hamiltonian
 structure of the CH equation was implied in the work of Fuchssteiner and
 Fokas \cite{14}.Complete integrability and infinity of conservation
 laws have been studied in \cite{2},\cite{3},\cite{33}. The inverse scattering transform
 was developed by Constantin, McKean,   Gerdjikov and Ivanov \cite{26},\cite{27}. Other
 progress in studying the CH equation includes 
 the existence of peaked solitons
 and geometry of multi-peakons \cite{2},\cite{34}, geometric formulations \cite{30},
and  waves breaking \cite{29}. The
 classic papers on the algebro-geometric solutions of the CH equation,
 or more appropriately, one is due to Qiao, the other
 is Gesztesy and Holden. Qiao obtained the algebro-geometric
 solution on a symplectic submainfold \cite{19}. Gesztesy and Holden
 derived the algebro-geometric solutions for the whole CH hierarchy
 by using the polynomial recursion method \cite{10}-\cite{12}.

 This paper is concerned with the following
 equation, called the modified Camassa-Holm (MCH) equation
     \begin{equation}\label{1.3}
        u_t-u_{xxt}+3u^2u_x-u_x^3=(4u-2u_{xx})u_xu_{xx}
        +(u^2-u_x^2)u_{xxx},
     \end{equation}
 where $u(x,t)$ is the function of spatial variable $x$ and time
 variable $t$. For our convenience, let us rewrite (1.3) as
     \begin{equation}\label{1.4}
        q_t+q_x(u^2-u_x^2)+2q^2u_x=0,
     \end{equation}
      \begin{equation}\label{1.5}
        q=u-u_{xx}.
      \end{equation}
 The MCH equation (1.3)  was derived as an integrable system by Fuchssteiner,  and
 Olver and Rosenau  by applying the tri-Hamiltonian method to
 the representation of the modified KdV equation \cite{13},\cite{36}. Later, Qiao
 recovered the MCH equation (\ref{1.3}) from the two-dimensional Euler
 equations by using the approximation procedure,  provided
the Lax pair and bi-Hamiltonian
 structure for the MCH equation, and first time proposed the W/M-shape solitons \cite{17}.
 Recently, Gui,
 Liu, Olver and Qu showed the wave-breaking and peaked
 traveling-wave solutions for the MCH equation \cite{35}. However,
 within the knowledge of the authors, the algebro-geometric solutions
 of the entire MCH hierarchy are not studied yet.

 The main task
 of this paper focuses on the algebro-geometric
 solutions of the whole MCH hierarchy in which (\ref{1.3}) is just the second member. The outline of the present paper is as follows.

 In section 2, based
 on the polynomial recursion formalism, we derive the MCH hierarchy, the associated sequences, and Lax pairs. A hyper-elliptic curve
 $\mathcal{K}_{n}$ with arithmetic genus $n$ is introduced with
 the help of the characteristic polynomial of Lax matrix $V_n$ for the
 stationary MCH hierarchy.

 In Section 3, we study a meromorphic
 function $\phi$ such that $\phi$ satisfies a nonlinear
 second-order differential equation. Then we study the properties of
the Baker-Akhiezer function $\psi$, and furthermore the stationary MCH
 equations are decomposed into a system of Dubrovin-type
 equations. The stationary trace formulas
   are obtained for the MCH hierarchy.

 In Section 4, we present the first set of our results, the explicit
 theta function representations of Baker-Akhiezer function, the
 meromorphic function  and
 the potentials $u$
 for the entire stationary MCH hierarchy. Furthermore, we study the
  initial value problem on an algebro-geometric curve for the stationary MCH
 hierarchy.

 In Sections 5 and 6, we extend the analysis in Sections 3 and 4 
 to the time-dependent case. Each equation in the MCH
 hierarchy is permitted to evolve in terms of an independent time
 parameter $t_r$. As an initial data we use a stationary solution of
 the $n$th equation
 and then construct a
 time-dependent solution of the $r$th equation in the MCH hierarchy.
 The Baker-Akhiezer function, the meromorphic function, the analogs
 of the Dubrovin-type equations, the trace formulas, and the theta function
 representation in Section 4 are all extended to the time-dependent
 case.

\section{The MCH hierarchy}
 In this section, let us derive the MCH hierarchy and the corresponding
 sequence of zero-curvature pairs by using a polynomial recursion formalism.
 Moreover, we introduce the hyper-elliptic curve connecting to
 the stationary MCH hierarchy.

 Throughout this section, let us make the following hypothesis.

\newtheorem{hyp1}{Hypothesis}[section]
\begin{hyp1}
      In the stationary case, let us assume
  \begin{equation}\label{2.1}
    \begin{split}
    & u\in C^\infty(\mathbb{C}),\ \
   \partial_x^ku\in L^\infty (\mathbb{C}),\qquad k\in \mathbb{N}_0.
    \end{split}
  \end{equation}
    In the time-dependent case, let us assume
  \begin{equation}
    \begin{split}\label{2.2}
    & u(\cdot,t)\in C^\infty (\mathbb{C}),\ \
     \partial_x^ku(\cdot,t)\in L^\infty
    (\mathbb{C}),\qquad k\in \mathbb{N}_0,~~ t\in \mathbb{C},\\
    & u(x,\cdot), u_{xx}(x,\cdot)\in C^1(\mathbb{C}).\\
  \end{split}
  \end{equation}
\end{hyp1}

Let us begin with the polynomial recursion formalism. Define
 $\{f_{2 l}\}_{l\in\mathbb{N}_{0}}$,
 $\{g_{2 l}\}_{l\in\mathbb{N}_{0}}$, and
 $\{h_{2 l}\}_{l\in\mathbb{N}_{0}}$
through the following recursive relations
    \begin{equation}\label{2.3}
      \begin{split}
        & f_{2l+1}=0, ~~ g_{2l+1}=0,~~ h_{2l+1}=0, ~~~~~l\in\mathbb{N}_0,\\
        & g_0=-1, ~~ \\
        &
          g_{2 l}=-\frac{1}{2} \partial^{-1} (u-u_{xx})~\delta_{2l,x},
           ~~~~~l\in\mathbb{N},\\
        & f_{2l,x}+f_{2l}=(u-u_{xx})g_{2l}, ~~~~~l\in\mathbb{N}_0,\\
        & h_{2l}=-2(u-u_{xx})^{-1}g_{2l+2,x}-f_{2l},~~~~~l\in\mathbb{N}_0.\\
      \end{split}
    \end{equation}
where $\delta_{2l}$ are determined by the following formulas (for details, see \cite{18})
    \begin{equation}\label{2.4}
      \begin{split}
       & \delta_{2}=2u,\\
       & \delta_{2l+2}=\mathcal {G} (\delta_{2l}), \qquad l \in \mathbb{N},\\
       & \mathcal {G}=-e^{-x} \partial^{-1} e^{2x} \partial^{-1}
          e^{-x} (u-u_{xx}) \partial^{-1} (u-u_{xx})\partial.
      \end{split}
    \end{equation}
Apparently, the first few can be computed as follows:
    \begin{equation}\label{2.5}
       \begin{split}
       & f_0=u_x-u,\\
       & f_2=-\frac{1}{2}(u-u_{xx})(u^2-u_x^2) + (u-u_{xx})c_1,\\
       & g_0=-1,\\
       & g_2=-\frac{1}{2}(u^2-u_x^2) +  c_1,\\
       & h_0= u+u_x,\\
       & h_2= \frac{1}{2}(u-u_{xx})(u^2-u_x^2) - (u-u_{xx})c_1, \ {\rm etc},
       \end{split}
    \end{equation}
where $\{c_l\}_{l\in\mathbb{N}_0}\subset\mathbb{C}$ are integration
constants.

Next, we introduce the corresponding homogeneous coefficients
$\hat{f}_{2l}, \hat{g}_{2l},$ and  $\hat{h}_{2l},$ defined through taking
$c_k=0$ for $k=1,\cdots,l,$
    \begin{equation}\label{2.6}
       \begin{split}
        & \hat{f}_0=u_x-u, \quad
           \hat{f}_{2}=-\frac{1}{2}(u-u_{xx})(u^2-u_x^2),
           \quad \hat{f}_{2l}=f_{2l}|_{c_k=0,~k=1,\ldots,l},\\
        & \hat{g}_0=-1, \quad
           \hat{g}_2=-\frac{1}{2}(u^2-u_x^2), \quad
            \hat{g}_{2l}=g_{2l}|_{c_k=0,~k=1,\ldots,l}, \\
        &   \hat{h}_0=u_x+u, \quad
           \hat{h}_{2}=\frac{1}{2}(u-u_{xx})(u^2-u_x^2),
           \quad \hat{h}_{2l}=h_{2l}|_{c_k=0,~k=1,\ldots,l}.
       \end{split}
    \end{equation}
It is easy to see 
    \begin{equation}\label{2.7}
      f_{2l}=\sum_{k=0}^{l}c_{l-k}\hat{f}_{2k}, \quad
      g_{2l}=\sum_{k=0}^{l}c_{l-k}\hat{g}_{2k}, \quad
      h_{2l}=\sum_{k=0}^{l}c_{l-k}\hat{h}_{2k}, \quad
       l\in \mathbb{N}_0,
    \end{equation}
where
    \begin{equation}\label{2.8}
        c_0=1.
    \end{equation}
Let us now consider the following  $2\times 2$ matrix iso-spectral
problem
     \begin{equation}\label{2.9}
       \psi_x=U(u,\lambda)\psi=
       \left(
         \begin{array}{cc}
           -\frac{1}{2} & \frac{1}{2} \lambda^{-1}(u-u_{xx}) \\
           -\frac{1}{2} \lambda^{-1}(u-u_{xx}) & \frac{1}{2} \\
         \end{array}
       \right)
       \psi
     \end{equation}
and an auxiliary problem
    \begin{equation} \label{2.10}
       \psi_{t_n}=V_n(\lambda)\psi,
    \end{equation}
where $V_n(\lambda)$ is defined by
    \begin{equation}\label{2.11}
      V_n(\lambda)=
      \left(
        \begin{array}{cc}
          -G_n & \lambda^{-1}F_n \\
          \lambda^{-1}H_n & G_n \\
        \end{array}
      \right),
    \qquad \lambda \in \mathbb{C} \setminus \{0\},
    \quad n\in\mathbb{N}_0.
    \end{equation}
Assume that $F_n$, $G_n$, and $H_n$ are polynomials of degree $2n$
with $C^\infty$ coefficients with respect to $x$.
Let us consider the
stationary zero-curvature equation
\begin{equation}\label{2.12}
   -V_{n,x}+[U,V_n]=0,
\end{equation}
that is
   \begin{eqnarray}\label{2.13}
      F_{n,x}&=&-F_n+(u-u_{xx})G_n, \\
      H_{n,x}&=& H_n+(u-u_{xx})G_n, \\
      \lambda^2G_{n,x}&=&-\frac{1}{2}(u-u_{xx})H_n
          -\frac{1}{2}(u-u_{xx})F_n.
    \end{eqnarray}
From (\ref{2.13})-(2.15), a direct calculation shows 
   \begin{equation}\label{2.16}
     \frac{d}{dx} \mathrm{det} (V_n(\lambda,x))=
     -\frac{1}{\lambda^2} \frac{d}{dx} \Big(
     \lambda^2G_n(\lambda,x)^2+F_n(\lambda,x)H_n(\lambda,x)
     \Big)=0,
   \end{equation}
and therefore $\lambda^2G_n^2+F_nH_n$ is $x$-independent to imply
   \begin{equation}\label{2.17}
    \lambda^2G_n^2+F_nH_n=R_{4n+2}
   \end{equation}
where the integration constant $R_{4n+2}$ is a monic polynomial of
degree $4n+2$ with respect to $\lambda$. Let
$\{E_m^2\}_{m=1,\cdots,2n+1}$ denote its zeros,   then
   \begin{equation}\label{2.18}
   R_{4n+2}(\lambda)=\prod_{m=1}^{2n+1}(\lambda^2-E_m^2),\qquad
   \{E_m^2\}_{m=1,\cdots,2n+1}\in\mathbb{C}.
   \end{equation}
In order to derive the corresponding hyper-elliptic curve, we compute
the characteristic polynomial $\mathrm{det}(yI-\lambda V_n)$ of the Lax
matrix $\lambda V_n$,
    \begin{eqnarray}\label{2.19}
      \mathrm{det}(yI-\lambda V_n)&=&
        y^2-\lambda^2G_n(\lambda)^2-F_n(\lambda)H_n(\lambda)
         \nonumber \\
      &=& y^2-R_{4n+2}(\lambda)=0.
    \end{eqnarray}
Equation (\ref{2.19}) naturally leads to the hyper-elliptic curve
$\mathcal{K}_n$, where
  \begin{eqnarray}\label{2.20}
     &&\mathcal {K}_n:\mathcal {F}_n(\lambda,y)=y^2-R_{4n+2}(\lambda)=0,\nonumber \\
    &&R_{4n+2}(\lambda)=\prod_{m=1}^{2n+1}(\lambda^2-E_m^2),
    \qquad \{E_m^2\}_{m=1,\cdots,2n+1}\in\mathbb{C}.
   \end{eqnarray}
Actually,
it is more convenient to introduce the
notations $z=\lambda^2, \widetilde{E}_m=E_m^2$, so that
$\mathcal{K}_n$ becomes the hyper-elliptic curve of genus $n \in
\mathbb{N}_0$ (possibly with a singular affine part), namely,
  \begin{eqnarray}\label{2.21}
     &&\mathcal {K}_n:\mathcal {F}_n(z,y)=y^2-R_{2n+1}(z)=0,\nonumber \\
    &&R_{2n+1}(z)=\prod_{m=1}^{2n+1}(z-\widetilde{E}_m),
    \qquad \{\widetilde{E}_m\}_{m=1,\cdots,2n+1}\in\mathbb{C}.
   \end{eqnarray}

The stationary zero-curvature equation
(\ref{2.12}) implies polynomial recursion relations (\ref{2.3}).
Let us introduce the following polynomials $F_n(\lambda), G_n(\lambda)$
and $H_n(\lambda)$ with respect to the spectral parameter $\lambda$,
   \begin{equation}\label{2.22}
     F_n(\lambda)=\sum_{l=0}^n f_{2l} \lambda^{2(n-l)},
   \end{equation}

   \begin{equation}\label{2.23}
     G_n(\lambda)=\sum_{l=0}^n g_{2l} \lambda^{2(n-l)},
   \end{equation}

   \begin{equation}\label{2.24}
     H_n(\lambda)=\sum_{l=0}^n h_{2l} \lambda^{2(n-l)}.
   \end{equation}
Inserting (\ref{2.22})-(\ref{2.24}) into (\ref{2.13})-(2.15)
yields the recursion relations (\ref{2.3}) for $f_{2l}$ and
$g_{2l}$, $l=0,\ldots, n.$ By using (2.15), we obtain the recursion formula
for $h_{2l}, l=0, \ldots, n-1$ in (\ref{2.3}) and
     \begin{equation}\label{2.25}
        h_{2n}=-f_{2n}.
     \end{equation}
Moreover, from (2.14), we have
   \begin{equation}\label{2.26}
      h_{2n,x}-h_{2n}-(u-u_{xx})g_{2n}=0,
      \qquad n\in \mathbb{N}_0.
   \end{equation}
Hence, inserting  the relations (\ref{2.25}) and
     \begin{equation}\label{2.27}
        f_{2n,x}+f_{2n}-(u-u_{xx})g_{2n}=0
     \end{equation}
into (\ref{2.26}), we obtain
     \begin{equation}\label{2.28}
        \mathrm{s}-\mathrm{MCH}_n(u)=-2f_{2n,x}=0,
        \qquad n\in \mathbb{N}_0.
     \end{equation}
The stationary MCH hierarchy is defined by (\ref{2.28}).
The first few equations are 
     \begin{equation}\label{2.29}
       \begin{split}
       & \mathrm{s-MCH}_0(u)=2u_x-2u_{xx}=0,\\
       &
       \mathrm{s-MCH}_1(u)=(u_x-u_{xxx})(u^2-u_x^2)+2(u-u_{xx})^2u_x
                            +(u_x-u_{xxx})c_1=0,
       \\
       & \mathrm{etc}.
       \end{split}
      \end{equation}
By definition, the set of solutions of (\ref{2.28}) represents a
class of algebro-geometric MCH solutions with $n$ ranging in
$\mathbb{N}_0$ and $c_l$ in $\mathbb{C},~l\in\mathbb{N}$. We call
the stationary algebro-geometric MCH solutions $u$ as MCH potentials
at times.

\newtheorem{rem2.2}[hyp1]{Remark}
  \begin{rem2.2}
    Here we emphasize that if $u$ satisfies one of the stationary MCH equations in
    $(\ref{2.28})$, then it must satisfy infinitely many such equations
    with order higher than $n$ for certain choices of integration
    constants $c_l$. This is a common characteristic of the general
    integrable soliton equations such as the KdV, AKNS and CH
    hierarchies \cite{12}.
  \end{rem2.2}

Next, we introduce the following homogeneous polynomials
$\widehat{F}_{l}, \widehat{G}_{l}, \widehat{H}_{l}$ defined by
    \begin{equation}\label{2.30}
      \widehat{F}_{l}(\lambda)=F_{l}(\lambda)|_{c_k=0,~k=1,\dots,l}
      =\sum_{k=0}^l \hat{f}_{2k} \lambda^{2(l-k)},
      \qquad l=0,\ldots,n,
    \end{equation}

 \begin{equation}\label{2.31}
      \widehat{G}_{l}(\lambda)=G_{l}(\lambda)|_{c_k=0,~k=1,\dots,l}
      =\sum_{k=0}^l \hat{g}_{2k} \lambda^{2(l-k)},
      \qquad l=0,\ldots,n,
    \end{equation}

\begin{equation}\label{2.32}
      \widehat{H}_{l}(\lambda)=H_{l}(\lambda)|_{c_k=0,~k=1,\dots,l}
      =\sum_{k=0}^l \hat{h}_{2k} \lambda^{2(l-k)},
      \qquad l=0,\ldots,n-1,
    \end{equation}

\begin{equation}\label{2.33}
      \widehat{H}_{n}(\lambda)= -\hat{f}_{2n}+
      \sum_{k=0}^{n-1} \hat{h}_{2k} \lambda^{2(n-k)}.
          \end{equation}
Then, the corresponding homogeneous formalism of (\ref{2.28}) are
given by
      \begin{equation}\label{2.34}
        \mathrm{s}-\widehat{\mathrm{MCH}}_n(u)=
        \mathrm{s-MCH}_n(u)|_{c_l=0,~l=1,\dots,n}=0,
        \qquad n\in\mathbb{N}_0.
      \end{equation}

Let us conclude this section with the time-dependent MCH hierarchy.
The "time-dependent" means that $u$ is a function of both space
and time. Let us introduce a deformation parameter $t_n \in \mathbb{C}$
in $u$, replacing $u(x)$ by $u(x,t_n)$, for each equation in the
hierarchy. In addition, the definitions (\ref{2.9}),
(\ref{2.11}) and (\ref{2.22})-(\ref{2.24}) of $U,$ $V_n$ and $F_n,
G_n$ and $H_n$ are still available. Then, the compatibility condition
yields the zero-curvature equation
   \begin{equation}\label{2.35}
   U_{t_n}-V_{n,x}+[U,V_n]=0, \qquad n\in \mathbb{N}_0,
   \end{equation}
namely,
    \begin{eqnarray}\label{2.36}
      \frac{1}{2}(u_{t_n}-u_{xxt_n})-F_{n,x}-F_n+(u-u_{xx})G_n=0, \\
      -\frac{1}{2}(u_{t_n}-u_{xxt_n})-H_{n,x}+ H_n+(u-u_{xx})G_n=0, \\
      \lambda^2G_{n,x}=-\frac{1}{2}(u-u_{xx})H_n
          -\frac{1}{2}(u-u_{xx})F_n.
    \end{eqnarray}
Inserting the polynomial expressions for $F_n$, $G_n$ and $H_n$ into
(\ref{2.36}) and (2.38), then we have the relations
(\ref{2.3}) for $f_{2l}$ and $g_{2l}$, $l=0,\ldots,n$. By using
(2.38), we obtain the recursion formula (\ref{2.3}) for $h_{2l}$,
$l=0,\ldots,n-1$ and
      \begin{equation}\label{2.39}
        h_{2n}=-f_{2n}.
      \end{equation}
Moreover, from (\ref{2.36}) and (2.37), we have 
     \begin{eqnarray}\label{2.40}
       \frac{1}{2}(u_{t_n}-u_{xxt_n})-f_{2n,x}-f_{2n}+
           (u-u_{xx})g_{2n}=0,\\
       -\frac{1}{2}(u_{t_n}-u_{xxt_n})-h_{2n,x}+h_{2n}+
           (u-u_{xx})g_{2n}=0.
     \end{eqnarray}
Hence, by (\ref{2.39}) and (\ref{2.40}), (2.41) can
be rewritten as
     \begin{equation}\label{2.42}
        \mathrm{MCH}_n(u)=u_{t_n}-u_{xxt_n}-2f_{2n,x}=0,
        \qquad (x,t_n)\in \mathbb{C}^2, ~n\in\mathbb{N}_0.
     \end{equation}
Varying $n\in\mathbb{N}_0$ in (\ref{2.42}) defines the
time-dependent MCH hierarchy. The first few equations
are 
    \begin{equation}\label{2.43}
       \begin{split}
       & \mathrm{MCH}_0(u)=u_{t_0}-u_{xxt_0}+2u_x-2u_{xx}=0,\\
       &
       \mathrm{MCH}_1(u)=u_{t_1}-u_{xxt_1}+(u_x-u_{xxx})(u^2-u_x^2)+2(u-u_{xx})^2u_x \\
       & ~~~~~~~~~~~~~+(u_x-u_{xxx})c_1=0,
       \\
       & \mathrm{etc}.
       \end{split}
      \end{equation}
The second equation $\mathrm{MCH}_1(u)=0$ (with $c_1=0$) in the
hierarchy represents the Modified Camassa-Holm (MCH) equation as discussed
in section 1. Similarly, one can introduce the corresponding
homogeneous MCH hierarchy by
    \begin{equation}\label{2.44}
        \widehat{\mathrm{MCH}}_n(u)=\mathrm{MCH}_n(u)|_{c_l=0,~l=1,\ldots,n}=0,
        \qquad n\in\mathbb{N}_0.
    \end{equation}
The MCH hierarchy  can also be defined in the form of $h_{2n}$ for
the relation $h_{2n}=-f_{2n}$. The integration
constants $c_l=0$ $l=1,\ldots,n$ are taken to derive the Modified Camassa-Holm
equation $\mathrm{MCH}_1(u)=0$.

In fact, since the Lenard operator recursion formalism is almost universally
adopted in the contemporary literature on the integrable soliton
equations, it might be worthwhile to adopt the alternative approach
using the polynomial recursion relations.

\section{The stationary MCH formalism}
 In this section, we focus our attention on the stationary case. By
 using the polynomial recursion formalism described in section 2, we
 define a fundamental meromorphic function $\phi(P,x)$ on a hyper-elliptic
 curve $\mathcal{K}_n$. Moreover, we study the properties of the
 Baker-Akhiezer function $\psi(P,x,x_0)$, Dubrovin-type equations,
 and trace formulas.

 We emphasize that the analysis about the stationary case described in section
 2 also holds here for the present context. 

 Let 
     \begin{equation}\label{3.1}
        z=\lambda^2, ~~~~
        \widetilde{E}_m=E_m^2,
     \end{equation}
then the hyper-elliptic curve $\mathcal{K}_n$ is given by (\ref{2.21}).
$\mathcal{K}_n$ is compactified by joining a point at infinity,
$P_\infty$, but for our convenience, the compactification is
still denoted by $\mathcal{K}_n$.
Point $P$ on
$$\mathcal{K}_{n} \setminus \{P_{\infty}\},$$
is referred to a pair type $P=(z,y(P))$, where $y(\cdot)$ is the
meromorphic function on $\mathcal{K}_{n}$ satisfying
$$\mathcal{F}_n(z,y(P))=0.$$
The complex structure on $\mathcal{K}_{n}$ is defined in the usual
way by introducing local coordinates
$$\zeta_{Q_0}:P\rightarrow(z-z_0)$$
near points
$$Q_0=(z_0,y(Q_0))\in \mathcal{K}_{n} \setminus P_0, \qquad  P_0=(0,y(P_0)),$$
which are neither branch nor singular points of $\mathcal{K}_{n}$.
Near $P_0$, the local coordinates are
   $$\zeta_{P_0}:P \rightarrow z^{1/2};$$
near the branch point $P_{\infty} \in \mathcal{K}_{n}$, the local
coordinates are
   $$\zeta_{P_{\infty}}:P \rightarrow z^{-1},$$
and also the similar case at branch and singular points of $\mathcal{K}_{n}.$
Thus, $\mathcal{K}_{n}$ becomes a two-sheeted hyper-elliptic Riemann
surface \cite{31} with genus $n \in \mathbb{N}_0$ (possibly with a singular
affine part) in a standard manner.

We also notice that fixing the zeros
$\widetilde{E}_1,\ldots,\widetilde{E}_{2n+1}$ of $R_{2n+1}$
discussed in (\ref{2.21}) leads to the curve $\mathcal{K}_{n}$ fixed.
Then the integration constants $c_1,\ldots,c_n$ in $f_{2n}$ are
uniquely determined, which are the symmetric functions of
$\widetilde{E}_1,\ldots,\widetilde{E}_{2n+1}$.

The holomorphic map
   $\ast,$ changing sheets, is defined by
       \begin{eqnarray}\label{3.2}
       && \ast: \begin{cases}
                        \mathcal{K}_{n}\rightarrow\mathcal{K}_{n},
                       \\
                       P=(z,y_j(z))\rightarrow
                       P^\ast=(z,y_{j+1(\mathrm{mod}~
                       2)})(z)), \quad j=0,1,
                      \end {cases}
                     \nonumber \\
      && P^{\ast \ast}:=(P^\ast)^\ast, \quad \mathrm{etc}.,
       \end{eqnarray}
   where $y_j(z),\, j=0,1$ denote the two branches of $y(P)$
   satisfying $\mathcal{F}_{n}(z,y)=0$.\\
   Finally, positive divisors on $\mathcal{K}_{n}$ of degree $n$
   are denoted by
        \begin{equation}\label{3.3}
          \mathcal{D}_{P_1,\ldots,P_{n}}:
             \begin{cases}
              \mathcal{K}_{n}\rightarrow \mathbb{N}_0,\\
              P\rightarrow \mathcal{D}_{P_1,\ldots,P_{n}}=
                \begin{cases}
                  \textrm{ $k$ if $P$ occurs $k$
                      times in $\{P_1,\ldots,P_{n}\},$}\\
                   \textrm{ $0$ if $P \notin
                     $$ \{P_1,\ldots,P_{n}\}.$}
                \end{cases}
             \end{cases}
        \end{equation}

  For the notational simplicity, assume $n
  \in \mathbb{N}$. The case $n=0$ is treated in
  Example 4.10.

  Next, we define the stationary Baker-Akhiezer function
  $\psi(P,x,x_0)$ on $\mathcal{K}_{n}\setminus \{P_{\infty},P_0\}$
  as follows
       \begin{equation}\label{3.4}
         \begin{split}
          & \psi(P,x,x_0)=\left(
                            \begin{array}{c}
                              \psi_1(P,x,x_0) \\
                              \psi_2(P,x,x_0) \\
                            \end{array}
                          \right), \\
          & \psi_x(P,x,x_0)=U(u(x),z(P))\psi(P,x,x_0),\\
           & z^\frac{1}{2}V_n(u(x),z(P))\psi(P,x,x_0)=y(P)\psi(P,x,x_0),\\
          & \psi_1(P,x_0,x_0)=1;
             \quad   P=(z,y)\in \mathcal{K}_{n}
           \setminus \{P_{\infty},P_0\},~x\in \mathbb{C}.
         \end{split}
       \end{equation}
  Closely related to $\psi(P,x,x_0)$ is the following meromorphic
  function $\phi(P,x)$ on $\mathcal{K}_{n}$ defined by
       \begin{equation}\label{3.5}
         \phi(P,x)=\frac{2z}{q} \Big(
         \frac{ \psi_{1,x}(P,x,x_0)}{\psi_1(P,x,x_0)}+\frac{1}{2} \Big),
         \quad P\in \mathcal{K}_{n},~ x\in \mathbb{C}
       \end{equation}
with 
       \begin{equation}\label{3.6}
          \begin{split}
         \psi_1(P,x,x_0)=\mathrm{exp}\left(z^{-1}\int_{x_0}^x
         \frac{1}{2}q(x^\prime)\phi(P,x^\prime) dx^\prime -
         \frac{1}{2}(x-x_0)\right),\\
         \quad P\in \mathcal{K}_{n}\setminus \{P_{\infty},P_0\},
         \end{split}
        \end{equation}
 where $q=u-u_{xx}$. \\
 Then, based on (\ref{3.4}) and (\ref{3.5}), a direct calculation shows that
    \begin{eqnarray}\label{3.7}
        \phi(P,x)&=&z^{1/2}\frac{y+z^{1/2}G_n(z,x)}{F_n(z,x)}
           \nonumber \\
        &=&
         \frac{z^{1/2}H_n(z,x)}{y-z^{1/2}G_n(z,x)},
    \end{eqnarray}
 and
      \begin{equation}\label{3.8}
        \psi_2(P,x,x_0)= \psi_1(P,x,x_0)\phi(P,x)/z^{1/2}.
      \end{equation}
 By inspection of the expression of $F_n$ and $H_n$ in (\ref{2.22})
 and (\ref{2.24}), we notice that $F_n$ and $H_n$ are even functions
 with respect to $\lambda$. Thus, if $\lambda$ is the root of $F_n$ or
 $H_n$, then $-\lambda$ is also their root. In this way, we can
 write $F_n$ and $H_n$ as the following finite products
        \begin{equation}\label{3.9}
          \begin{split}
            & F_n(\lambda)=f_0\prod_{j=1}^{n}(\lambda^2-\bar{\mu}_j^2),\\
            &
            H_n(\lambda)=h_0\prod_{j=1}^{n}(\lambda^2-\bar{\nu}_j^2).
          \end{split}
        \end{equation}
 For our convenience, 
 let 
          \begin{equation}\label{3.10}
            z=\lambda^2,\quad
            \mu_j=\bar{\mu}_j^2, \quad
             \nu_j=\bar{\nu}_j^2,
          \end{equation}
 then $F_n$ and $H_n$ can be rewritten as the following formalism,
          \begin{equation}\label{3.11}
            F_n(z)=f_0\prod_{j=1}^{n}(z-\mu_j)=
             (u_x-u)\prod_{j=1}^{n}(z-\mu_j),
          \end{equation}
           \begin{equation}\label{3.12}
            H_n(z)=h_0\prod_{j=1}^{n}(z-\nu_j)=
             (u_x+u)\prod_{j=1}^{n}(z-\nu_j).
          \end{equation}
 Moreover, defining
      \begin{equation}\label{3.13}
        \hat{\mu}_j(x)
            =\left(\mu_j(x),-\sqrt{\mu_j(x)}G_n(\mu_j(x),x)\right)
            \in \mathcal{K}_{n}, ~
            j=1,\ldots,n,~x\in\mathbb{C},
      \end{equation}

       \begin{equation}\label{3.14}
        \hat{\nu}_j(x)
            =\left(\nu_j(x),\sqrt{\nu_j(x)}G_n(\nu_j(x),x)\right)
            \in \mathcal{K}_{n}, ~
            j=1,\ldots,n,~x\in\mathbb{C},
      \end{equation}
and taking $z=0$ in (\ref{2.17}) yields
         \begin{equation}\label{3.15}
            f_{2n}h_{2n}=-f_{2n}^2=-\prod_{j=1}^{2n+1}
            \widetilde{E}_m.
         \end{equation}
 So, we can choose
 \begin{equation}\label{3.16}
    P_0=(0,f_{2n})=(0,\prod_{j=1}^{2n+1}
            \widetilde{E}_m^{1/2}).
 \end{equation}

 Due to (\ref{2.1}), $u$ is smooth and bounded, and therefore
 $F_n(z,x)$ and $H_n(z,x)$ share the same property. Thus, we have 
    \begin{equation}\label{3.17}
        \mu_j,\nu_j \in C(\mathbb{R}), \quad
        j=1,\dots,n,
    \end{equation}
 where $\mu_j,\nu_j$ may have appropriate multiplicities.

 The branch of $y(\cdot)$ near $P_\infty$ is fixed according to
       \begin{equation}\label{3.18}
         \underset{|z(P)| \rightarrow \infty \atop P \rightarrow P_\infty}{\mathrm{lim}}
         \frac{y(P)}{z^{1/2}G_n(z,x)}=1.
       \end{equation}
 Also by (\ref{3.7}), the divisor
   $(\phi(P,x))$ of $\phi(P,x)$ is given by
         \begin{equation}\label{3.19}
           (\phi(P,x))=\mathcal{D}_{P_0,\hat{\nu}_1(x),\ldots,\hat{\nu}_{n}(x)}(P)
           -\mathcal{D}_{P_{\infty},\hat{\mu}_1(x),
           \ldots,\hat{\mu}_{n}(x)}(P).
         \end{equation}
 That means, $P_0,\hat{\nu}_1(x),\ldots,\hat{\nu}_{n}(x)$ are the
 $n+1$ zeros of $\phi(P,x)$ and
 $P_{\infty},\hat{\mu}_1(x), \ldots,$ $\hat{\mu}_{n}(x)$ are its $n+1$ poles.
 These zeros and poles can be abbreviated in the following form
            \begin{equation}\label{3.20}
                \underline{\hat{\mu}}=\{\hat{\mu}_1,\ldots,\hat{\mu}_n\},
                \quad
                \underline{\hat{\nu}}=\{\hat{\nu}_1,\ldots,\hat{\nu}_n\}
                \in \mathrm{Sym}^n(\mathcal{K}_n).
            \end{equation}
 Let us recall the holomorphic map (\ref{3.2}),
 \begin{eqnarray}\label{3.21}
       && \ast: \begin{cases}
                        \mathcal{K}_{n}\rightarrow\mathcal{K}_{n},
                       \\
                       P=(z,y_j(z))\rightarrow
                       P^\ast=(z,y_{j+1(\mathrm{mod}~
                       2)})(z)), \quad j=0,1,
                      \end {cases}
                     \nonumber \\
      && P^{\ast \ast}:=(P^\ast)^\ast, \quad \mathrm{etc}.,
       \end{eqnarray}
where $y_j(z),\, j=0,1$ satisfy $\mathcal{F}_{n}(z,y)=0$, namely,
        \begin{equation}\label{3.22}
          (y-y_0(z))(y-y_1(z))
          =y^2-R_{2n+1}(z)=0.
        \end{equation}
From (\ref{3.22}), we can easily get
        \begin{equation}\label{3.23}
           \begin{split}
             & y_0+y_1=0,\\
             & y_0y_1=-R_{2n+1}(z),\\
             & y_0^2+y_1^2=2R_{2n+1}(z).\\
           \end{split}
        \end{equation}
Further properties of $\phi(P,x)$ are summarized as follows.

 \newtheorem{lem3.1}{Lemma}[section]
   \begin{lem3.1}
    Under the assumption $(\ref{2.1})$, let $P=(z,y)\in \mathcal{K}_{n}\setminus \{P_{\infty},P_0\}$ and
    $(z,x)\in \mathbb{C}^2$, and $u$ satisfy the $n$th
    stationary MCH equation $(\ref{2.28})$.  Then
     \begin{equation}\label{3.24}
        \phi_x(P)+\frac{1}{2}(u-u_{xx})z^{-1}\phi(P)^2
        -\phi(P)=-\frac{1}{2}(u-u_{xx}),
     \end{equation}
     \begin{equation}\label{3.25}
        \phi(P)\phi(P^\ast)=-\frac{zH_n(z)}{F_n(z)},
     \end{equation}
     \begin{equation}\label{3.26}
        \phi(P)+\phi(P^\ast)=2\frac{zG_n(z)}{F_n(z)},
     \end{equation}
     \begin{equation}\label{3.27}
        \phi(P)-\phi(P^\ast)=z^{1/2}\frac{2y}{F_n(z)}.
     \end{equation}
   \end{lem3.1}
\textbf{Proof.}~~A direct calculation shows that (\ref{3.24}) holds.
 Let us now prove (\ref{3.25})-(\ref{3.27}). Without loss of generality,
 let $y_0(P)=y(P)$.
 From (\ref{3.7}), (\ref{2.17}) and (\ref{3.23}), we arrive at
    \begin{eqnarray}\label{3.28}
      \phi(P)\phi(P^\ast) &=& \sqrt{z}\frac{y_0+\sqrt{z}G_n}{F_n}
      ~\times ~ \sqrt{z}\frac{y_1+\sqrt{z}G_n}{F_n}
        \nonumber \\
      &=& z\frac{y_0y_1+(y_0+y_1)\sqrt{z}G_n+zG_n^2}{F_n^2}
        \nonumber \\
      &=& z\frac{-R_{2n+1}+zG_n^2}{F_n^2}
         \nonumber \\
      &=& z\frac{-F_nH_n}{F_n^2}
         \nonumber \\
      &=& -\frac{zH_n}{F_n},
    \end{eqnarray}

   \begin{eqnarray}\label{3.29}
    \phi(P)+\phi(P^\ast) &=&  \sqrt{z}\frac{y_0+\sqrt{z}G_n}{F_n}
      ~+ ~ \sqrt{z}\frac{y_1+\sqrt{z}G_n}{F_n}
          \nonumber \\
      &=& \sqrt{z} \frac{(y_0+y_1)+2\sqrt{z}G_n}{F_n}
           \nonumber \\
      &=& \frac{2zG_n}{F_n},
   \end{eqnarray}

  \begin{eqnarray}\label{3.30}
    \phi(P)-\phi(P^\ast) &=&  \sqrt{z}\frac{y_0+\sqrt{z}G_n}{F_n}
      ~-~ \sqrt{z}\frac{y_1+\sqrt{z}G_n}{F_n}
          \nonumber \\
      &=& \sqrt{z} \frac{(y_0-y_1)}{F_n}
           \nonumber \\
      &=& \sqrt{z}\frac{2y_0}{F_n}= \sqrt{z}\frac{2y}{F_n}.
   \end{eqnarray}
Let us discuss properties of $\psi(P,x,x_0)$ below.

\newtheorem{lem3.2}[lem3.1]{Lemma}
  \begin{lem3.2}
    Under the assumption $(\ref{2.1})$, let $P=(z,y)\in \mathcal{K}_{n}\setminus \{P_{\infty},P_0\}$,
    $(x,x_0)\in \mathbb{C}^2$,  $u$ satisfy the $n$th
    stationary MCH equation $(\ref{2.28})$.
    Then
    \begin{equation}\label{3.31}
        \psi_1(P,x,x_0)=\Big(\frac{F_n(z,x)}{F_n(z,x_0)}\Big)^{1/2}
         \mathrm{exp}\Big(\frac{y}{2\sqrt{z}}
          \int_{x_0}^x q(x^\prime)F_n(z,x^\prime)^{-1} dx^\prime
          \Big),
    \end{equation}
    \begin{equation}\label{3.32}
        \psi_1(P,x,x_0)\psi_1(P^\ast,x,x_0)=\frac{F_n(z,x)}{F_n(z,x_0)},
    \end{equation}
    \begin{equation}\label{3.33}
        \psi_2(P,x,x_0)\psi_2(P^\ast,x,x_0)=-\frac{H_n(z,x)}{F_n(z,x_0)},
    \end{equation}
    \begin{equation}\label{3.34}
        \psi_1(P,x,x_0)\psi_2(P^\ast,x,x_0)
        +\psi_1(P^\ast,x,x_0)\psi_2(P,x,x_0)
        =2\frac{\sqrt{z}G_n(z,x)}{F_n(z,x_0)},
    \end{equation}
    \begin{equation}\label{3.35}
        \psi_1(P,x,x_0)\psi_2(P^\ast,x,x_0)
        -\psi_1(P^\ast,x,x_0)\psi_2(P,x,x_0)
        =-\frac{2y}{F_n(z,x_0)}.
    \end{equation}
  \end{lem3.2}
\textbf{Proof.}~~Equation (\ref{3.31}) can be proven through the following procedure.
Using (\ref{2.13}), the expression of $\psi_1$, (\ref{3.6}) and
(\ref{3.7}), we obtain
      \begin{eqnarray}\label{3.36}
        \psi_1(P,x,x_0) &=& \mathrm{exp}\left(z^{-1}\int_{x_0}^x
         \frac{1}{2}q(x^\prime)  \frac{\sqrt{z}(y+\sqrt{z}G_n)}{F_n} dx^\prime -
         \frac{1}{2}(x-x_0)\right)
          \nonumber \\
        &=& \mathrm{exp}\left(\frac{1}{\sqrt{z}}\int_{x_0}^x
         \frac{1}{2}q(x^\prime)  \frac{y+\sqrt{z}G_n}{F_n} dx^\prime -
         \frac{1}{2}(x-x_0)\right)
           \nonumber \\
        &=& \mathrm{exp}\left(\frac{1}{\sqrt{z}}\int_{x_0}^x
         \Big[ \frac{1}{2}q(x^\prime)  \frac{y}{F_n}
         +\frac{1}{2}\frac{F_{n,x^\prime}}{F_n} \Big] dx^\prime \right)
        \end{eqnarray}
which implies (\ref{3.31}). Moreover, (\ref{3.6}) together with
(\ref{3.28})-(\ref{3.30}) leads to
    \begin{eqnarray}\label{3.37}
      \psi_1(P)\psi_1(P^\ast) &=&
       \mathrm{exp}\left(z^{-1}\int_{x_0}^x
         \frac{1}{2}q(x^\prime) (\phi(P)+\phi(P^\ast)) dx^\prime -
         (x-x_0)\right)
          \nonumber \\
       &=& \mathrm{exp}\left(z^{-1}\int_{x_0}^x
         \frac{1}{2}q(x^\prime) (\frac{2zG_n}{F_n}) dx^\prime -
         (x-x_0)\right)
          \nonumber \\
       &=& \mathrm{exp}\left(\int_{x_0}^x
          \frac{F_{n,x^\prime}}{F_n} dx^\prime
         \right)
          \nonumber \\
       &=& \frac{F_n(z,x)}{F_n(z,x_0)},
    \end{eqnarray}
    \begin{eqnarray}\label{3.38}
       \psi_2(P,x,x_0)\psi_2(P^\ast,x,x_0) &=& z^{-1}
       \psi_1(P,x,x_0)\phi(P,x)\psi_1(P^\ast,x,x_0)\phi(P^\ast,x)
         \nonumber \\
       &=& z^{-1} \frac{F_n(z,x)}{F_n(z,x_0)}\frac{(-zH_n(z,x))}{F_n(z,x)}
          \nonumber \\
       &=& -\frac{H_n(z,x)}{F_n(z,x_0)},
    \end{eqnarray}
    \begin{eqnarray}\label{3.39}
     &&
      \psi_1(P,x,x_0)\psi_2(P^\ast,x,x_0)+\psi_1(P^\ast,x,x_0)\psi_2(P,x,x_0)
        \nonumber \\
    && ~~~~~~~~
       =\psi_1(P)\psi_1(P^\ast)\phi(P^\ast)/z^{1/2}+
        \psi_1(P^\ast)\psi_1(P)\phi(P)/z^{1/2}
         \nonumber \\
    && ~~~~~~~~
       =\psi_1(P)\psi_1(P^\ast)(\phi(P)+\phi(P^\ast))/z^{1/2}
          \nonumber \\
    && ~~~~~~~~
       =\frac{F_n(z,x)}{F_n(z,x_0)}\frac{2zG_n(z,x)}{F_n(z,x)}/z^{1/2}
         \nonumber \\
    && ~~~~~~~~
       =z^{1/2}\frac{2G_n(z,x)}{F_n(z,x_0)},
    \end{eqnarray}
    \begin{eqnarray}\label{3.40}
    &&
      \psi_1(P,x,x_0)\psi_2(P^\ast,x,x_0)-\psi_1(P^\ast,x,x_0)\psi_2(P,x,x_0)
        \nonumber \\
    && ~~~~~~~~
       =\psi_1(P)\psi_1(P^\ast)\phi(P^\ast)/z^{1/2}-
        \psi_1(P^\ast)\psi_1(P)\phi(P)/z^{1/2}
         \nonumber \\
    && ~~~~~~~~
       =\psi_1(P)\psi_1(P^\ast)(\phi(P^\ast)-\phi(P))/z^{1/2}
          \nonumber \\
    && ~~~~~~~~
       =\frac{F_n(z,x)}{F_n(z,x_0)}\frac{-2z^{1/2}y}{F_n(z,x)}/z^{1/2}
         \nonumber \\
    && ~~~~~~~~
       =-\frac{2y}{F_n(z,x_0)}.
    \end{eqnarray}

In Lemma 3.2, if we choose
      $$\psi_1(P)=\psi_{1,+}, ~ \psi_1(P^\ast)=\psi_{1,-},~
       \psi_2(P)=\psi_{2,+}, ~ \psi_2(P^\ast)=\psi_{2,-},$$
then (\ref{3.32})-(\ref{3.35}) imply 
     \begin{equation}\label{3.41}
        (\psi_{1,+}\psi_{2,-}-\psi_{1,-}\psi_{2,+})^2=
          (\psi_{1,+}\psi_{2,-}+\psi_{1,-}\psi_{2,+})^2
          -4\psi_{1,+}\psi_{2,-}\psi_{1,-}\psi_{2,+},
     \end{equation}
which is equivalent to the basic identity (\ref{2.17}),
$zG_n^2+F_nH_n=R_{2n+1}$. This fact reveals the relations between
our approach and the algebro-geometric solutions of the MCH
hierarchy.

\newtheorem{rem3.3}[lem3.1]{Remark}
  \begin{rem3.3}
    The definition of stationary Baker-Akhiezer function $\psi$ of
    the MCH hierarchy is analogous to that in the context of KdV or AKNS
    hierarchies. But the crucial difference is that $P_0$ is an
    essential singularity of $\psi$ in the MCH hierarchy, which is the same as in
    the CH hierarchy, but different from the KdV or AKNS
    hierarchy. This fact will be shown in the asymptotic expansions
    of $\psi$ in next section.
  \end{rem3.3}

   Furthermore, we derive Dubrovin-type equations, which are first-order
   coupled systems of differential equations and govern the dynamics
   of the zeros $\mu_j(x)$ and $\nu_j(x)$ of $F_n(z,x)$
   and $H_n(z,x)$ with respect to $x$. We recall that the affine part of
   $\mathcal{K}_n$ is nonsingular if
     \begin{equation}\label{3.42}
        \{\widetilde{E}_m\}_{m=1,\ldots,2n+1} \subset \mathbb{C},
        \quad \widetilde{E}_m \neq \widetilde{E}_{m^\prime}
        \quad \textrm{for $m \neq m^\prime,~
        m,m^\prime=1,\ldots,2n+1$}.
     \end{equation}

 \newtheorem{lem3.4}[lem3.1]{Lemma}
    \begin{lem3.4}
    Assume that $(\ref{2.1})$ holds and $u$ satisfies the $n$th
    stationary MCH equation $(\ref{2.28})$.\\

        $(\mathrm{i})$
     If the zeros $\{\mu_j(x)\}_{j=1,\ldots,n}$
     of $F_n(z,x)$ remain distinct for $ x \in
     \Omega_\mu,$ where $\Omega_\mu \subseteq \mathbb{C}$ is open
     and connected, then
     $\{\mu_j(x)\}_{j=1,\ldots,n}$ satisfy the system of
     differential equations
    \begin{equation}\label{3.43}
           \mu_{j,x}=\frac{(u-u_{xx})y(\hat{\mu}_j)}{(u_x-u)\sqrt{\mu_j}}
            \prod_{\scriptstyle k=1 \atop \scriptstyle k \neq j }^{n}
            (\mu_j(x)-\mu_k(x))^{-1}, \quad j=1,\ldots,n,
        \end{equation}
   with initial conditions
       \begin{equation}\label{3.44}
         \{\hat{\mu}_j(x_0)\}_{j=1,\ldots,n}
         \in \mathcal{K}_{n},
       \end{equation}
   for some fixed $x_0 \in \Omega_\mu$. The initial value
   problems $(\ref{3.43})$, $(\ref{3.44})$ have a unique solution
   satisfying
        \begin{equation}\label{3.45}
         \hat{\mu}_j \in C^\infty(\Omega_\mu,\mathcal{K}_{n}),
         \quad j=1,\ldots,n.
        \end{equation}

        $(\mathrm{ii})$
     If the zeros $\{\nu_j(x)\}_{j=1,\ldots,n}$
     of $H_n(z,x)$ remain distinct for $ x \in
     \Omega_\nu,$ where $\Omega_\nu \subseteq \mathbb{C}$ is open
     and connected, then
     $\{\nu_j(x)\}_{j=1,\ldots,n}$ satisfy the system of
     differential equations
    \begin{equation}\label{3.46}
           \nu_{j,x}=-\frac{(u-u_{xx})y(\hat{\nu}_j)}{(u_x+u)\sqrt{\nu_j}}
            \prod_{\scriptstyle k=1 \atop \scriptstyle k \neq j }^{n}
            (\nu_j(x)-\nu_k(x))^{-1}, \quad j=1,\ldots,n,
        \end{equation}
   with initial conditions
       \begin{equation}\label{3.47}
         \{\hat{\nu}_j(x_0)\}_{j=1,\ldots,n}
         \in \mathcal{K}_{n},
       \end{equation}
   for some fixed $x_0 \in \Omega_\nu$. The initial value
   problems $(\ref{3.46})$, $(\ref{3.47})$ have a unique solution
   satisfying
        \begin{equation}\label{3.48}
         \hat{\nu}_j \in C^\infty(\Omega_\nu,\mathcal{K}_{n}),
         \quad j=1,\ldots,n.
        \end{equation}
    \end{lem3.4}
\textbf{Proof.}~~For our convenience, let us 
focus on
(\ref{3.43}) and (\ref{3.45}). The proof of (\ref{3.46}) and
(\ref{3.48}) follows in an identical manner. The derivatives of
(\ref{3.11}) with respect to $x$ take on
          \begin{equation}\label{3.49}
            F_{n,x}(\mu_j)= -(u_x-u)\mu_{j,x}
             \prod_{\scriptstyle k=1 \atop \scriptstyle k \neq j }^{n}
            (\mu_j(x)-\mu_k(x)).
          \end{equation}
On the other hand, inserting $z=\mu_j$ into equation (\ref{2.13}) leads to
       \begin{equation}\label{3.50}
        F_{n,x}(\mu_j)=(u-u_{xx})G_n(\mu_j)
        = (u-u_{xx})\frac{y(\hat{\mu}_j)}{-\sqrt{\mu_j}}.
       \end{equation}
Comparing (\ref{3.49}) with (\ref{3.50}) gives  ({\ref{3.43}). The
  smoothness assertion (\ref{3.45}) is clear as long as
  $\hat{\mu}_j$ stays away from the branch points $(\widetilde{E}_m,0)$. In case
  $\hat{\mu}_j$ hits such a branch point, one can use the local
  chart around $(\widetilde{E}_m,0)$
  ($\zeta=\sigma(z-\widetilde{E}_m)^{1/2}$, $\sigma=\pm 1$)
  to verify (\ref{3.45}). \quad $\square$ \\

 Let us now turn to the trace formulas of the MCH invariants, which are
 the expressions of $f_{2l}$ and $h_{2l}$ in terms of symmetric
 functions of the zeros $\mu_j$ and $\nu_j$ of $F_n$ and $H_n$.
 Here, we just consider the simplest case.

 \newtheorem{lem3.5}[lem3.1]{Lemma}
   \begin{lem3.5}
    If $(\ref{2.1})$  holds and $u$ satisfies the $n$th
    stationary MCH equation $(\ref{2.28})$, then
      \begin{equation}\label{3.51}
        -\frac{1}{2}(u-u_{xx})(u^2-u_x^2)
        +\frac{1}{2}(u-u_{xx})\sum_{m=1}^{2n+1}\widetilde{E}_m=
        (u-u_x)\sum_{j=1}^n \mu_j,
      \end{equation}
      \begin{equation}\label{3.52}
         \frac{1}{2}(u-u_{xx})(u^2-u_x^2)
        -\frac{1}{2}(u-u_{xx})\sum_{m=1}^{2n+1}\widetilde{E}_m=
        -(u+u_x)\sum_{j=1}^n \nu_j,
      \end{equation}
   \end{lem3.5}
\textbf{Proof.}~~ By comparison of the coefficients of $z^{n-1}$
($\lambda^{2n-2}$) in (\ref{2.22}) and
(\ref{2.24}), taking account into (\ref{3.11}) and
(\ref{3.12}) yields
       \begin{equation}\label{3.53}
         -\frac{1}{2}(u-u_{xx})(u^2-u_x^2)
        +(u-u_{xx})c_1=
        (u-u_x)\sum_{j=1}^n \mu_j,
       \end{equation}
        \begin{equation}\label{3.54}
         \frac{1}{2}(u-u_{xx})(u^2-u_x^2)
        -(u-u_{xx})c_1=-
        (u+u_x)\sum_{j=1}^n \nu_j,
       \end{equation}
On the other hand, considering the coefficient of $z^{2n}$
($\lambda^{4n}$) in  $\lambda^2G_n^2+F_nH_n=R_{2n+1}$ leads
to
      \begin{equation}\label{3.55}
        2g_0g_2+h_0f_0=-\sum_{m=1}^{2n+1}\widetilde{E}_m,
      \end{equation}
which implies
       \begin{equation}\label{3.56}
        c_1=\frac{1}{2}\sum_{m=1}^{2n+1}\widetilde{E}_m.
       \end{equation}
The trace formulas in the MCH hierarchy are implicit for the
potential $u$, which is apposed to the general integrable soliton
equations such as KdV, AKNS and CH hierarchies. But this dose not
affect to obtain the algebro-geometric solutions of the MCH
hierarchy, and we can still construct $u$ from trace formulas.

\section{Algebro-geometric solutions of the stationary MCH hierarchy}
In this section, we continue our study of the stationary MCH
hierarchy, and will obtain explicit Riemann theta function
representations for the meromorphic function $\phi$, and in particular
for the potentials $u$ of the stationary MCH hierarchy.

Let us begin with the asymptotic properties of $\phi$ and
$\psi_j,j=1,2$.

\newtheorem{lem4.1}{Lemma}[section]
   \begin{lem4.1}
    Assume that $(\ref{2.1})$ holds and $u$ satisfies the $n$th
    stationary MCH equation $(\ref{2.28})$. Moreover, let $P=(z,y)
    \in \mathcal{K}_n \setminus \{P_\infty,P_0\},$ $(x,x_0) \in
    \mathbb{C}^2$. Then
     \begin{equation}\label{4.1}
        \phi(P)\underset{\zeta \rightarrow 0}{=}
        \frac{2}{u-u_x}\zeta^{-1}+O(1),
        \qquad P \rightarrow P_\infty, \qquad \zeta=z^{-1},
     \end{equation}
     \begin{equation}\label{4.2}
        \phi(P)\underset{\zeta \rightarrow 0}{=}
         i\zeta+O(\zeta^2),
        \qquad P \rightarrow P_0, \qquad \zeta=z^{1/2},
     \end{equation}
      \begin{equation}\label{4.3}
         \begin{split}
        & \psi_1(P,x,x_0)\underset{\zeta \rightarrow 0}{=}
        \mathrm{exp} \Big( \int_{x_0}^{x}
         (\frac{u-u_{xx}}{u-u_x} + O(\zeta) )~dx^\prime -\frac{1}{2}(x-x_0)
         \Big),  \\
         & ~~~~~~~~~~~~~~~~~~~~~~~~~~~~~~~~~~
          P \rightarrow P_\infty, \quad \zeta=z^{-1},
         \end{split}
      \end{equation}
      \begin{equation}\label{4.4}
         \begin{split}
      & \psi_2(P,x,x_0)\underset{\zeta \rightarrow 0}{=}
        \Big(\frac{2}{u-u_x}\zeta^{-1/2}+O(\zeta^{1/2})\Big)
        ~\mathrm{exp} \Big( \int_{x_0}^{x}
         (\frac{u-u_{xx}}{u-u_x} + O(\zeta) )~dx^\prime \\
         & ~~~~~~~~~~~~~~~~~~~-\frac{1}{2}(x-x_0) \Big), \qquad
         P \rightarrow P_\infty, \quad \zeta=z^{-1},
         \end{split}
      \end{equation}
  and
       \begin{equation}\label{4.5}
        \psi_1(P,x,x_0)\underset{\zeta \rightarrow 0}{=}
         \mathrm{exp} \Big( ~\frac{1}{\zeta} \int_{x_0}^x
          \frac{1}{2}(u-u_{xx})(i+O(1))~ dx^\prime +O(1) \Big),
         \end{equation}
         \begin{equation*}
             P \rightarrow P_0, \quad \zeta=z^{1/2},
         \end{equation*}
        \begin{equation}\label{4.6}
          \psi_2(P,x,x_0)\underset{\zeta \rightarrow 0}{=}
          (i+O(\zeta))
           ~\mathrm{exp} \Big( ~\frac{1}{\zeta} \int_{x_0}^x
          \frac{1}{2}(u-u_{xx})(i+O(1))~ dx^\prime +O(1) \Big),
             \end{equation}
          \begin{equation*}
           P \rightarrow P_0, \quad \zeta=z^{1/2}.
        \end{equation*}
   \end{lem4.1}
\textbf{Proof.}~~Under the local coordinates $\zeta=z^{-1}$
near $P_\infty$ and $\zeta=z^{1/2}$ near $P_0$, the existence of the
asymptotic expansion of $\phi$ is clear from its explicit
expression in (\ref{3.7}). Next, we use the Riccati-type equation
(\ref{3.24}) to compute the explicit expansion coefficients.
Inserting the ansatz
  \begin{equation}\label{4.7}
     \phi \underset{z \rightarrow \infty}{=}
       \phi_{-1} z + \phi_0 +O(z^{-1})
  \end{equation}
into (\ref{3.24}) and comparing the powers of $z$, then we have 
(\ref{4.1}). Similarly, inserting the ansatz
  \begin{equation}\label{4.8}
     \phi \underset{z \rightarrow 0}{=}
       \phi_{1} z^{1/2} + \phi_2 z  +O(z^{3/2})
  \end{equation}
into (\ref{3.24}) and comparing the power of $z^0$, we obtain
(\ref{4.2}). Finally, expansions (\ref{4.3})-(\ref{4.6}) follow up by 
(\ref{3.6}), (\ref{3.8}), (\ref{4.1}) and (\ref{4.2}). \qquad
$\square$

\newtheorem{rem4.2}[lem4.1]{Remark}
  \begin{rem4.2}
    We notice the unusual fact: $P_0$ is the essential singularity
    of $\psi_j$, $j=1,2$. From $(\ref{4.5})$ and $(\ref{4.6})$, the
    leading-order exponential term $\psi_j$, $j=1,2,$ near $P_0$ is
    $x$-dependent, which makes the problem worse. One can obtain the
    analogous result near $P_\infty$ as displayed in $(\ref{4.3})$
    and $(\ref{4.4})$. This is in sharp contrast to standard
    Baker-Akhiezer functions that typically feature a linear
    behavior with respect to $x$, such as
    $\mathrm{exp}(c(x-x_0))$ and $\mathrm{exp}(c(x-x_0)\zeta^{-1})$
    near $P_\infty$ and $P_0$, respectively.
  \end{rem4.2}
Let us now introduce the holomorphic differentials $\eta_l(P)$ on
$\mathcal{K}_{n}$ 
      \begin{equation}\label{4.9}
        \eta_l(P)=\frac{z^{l-1}}{\sqrt{z}y(P)} dz,
        \qquad l=1,\ldots,n,
      \end{equation}
and choose a homology basis $\{a_j,b_j\}_{j=1}^{n}$ on
$\mathcal{K}_{n}$ in such a way that the intersection matrix of the
cycles satisfies
$$a_j \circ b_k =\delta_{j,k},\quad a_j \circ a_k=0, \quad
b_j \circ   b_k=0, \quad j,k=1,\ldots, n.$$ Define an invertible
matrix $E \in GL(n, \mathbb{C})$ as
    follows
       \begin{equation}\label{4.10}
          \begin{split}
        & E=(E_{j,k})_{n \times n}, \quad E_{j,k}=
           \int_{a_k} \eta_j, \\
        &  \underline{c}(k)=(c_1(k),\ldots, c_{n}(k)), \quad
           c_j(k)=(E^{-1})_{j,k},
           \end{split}
       \end{equation}
and the normalized holomorphic differentials as follows
        \begin{equation}\label{4.11}
          \omega_j= \sum_{l=1}^{n} c_j(l)\eta_l, \quad
          \int_{a_k} \omega_j = \delta_{j,k}, \quad
          \int_{b_k} \omega_j= \tau_{j,k}, \quad
          j,k=1, \ldots ,n.
        \end{equation}
Apparently, 
the matrix $\tau$ is symmetric and has a
positive-definite imaginary part.

The symmetric functions $\Phi_{n-1}^{(j)}(\underline{\mu})$ and
$\Psi_n(\underline{\mu})$ are defined by
     \begin{equation}\label{4.12}
        \Phi_{n-1}^{(j)}(\underline{\mu})=(-1)^{n-1}
          \prod_{\scriptstyle p=1 \atop \scriptstyle p \neq j}^n \mu_p,
     \end{equation}
     \begin{equation}\label{4.13}
        \Psi_n(\underline{\mu})=(-1)^n \prod_{p=1}^n \mu_p.
     \end{equation}

Let us present our results in the following theorem.

\newtheorem{the4.3}[lem4.1]{Theorem}
  \begin{the4.3}
    Assume that $(\ref{2.1})$ holds.

     $(\mathrm{i})$
     Suppose that
    $\{\hat{\mu}_j\}_{j=1,\ldots,n}$ satisfy the stationary
    Dubrovin equations $(\ref{3.43})$ on $\Omega_\mu$ and
    remain distinct for $ x \in
     \Omega_\mu,$ where $\Omega_\mu \subseteq \mathbb{C}$ is open
     and connected. Let the associated divisor be
       \begin{equation}\label{4.14}
        \mathcal{D}_{\underline{\hat{\mu}}(x)} \in \mathrm{Sym}^n
        (\mathcal{K}_n),
        \qquad
        \underline{\hat{\mu}}=\{\hat{\mu}_1,\ldots,\hat{\mu}_n\}
        \in  \mathrm{Sym}^n (\mathcal{K}_n).
       \end{equation}
 Then
        \begin{equation}\label{4.15}
            \partial_x
            \underline{\alpha}_{Q_0}(\mathcal{D}_{\underline{\hat{\mu}}(x)})
            =-\frac{u-u_{xx}}{u_x-u}
             \frac{1}{ \Psi_n(\underline{\mu}(x))}
             \underline{c}(1),
            \qquad    x \in \Omega_\mu.
        \end{equation}
 In particular, the Abel map does not linearize the divisor
 $\mathcal{D}_{\underline{\hat{\mu}}(x)}$ on $\Omega_\mu$.

  $(\mathrm{ii})$
     Suppose that
    $\{\hat{\nu}_j\}_{j=1,\ldots,n}$ satisfy the stationary
    Dubrovin equations $(\ref{3.46})$ on $\Omega_\nu$ and
    remain distinct for $ x \in
     \Omega_\nu,$ where $\Omega_\nu \subseteq \mathbb{C}$ is open
     and connected. Let the associated divisor be
       \begin{equation}\label{4.16}
        \mathcal{D}_{\underline{\hat{\nu}}(x)} \in \mathrm{Sym}^n
        (\mathcal{K}_n),
        \qquad
        \underline{\hat{\nu}}=\{\hat{\nu}_1,\ldots,\hat{\nu}_n\}
        \in  \mathrm{Sym}^n (\mathcal{K}_n).
       \end{equation}
 Then
        \begin{equation}\label{4.17}
            \partial_x
            \underline{\alpha}_{Q_0}(\mathcal{D}_{\underline{\hat{\nu}}(x)})
            =\frac{u-u_{xx}}{u_x+u}
             \frac{1}{ \Psi_n(\underline{\nu}(x))}
             \underline{c}(1),
            \qquad    x \in \Omega_\nu.
        \end{equation}
 In particular, the Abel map does not linearize the divisor
 $\mathcal{D}_{\underline{\hat{\nu}}(x)}$ on $\Omega_\nu$.
  \end{the4.3}
\textbf{Proof.}~~It is easy to see that
        \begin{equation}\label{4.18}
 \frac{1}{\mu_j}=
        \frac{\prod_{\scriptstyle p=1 \atop \scriptstyle p \neq j}^n \mu_p}
            {\prod_{p=1}^n \mu_p}
        =-\frac{
        \Phi_{n-1}^{(j)}(\underline{\mu})}{\Psi_n(\underline{\mu})},
        \quad j=1,\ldots,n.
        \end{equation}
Let
        \begin{equation}\label{4.19}
        \underline{\omega}=(\omega_1,\ldots,\omega_n),
        \end{equation}
and choose an appropriate base point $Q_0$. Then we have
      \begin{eqnarray}\label{4.20}
 \partial_x  \underline{\alpha}_{Q_0}(\mathcal{D}_{\underline{\hat{\mu}}(x)})
 &=&
 \partial_x  \Big(\sum_{j=1}^n \int_{Q_0}^{\hat{\mu}_j} \underline{\omega} \Big)
 =\sum_{j=1}^n \mu_{j,x} \sum_{k=1}^n \underline{c}(k)
  \frac{\mu_j^{k-1}}{\sqrt{\mu_j}y(\hat{\mu}_j)}
     \nonumber \\
  &=&
    \sum_{j=1}^n \sum_{k=1}^n \frac{u-u_{xx}}{u_x-u}
    \frac{\mu_j^{k-1}}{\mu_j}
    \frac{1}
    {\prod_{\scriptstyle l=1 \atop \scriptstyle l \neq j }^{n}(\mu_j-\mu_l)}
    \underline{c}(k)
     \nonumber \\
  &=&
    \sum_{j=1}^n \sum_{k=1}^n \frac{u-u_{xx}}{u_x-u}
        \frac{\mu_j^{k-2}}
    {\prod_{\scriptstyle l=1 \atop \scriptstyle l \neq j }^{n}(\mu_j-\mu_l)}
    \underline{c}(k)
     \nonumber \\
  &=&
    -\frac{1}{\Psi_n(\underline{\mu})} \frac{u-u_{xx}}{u_x-u}
     \sum_{j=1}^n \sum_{k=1}^n \underline{c}(k)
       \frac{\mu_j^{k-1}}
    {\prod_{\scriptstyle l=1 \atop \scriptstyle l \neq j }^{n}(\mu_j-\mu_l)}
     \Phi_{n-1}^{(j)}(\underline{\mu})
       \nonumber \\
  &=&
  -\frac{1}{\Psi_n(\underline{\mu})} \frac{u-u_{xx}}{u_x-u}
     \sum_{j=1}^n \sum_{k=1}^n \underline{c}(k)
     (U_n(\underline{\mu}))_{k,j}(U_n(\underline{\mu}))_{j,1}^{-1}
       \nonumber \\
  &=&
    -\frac{1}{\Psi_n(\underline{\mu})} \frac{u-u_{xx}}{u_x-u}
      \sum_{k=1}^n \underline{c}(k) \delta_{k,1}
      \nonumber \\
  &=&
      -\frac{1}{\Psi_n(\underline{\mu})} \frac{u-u_{xx}}{u_x-u}
        \underline{c}(1),
      \end{eqnarray}
where we used (E.25) and (E.26) in \cite{12},
    \begin{equation}\label{4.21}
      (U_n(\underline{\mu}))=
       \Big( \frac{\mu_j^{k-1}}
    {\prod_{\scriptstyle l=1 \atop \scriptstyle l \neq j }^{n}(\mu_j-\mu_l)}
    \Big)_{j,k=1}^n,
    \quad
     (U_n(\underline{\mu}))^{-1}=
     \Big(\Phi_{n-1}^{(j)}(\underline{\mu})\Big)_{j=1}^n.
    \end{equation}
The analogous results hold for the corresponding divisor
$\mathcal{D}_{\underline{\hat{\nu}}(x)}$, which can be obtained in the same
way. \quad $\square$ \\

Next, we give the special forms for Theorem 4.3, which can be used
to provide the proper change of variables to linearize the divisors
$\mathcal{D}_{\underline{\hat{\mu}}(x)}$ and
$\mathcal{D}_{\underline{\hat{\nu}}(x)}$ associated with
$\phi(P,x)$.
We only consider the case about the divisor
$\mathcal{D}_{\underline{\hat{\mu}}(x)}$. One may conclude that the
analogous results also hold for the other divisor
$\mathcal{D}_{\underline{\hat{\nu}}(x)}$.

Let us introduce \footnote{ Here we choose the same path of integration
from $Q_0$ and $P$ for all integrals in (\ref{4.22}) \\ and
(\ref{4.23}).}
   \begin{equation}\label{4.22}
     \begin{split}
    & \underline{\widehat{B}}_{Q_0}: \mathcal{K}_n \setminus
    \{P_\infty,P_0\} \rightarrow \mathbb{C}^n, \\
    & P \mapsto \underline{\widehat{B}}_{Q_0}(P)
    =(\widehat{B}_{Q_0,1},\ldots,\widehat{B}_{Q_0,n}) \\
    & ~~~~~~~~~~~~~~~~~
     =
      \begin{cases}
       \int_{Q_0}^P \tilde{\omega}_{P_\infty,P_0}^{(3)},
         ~~~~~~~~~~~~~n=1 \\
       \Big(\int_{Q_0}^P \eta_2,\ldots, \int_{Q_0}^P \eta_n,
           \int_{Q_0}^P \tilde{\omega}_{P_\infty,P_0}^{(3)}
       \Big),~~~~~~~~~~~ n \geq 2,
      \end{cases}
     \end{split}
   \end{equation}
where
$$\tilde{\omega}_{P_\infty,P_0}^{(3)}=
\frac{z^{n}}{\sqrt{z}y(P)}dz.$$ and
   \begin{equation}\label{4.23}
     \begin{split}
       & \underline{\hat{\beta}}_{Q_0}:
         \mathrm{Sym}^n(
       \mathcal{K}_n \setminus
    \{P_\infty,P_0\}) \rightarrow \mathbb{C}^n, \\
      & \mathcal{D}_{\underline{Q}} \mapsto
      \underline{\hat{\beta}}_{Q_0}(\mathcal{D}_{\underline{Q}})
      =\sum_{j=1}^n \underline{\widehat{B}}_{Q_0}(Q_j), \\
      &
      \underline{Q}=\{Q_1,\ldots,Q_n\}
      \in \mathrm{Sym}^n( \mathcal{K}_n \setminus
    \{P_\infty,P_0\}).
     \end{split}
   \end{equation}

\newtheorem{the4.4}[lem4.1]{Theorem}
  \begin{the4.4}
    Assume that $(\ref{2.1})$ holds, and
    the statements of $\mu_j$ in Theorem $4.3$ are all true. 
    Then
    \begin{equation}\label{4.24}
        \partial_x \sum_{j=1}^n \int_{Q_0}^{\hat{\mu}_j(x)} \eta_1
        = -\frac{1}{\Psi_n(\underline{\mu}(x))} \frac{u-u_{xx}}{u_x-u},
        \qquad x\in \Omega_\mu,
    \end{equation}
    \begin{equation}\label{4.25}
        \partial_x \underline{\hat{\beta}}
        (\mathcal{D}_{\underline{\hat{\mu}}(x)})
        =
        \begin{cases}
         \frac{u-u_{xx}}{u_x-u},
         ~~~~~~~~n=1,\\
          \frac{u-u_{xx}}{u_x-u}(0,\ldots,0,1),
          ~~~~~~~~n\geq 2,
        \end{cases}
        \quad  x\in \Omega_\mu.
    \end{equation}
  \end{the4.4}
\textbf{Proof.}~~ Equation (\ref{4.24}) is a special case
of (\ref{4.15}), and (\ref{4.25}) follows from (\ref{4.20}).
Alternatively, one can follow the same way as shown in Theorem 4.3 to derive
(\ref{4.24}) and (\ref{4.25}). \quad   $\square$ \\

Let $\theta(\underline{z})$ denote the Riemann theta function
associated with $\mathcal{K}_{n}$ and an appropriately fixed
homology basis. We assume $\mathcal{K}_{n}$ to be nonsingular.
Next, choosing a convenient base
point $Q_0 \in \mathcal{K}_{n} \setminus \{P_{\infty},P_0\}$, the
vector of Riemann constants $\underline{\Xi}_{Q_0}$ is given by
(A.66) \cite{12}, and the Abel maps $\underline{A}_{Q_0}(\cdot) $ and
$\underline{\alpha}_{Q_0}(\cdot)$ are defined by
  \begin{equation}\label{4.26}
    \begin{split}
    &
   \underline{A}_{Q_0}:\mathcal{K}_{n} \rightarrow
   J(\mathcal{K}_{n})=\mathbb{C}^{n}/L_{n}, \\
   &
   P \mapsto \underline{A}_{Q_0} (P)=(\underline{A}_{Q_0,1}(P),\ldots,
  \underline{A}_{Q_0,n} (P))
    =\left(\int_{Q_0}^P\omega_1,\ldots,\int_{Q_0}^P\omega_{n}\right)
  (\mathrm{mod}~L_{n})
  \end{split}
  \end{equation}
   and
   \begin{equation}\label{4.27}
     \begin{split}
   & \underline{\alpha}_{Q_0}:
   \mathrm{Div}(\mathcal{K}_{n}) \rightarrow
   J(\mathcal{K}_{n}),\\
   & \mathcal{D} \mapsto \underline{\alpha}_{Q_0}
   (\mathcal{D})= \sum_{P\in \mathcal{K}_{n}}
    \mathcal{D}(P)\underline{A}_{Q_0} (P),
    \end{split}
    \end{equation}
where $L_{n}=\{\underline{z}\in \mathbb{C}^{n}|
           ~\underline{z}=\underline{N}+\tau\underline{M},
           ~\underline{N},~\underline{M}\in \mathbb{Z}^{n}\}.$ \\

Let
   \begin{equation}\label{4.28}
    \omega_{P_{\infty} P_0}^{(3)}(P)=
     \frac{1}{\sqrt{z}y} \prod_{j=1}^n(z-\lambda_j) dz
    \end{equation}
be the normalized differential of the third kind  holomorphic on
$\mathcal{K}_{n} \setminus \{P_{\infty},P_0\}$ with simple poles at
$P_{\infty}$ and $P_0$ with residues $\pm 1$, 
that is,
 \begin{equation}\label{4.29}
            \begin{split}
              & \omega_{P_{\infty} P_0}^{(3)}(P) \underset
              {\zeta \rightarrow 0}{=} (\zeta^{-1}+O(1))d \zeta,
              \quad \textrm{as $P \rightarrow P_{\infty},$}\\
              & \omega_{P_{\infty} P_0}^{(3)}(P) \underset
              {\zeta \rightarrow 0}{=} (-\zeta^{-1}+O(1))d \zeta,
              \quad \textrm{as $P \rightarrow P_0,$}
            \end{split}
         \end{equation}
where the local coordinate are given by
    \begin{equation}\label{4.30}
        \zeta=z^{-1}
        ~~ \textrm{for $P$ near $P_\infty$},
        \qquad
        \zeta=z^{1/2}
        ~~\textrm{for $P$ near $P_0$},
    \end{equation}
and the constants $\{\lambda_j\}_{j=1,\ldots,n}$ are determined by
the normalization condition
      $$\int_{a_k} \omega_{P_{\infty} P_0}^{(3)}=0,
      \qquad k=1,\ldots,n.$$
Then, we have
 \begin{equation} \label{4.31}
   \int_{Q_0}^P \omega_{P_{\infty} P_0}^{(3)}(P) \underset
  {\zeta \rightarrow 0}{=} \mathrm{ln} \zeta +
  e_0+O(\zeta),
  \quad \textrm{as $P \rightarrow P_{\infty},$}
  \end{equation}
  \begin{equation}\label{4.32}
    \int_{Q_0}^P \omega_{P_{\infty} P_0}^{(3)}(P) \underset
    {\zeta \rightarrow 0}{=} -\mathrm{ln} \zeta +
    d_0+O(\zeta),
     \quad \textrm{as $P \rightarrow P_0,$}
    \end{equation}
where constants $e_0,d_0 \in \mathbb{C}$ arise from the
integrals at their lower limits $Q_0$. We also notice 
  \begin{equation}\label{4.33}
   \underline{A}_{Q_0}(P)-\underline{A}_{Q_0}(P_\infty)
   \underset {\zeta \rightarrow 0}{=}
   \underline{U}\zeta+O(\zeta^2),
   \quad \textrm{as $P \rightarrow P_{\infty},$}
   \quad \underline{U}=\underline{c}(n),
  \end{equation}
  \begin{equation}\label{4.34}
   \underline{A}_{Q_0}(P)-\underline{A}_{Q_0}(P_0)
   \underset {\zeta \rightarrow 0}{=}
   -2\underline{U}\zeta+O(\zeta^2),
   \quad \textrm{as $P \rightarrow P_{0},$}
   \quad \underline{U}=\underline{c}(n).
  \end{equation}
The following abbreviations are used for our convenience:
    \begin{eqnarray}\label{4.35}
         &&
           \underline{z}(P,\underline{Q})= \underline{\Xi}_{Q_0}
           -\underline{A}_{Q_0}(P)+\underline{\alpha}_{Q_0}
             (\mathcal{D}_{\underline{Q}}), \nonumber \\
          &&
           P\in \mathcal{K}_{n},\,
           \underline{Q}=(Q_1,\ldots,Q_{n})\in
           \mathrm{Sym}^{n}(\mathcal{K}_{n}),
         \end{eqnarray}
where $\underline{z}(\cdot,\underline{Q}) $ is independent of the
choice of base point $Q_0$. \\

Moreover, from Theorems 4.3 and 4.4 we note that the Abel map
dose not linearize the divisor
$\mathcal{D}_{\underline{\hat{\mu}}(x)}$. However,
the change of variables
     \begin{equation}\label{4.36}
        x \mapsto \tilde{x}= \int^x dx^\prime
        \Big( \frac{1}{\Psi_n(\underline{\mu}(x^\prime))}
         \frac{u-u_{x^\prime x^\prime}}{u_{x^\prime}-u}
        \Big)
     \end{equation}
linearizes the Abel map
$\underline{A}_{Q_0}(\mathcal{D}_{\underline{\hat{\tilde{\mu}}}(\tilde{x})}),$
$\tilde{\mu}_j(\tilde{x})=\mu_j(x),j=1,\ldots,n.$ The intricate
relation between the variable $x$ and $\tilde{x}$ is discussed detailedly in
Theorem 4.5.

Based on the above all these preparations, let us now give an explicit representation
for the meromorphic function $\phi$ and the  solution
$u$ of the stationary MCH equations in terms of the Riemann theta function associated with
$\mathcal{K}_{n}$. Here we assume the affine part of $\mathcal{K}_{n}$ ($n\in \mathbb{N}$) to
be nonsingular.

\newtheorem{the4.5}[lem4.1]{Theorem}
 \begin{the4.5}
 Assume that the curve $\mathcal{K}_{n}$ is nonsingular,
$(\ref{2.1})$ holds, and $u$ satisfies the $n$th
stationary MCH equation $(\ref{2.28})$ on $\Omega$. Let
$P=(z,y) \in \mathcal{K}_n \setminus \{P_\infty,P_0\},$ and $x \in
\Omega$, where $ \Omega \subseteq \mathbb{C}$ is open and connected.
In addition, suppose that $\mathcal{D}_{\underline{\hat{\mu}}(x)}$,
or equivalently $\mathcal{D}_{\underline{\hat{\nu}}(x)}$ is
nonspecial for $x\in \Omega$. Then, $\phi$ and $u$ have the following
representations
  \begin{equation}\label{4.37}
    \phi(P,x)=i
    \frac{\theta(\underline{z}(P,\underline{\hat{\nu}}(x)))
            \theta(\underline{z}(P_{0},\underline{\hat{\mu}}(x)))}
            {\theta(\underline{z}(P_{0},\underline{\hat{\nu}}(x)))
            \theta(\underline{z}(P,\underline{\hat{\mu}}(x)))}
            \mathrm{exp}\left(d_0
            -\int_{Q_0}^P \omega_{P_{\infty} P_0}^{(3)}\right),
  \end{equation}
  \begin{eqnarray}\label{4.38}
   u(x)&=& i
    \frac{\theta(\underline{z}(P_0,\underline{\hat{\nu}}(x)))
            \theta(\underline{z}(P_{\infty},\underline{\hat{\mu}}(x)))}
            {\theta(\underline{z}(P_{\infty},\underline{\hat{\nu}}(x)))
            \theta(\underline{z}(P_0,\underline{\hat{\mu}}(x)))}
             \\
    &\times&
        \left(
    \frac{
    \sum_{j=1}^n \lambda_j -\sum_{j=1}^nU_j \partial_{\omega_j}
          \mathrm{ln}
\left(\frac{\theta(\underline{z}(P_{\infty},\underline{\hat{\mu}}(x))+\underline{\omega})}
 {\theta(\underline{z}(P_{0},\underline{\hat{\mu}}(x))+\underline{\omega})}\right)
     \Big|_{\underline{\omega}=0} }
     {
     \sum_{j=1}^n \lambda_j -\sum_{j=1}^nU_j \partial_{\omega_j}
          \mathrm{ln}
\left(\frac{\theta(\underline{z}(P_{\infty},\underline{\hat{\nu}}(x))+\underline{\omega})}
 {\theta(\underline{z}(P_{0},\underline{\hat{\nu}}(x))+\underline{\omega})}\right)
     \Big|_{\underline{\omega}=0}
     }
      +1 \right). \nonumber
  \end{eqnarray}
Moreover, let $\mu_j,$ $j=1,\ldots,n$ be not vanishing on $\Omega$.
Then, we have the following constraint
     \begin{eqnarray}\label{4.39}
        \int_{x_0}^x dx^\prime  \frac{u-u_{x^\prime x^\prime}}{u_{x^\prime}-u}
        &=&-(\tilde{x}-\tilde{x}_0) \sum_{j=1}^n
        \Big(\int_{a_j} \tilde{\omega}_{P_{\infty} P_0}^{(3)} \Big)
        c_j(1) \nonumber \\
        &+&
         \mathrm{ln}
        \left(
        \frac{\theta(\underline{z}(P_0,\underline{\hat{\mu}}(x_0)))
            \theta(\underline{z}(P_{\infty},\underline{\hat{\mu}}(x)))}
            {\theta(\underline{z}(P_{\infty},\underline{\hat{\mu}}(x_0)))
            \theta(\underline{z}(P_0,\underline{\hat{\mu}}(x)))}
        \right),
     \end{eqnarray}
and
     \begin{eqnarray}\label{4.40}
      &&
       \underline{\hat{z}}(P_\infty,\underline{\hat{\mu}}(x))=
       \underline{\widehat{\Xi}}_{Q_0}
       -\underline{\widehat{A}}_{Q_0}(P_\infty)
       +\underline{\hat{\alpha}}_{Q_0}(\mathcal{D}_{\underline{\hat{\mu}}(x)})
        \nonumber \\
      && =
      \underline{\widehat{\Xi}}_{Q_0}
       -\underline{\widehat{A}}_{Q_0}(P_\infty)
       +\underline{\hat{\alpha}}_{Q_0}(\mathcal{D}_{\underline{\hat{\mu}}(x_0)})
       -\int_{x_0}^x
       \frac{u-u_{x^\prime x^\prime}}{u_{x^\prime}-u}
             \frac{dx^\prime}{ \Psi_n(\underline{\mu}(x^\prime))}
             \underline{c}(1)
          \nonumber \\
       && =
        \underline{\widehat{\Xi}}_{Q_0}
       -\underline{\widehat{A}}_{Q_0}(P_\infty)
       +\underline{\hat{\alpha}}_{Q_0}(\mathcal{D}_{\underline{\hat{\mu}}(x_0)})
       -\underline{c}(1)(\tilde{x}-\tilde{x}_0),
     \end{eqnarray}
    \begin{eqnarray}\label{4.41}
      &&
       \underline{\hat{z}}(P_0,\underline{\hat{\mu}}(x))=
       \underline{\widehat{\Xi}}_{Q_0}
       -\underline{\widehat{A}}_{Q_0}(P_0)
       +\underline{\hat{\alpha}}_{Q_0}(\mathcal{D}_{\underline{\hat{\mu}}(x)})
        \nonumber \\
      && =
      \underline{\widehat{\Xi}}_{Q_0}
       -\underline{\widehat{A}}_{Q_0}(P_0)
       +\underline{\hat{\alpha}}_{Q_0}(\mathcal{D}_{\underline{\hat{\mu}}(x_0)})
       -\int_{x_0}^x
       \frac{u-u_{x^\prime x^\prime}}{u_{x^\prime}-u}
             \frac{dx^\prime}{ \Psi_n(\underline{\mu}(x^\prime))}
             \underline{c}(1)
          \nonumber \\
       && =
        \underline{\widehat{\Xi}}_{Q_0}
       -\underline{\widehat{A}}_{Q_0}(P_0)
       +\underline{\hat{\alpha}}_{Q_0}(\mathcal{D}_{\underline{\hat{\mu}}(x_0)})
       -\underline{c}(1)(\tilde{x}-\tilde{x}_0).
     \end{eqnarray}
 \end{the4.5}
\textbf{Proof.}~~First, let us assume 
    \begin{equation}\label{4.42}
      \mu_j(x)\neq \mu_{j^\prime}(x), \quad \nu_k(x)\neq \nu_{k^\prime}(x)
      \quad \textrm{for $j\neq j^\prime, k\neq k^\prime$ and
      $x\in\widetilde{\Omega}$},
      \end{equation}
 where $\widetilde{\Omega}\subseteq\Omega$ is open and connected. 
 From (\ref{3.19}), $\mathcal
 {D}_{P_{0}\underline{\hat{\nu}}}\sim
 \mathcal {D}_{P_{\infty}\underline{\hat{\mu}}}$, and
 $P_{\infty}=(P_{\infty})^\ast \notin\{\hat{\mu}_1,\cdots,\hat{\mu}_n \}$,
 based on some hypothesis of Theorem A. 31 \cite{12}, one can
 have  $\mathcal {D}_{\underline{\hat{\nu}}}
 \in \textrm{Sym}^n(\mathcal {K}_n)$ is nonspecial.
 This argument is 
 symmetric with respect to
 $\underline{\hat{\mu}}$ and $\underline{\hat{\nu}}$. Thus, $\mathcal
 {D}_{\underline{\hat{\mu}}}$ is nonspecial if and only
 if $\mathcal{D}_{\underline{\hat{\nu}}}$ is.

 Next, we derive the
 representations of $\phi$ and $u$ in terms of the Riemann theta
 functions. A special case of Riemann's vanishing theorem (Theorem A. 26 \cite{12})
 yields
  \begin{equation}\label{4.43}
        \theta(\underline{\Xi}_{Q_0}-\underline{A}_{Q_0}(P)+\underline{\alpha}_{Q_0}(\mathcal
        {D}_{\underline{Q}}))=0 \quad \textrm{ if and only if $P\in
        \{Q_1,\cdots,Q_n\}$}.
  \end{equation}
Therefore, the divisor (\ref{3.19}) of $\phi(P,x)$ suggests considering
expression of the following type
\begin{equation}\label{4.44}
C(x)\frac{
\theta(\underline{\Xi}_{Q_0}-\underline{A}_{Q_0}(P)+\underline{\alpha}_{Q_0}(\mathcal
{D}_{\underline{\hat{\nu}}(x)}))}{
\theta(\underline{\Xi}_{Q_0}-\underline{A}_{Q_0}(P)+\underline{\alpha}_{Q_0}(\mathcal
{D}_{\underline{\hat{\mu}}(x)}))} \mathrm{exp}\Big(\int_{Q_0}^P
\omega_{P_{\infty}P_{0}}^{(3)}\Big),
\end{equation}
where $C(x)$ is independent of $P\in\mathcal {K}_n$. So, together
with the asymptotic expansion of $\phi(P,x)$ near $P_0$ in
(\ref{4.2}), we are able to obtain (\ref{4.37}).

Let us now construct $u$ from trace formulas (\ref{3.51}) and
(\ref{3.52}). By comparing (\ref{3.51}) with (\ref{3.52}), we
have
  \begin{equation}\label{4.45}
    (u-u_x) \sum_{j=1}^n \mu_j= (u+u_x)\sum_{j=1}^n \nu_j,
  \end{equation}
namely,
  \begin{equation}\label{4.46}
    u+u_x=\frac{\sum_{j=1}^n \mu_j}{\sum_{j=1}^n \nu_j} (u-u_x).
  \end{equation}
Therefore,
   \begin{equation}\label{4.47}
    u=\frac{1}{2}(u+u_x) +\frac{1}{2}(u-u_x)
    =\frac{1}{2}\Big(\frac{\sum_{j=1}^n \mu_j}{\sum_{j=1}^n \nu_j}
    +1 \Big) (u-u_x).
   \end{equation}
On the other hand, the asymptotic expansion of $\phi$ near
$P_\infty$ with taking account into (\ref{4.37}) and comparing the
coefficient of $\zeta^{-1}$ yields
   \begin{equation}\label{4.48}
    u-u_x=2i
\frac{\theta(\underline{z}(P_0,\underline{\hat{\nu}}(x)))
 \theta(\underline{z}(P_{\infty},\underline{\hat{\mu}}(x)))}
 {\theta(\underline{z}(P_{\infty},\underline{\hat{\nu}}(x)))
 \theta(\underline{z}(P_0,\underline{\hat{\mu}}(x)))}.
   \end{equation}
Inserting the expressions of $\sum_{j=1}^n \mu_j$, $\sum_{j=1}^n
\nu_j$ (F.88 \cite{12}), and (\ref{4.48}) into (\ref{4.47}) leads to
(\ref{4.38}).

Moreover, from (\ref{4.24}) and (\ref{4.25}), one can readily get 
   \begin{eqnarray}\label{4.49}
   &&
     \sum_{j=1}^n \int_{Q_0}^{\hat{\mu}_j(x)}
     \tilde{\omega}_{P_\infty,P_0}^{(3)}
     -\sum_{j=1}^n \int_{Q_0}^{\hat{\mu}_j(x_0)}
     \tilde{\omega}_{P_\infty,P_0}^{(3)}
     \nonumber \\
     && =\int_{x_0}^x \partial_{x^\prime} \Big(
     \sum_{j=1}^n \int_{Q_0}^{\hat{\mu}_j(x^\prime)}
     \tilde{\omega}_{P_\infty,P_0}^{(3)}\Big) dx^\prime
      \nonumber \\
     && =\int_{x_0}^x dx^\prime
       \frac{u-u_{x^\prime x^\prime}}{u_{x^\prime}-u},
   \end{eqnarray}
By (F.88) \cite{12}, we may arrive at
     \begin{eqnarray}\label{4.50}
      &&
       \sum_{j=1}^n \int_{Q_0}^{\hat{\mu}_j(x)}
     \tilde{\omega}_{P_\infty,P_0}^{(3)}
     -\sum_{j=1}^n \int_{Q_0}^{\hat{\mu}_j(x_0)}
     \tilde{\omega}_{P_\infty,P_0}^{(3)}
     = \sum_{j=1}^n \int_{a_j} \tilde{\omega}_{P_\infty,P_0}^{(3)}
       \nonumber \\
     && ~~~~~~~~~\times ~
      \left(
      \sum_{k=1}^n \int_{Q_0}^{\hat{\mu}_k(x)} \omega_j
     -\sum_{k=1}^n \int_{Q_0}^{\hat{\mu}_k(x_0)} \omega_j
      \right)
      + \mathrm{ln} \left(
       \frac{\theta(\underline{z}(P_{\infty},\underline{\hat{\mu}}(x)))}
       {\theta(\underline{z}(P_{0},\underline{\hat{\mu}}(x)))}
      \right)
       \nonumber \\
      && ~~~~~~~~~ - ~
      \mathrm{ln} \left(
       \frac{\theta(\underline{z}(P_{\infty},\underline{\hat{\mu}}(x_0)))}
       {\theta(\underline{z}(P_{0},\underline{\hat{\mu}}(x_0)))}
      \right)
        \nonumber \\
      &&
      = \sum_{j=1}^n \int_{a_j} \tilde{\omega}_{P_\infty,P_0}^{(3)}
      \int_{x_0}^x \partial_{x^\prime} \Big(
     \sum_{k=1}^n \int_{Q_0}^{\hat{\mu}_k(x^\prime)}
     \omega_j \Big)dx^\prime
           \nonumber \\
     && ~~~~~~~~~
     + ~\mathrm{ln}
        \left(
        \frac{\theta(\underline{z}(P_0,\underline{\hat{\mu}}(x_0)))
            \theta(\underline{z}(P_{\infty},\underline{\hat{\mu}}(x)))}
            {\theta(\underline{z}(P_{\infty},\underline{\hat{\mu}}(x_0)))
            \theta(\underline{z}(P_0,\underline{\hat{\mu}}(x)))}
        \right)
        \nonumber \\
     &&
     =-\int_{x_0}^x
     \Big( \frac{u-u_{x^\prime x^\prime}}{u_{x^\prime}-u}
     \frac{1}{\prod_{k=1}^n \mu_k(x^\prime)} \Big) dx^\prime
      \sum_{j=1}^n \Big( \int_{a_j}
      \tilde{\omega}_{P_\infty,P_0}^{(3)} \Big)
      c_j(1)
       \nonumber \\
      && ~~~~~~~~~
     + ~\mathrm{ln}
        \left(
        \frac{\theta(\underline{z}(P_0,\underline{\hat{\mu}}(x_0)))
            \theta(\underline{z}(P_{\infty},\underline{\hat{\mu}}(x)))}
            {\theta(\underline{z}(P_{\infty},\underline{\hat{\mu}}(x_0)))
            \theta(\underline{z}(P_0,\underline{\hat{\mu}}(x)))}
        \right).
     \end{eqnarray}
Hence, inserting the condition of changing of variables (\ref{4.36})
into (\ref{4.50}) and comparing with (\ref{4.49}) can give the
constraint (\ref{4.39}). Equations (\ref{4.40}) and (\ref{4.41}) are
clear from (\ref{4.15}).
The extension of all results from $x \in
\widetilde{\Omega}$ to $x \in   \Omega$ follows by the
continuity of $\underline{\alpha}_{Q_0}$ and the hypothesis of
$\mathcal{D}_{\underline{\hat{\mu}}(x)}$ being nonspecial for $x \in
\Omega$. \quad $\square$

\newtheorem{rem4.6}[lem4.1]{Remark}
  \begin{rem4.6}
  The stationary MCH solution $u$ in $(\ref{4.38})$ is 
  a quasi-periodic function with respect to the new variable
  $\tilde{x}$ in $(\ref{4.36})$. The Abel map in $(\ref{4.40})$ and
  $(\ref{4.41})$  linearizes the divisor
  $\mathcal{D}_{\underline{\hat{\mu}}(x)}$ on $\Omega$ with respect
  to $\tilde{x}$.
  \end{rem4.6}

\newtheorem{rem4.7}[lem4.1]{Remark}
  \begin{rem4.7}
   The similar results to $(\ref{4.40})$ and $(\ref{4.41})$ (i.e.
   the Abel map also linearizes the divisor
   $\mathcal{D}_{\underline{\hat{\nu}}(x)}$ on $\Omega$ with respect
   to $\bar{x}$) hold for the divisor $\mathcal{D}_{\underline{\hat{\nu}}(x)}$
   associated with $\phi(P,x)$. The change of variables is
   \begin{equation}\label{4.51}
        x \mapsto \bar{x}= \int^x dx^\prime
        \Big( \frac{1}{\Psi_n(\underline{\nu}(x^\prime))}
         \frac{u-u_{x^\prime x^\prime}}{u_{x^\prime}+u}
        \Big).
     \end{equation}
  \end{rem4.7}

\newtheorem{rem4.8}[lem4.1]{Remark}
  \begin{rem4.8}
    Since  
    $\mathcal
 {D}_{P_{0}\underline{\hat{\nu}}}$ and
 $\mathcal {D}_{P_{\infty}\underline{\hat{\mu}}}$
 are linearly equivalent, that is,
    \begin{equation}\label{4.52}
    \underline{A}_{Q_0}(P_\infty) + \underline{\alpha}_{Q_0}
    (\mathcal{D}_{\underline{\hat{\mu}}(x)})
    =\underline{A}_{Q_0}(P_0)+ \underline{\alpha}_{Q_0}
    (\mathcal{D}_{\underline{\hat{\nu}}(x)}),
    \end{equation}
we have
    \begin{equation}\label{4.53}
        \underline{\alpha}_{Q_0}
    (\mathcal{D}_{\underline{\hat{\nu}}(x)})
    =\underline{\Delta}+ \underline{\alpha}_{Q_0}
    (\mathcal{D}_{\underline{\hat{\mu}}(x)}),
    \qquad \underline{\Delta}=\underline{A}_{P_0}(P_\infty).
    \end{equation}
Hence
   \begin{equation}\label{4.54}
    \underline{z}(P,\underline{\hat{\nu}})=
    \underline{z}(P,\underline{\hat{\mu}})+\underline{\Delta},
    \qquad P\in \mathcal{K}_n.
   \end{equation}
The representations of $\phi$ and $u$ in $(\ref{4.37})$ and
$(\ref{4.38})$ can be rewritten in terms of
$\mathcal{D}_{\underline{\hat{\nu}}(x)}$, respectively.
  \end{rem4.8}

\newtheorem{rem4.9}[lem4.1]{Remark}
  \begin{rem4.9}
   We have emphasized in Remark $4.2$ that the Baker-Akhiezer
   functions $\psi$ in $(\ref{3.6})$ and $(\ref{3.8})$ for the MCH hierarchy
   enjoy very difference from standard Baker-Akhiezer functions.
   Hence, one may not expect the usual theta function representations
   of $\psi_j$, $j=1,2,$ in terms of ratios of theta functions
   times a exponential term including $(x-x_0)$ multiplying a
   meromorphic differential with a pole with the essential singularity
   of $\psi_j$. However, using the properties of symmetric
   function and $(F.89)~[13]$, we obtain
   \begin{eqnarray}\label{4.55}
   F_n(z)&=&(u_x-u)(z^n+ \prod_{j=1}^{n-1} (z-\mu_j))
    = (u_x-u)(z^n+ \sum_{l=1}^{n-1} \Psi_{n-l}(\underline{\mu}) z^l)
      \nonumber \\
    &=&
    (u_x-u)\Big(z^n+\sum_{k=1}^n \Big( \Psi_{n+1-k}(\underline{\lambda})
      \nonumber \\
    &&
    -\sum_{j=1}^nc_j(k) \partial_{\omega_j}
    \mathrm{ln}
    \left(
    \frac{\theta(\underline{z}(P_\infty,\underline{\hat{\mu}})+\underline{\omega})}
    {\theta(\underline{z}(P_0,\underline{\hat{\mu}})+\underline{\omega})}
    \right)
    \Big|_{\underline{\omega}=0} z^{k-1} \Big) \Big)
       \nonumber \\
    &=&
     (u_x-u)\Big( \prod_{j=1}^n(z-\lambda_j)
       \nonumber \\
     &&
       -\sum_{j=1}^n \sum_{k=1}^n
       c_j(k) \partial_{\omega_j}
    \mathrm{ln}
    \left(
    \frac{\theta(\underline{z}(P_\infty,\underline{\hat{\mu}})+\underline{\omega})}
    {\theta(\underline{z}(P_0,\underline{\hat{\mu}})+\underline{\omega})}
    \right)
    \Big|_{\underline{\omega}=0} z^{k-1} \Big),
   \end{eqnarray}
and by inserting $(\ref{4.55})$ into $(\ref{3.31})$, we obtain the
theta function representation of $\psi_1$. Then, the corresponding
theta function representation of $\psi_2$ follows by
$(\ref{3.8})$ and $(\ref{4.37})$.
  \end{rem4.9}

Let us now discuss the trivial case $n=0$ excluded in Theorem
4.5.

\newtheorem{exa4.10}[lem4.1]{Example}
  \begin{exa4.10}
    Let $n=0$, $P=(z,y) \in \mathcal{K}_0 \setminus
    \{P_\infty,P_0\},$ and $(x,x_0)\in \mathbb{C}^2$. Then, we have
     \begin{eqnarray}\label{4.56}
       \begin{split}
        & \mathcal{K}_0: \mathcal{F}_0(z,y)=y^2-R_1(z)
        =y^2-(z-\widetilde{E}_1)=0,
        \quad \widetilde{E}_1 \in \mathbb{C}, \\
        & u=\widetilde{E}_1^{1/2}, \\
        & \mathrm{s-MCH}_0(u)=2u_x-2u_{xx}=0, \\
        & F_0(z,x)=f_0=u_x-u, \quad G_0(z,x)=g_0=-1, \quad
          H_0(z,x)=h_0=u_x+u, \\
        & \phi(P,x)=\frac{z^{1/2}y-z}{u_x-u}
                  =\frac{u_x+u}{z^{-1/2}y+1},\\
        & \psi_1(P,x,x_0)= \mathrm{exp}\Big( \int_{x_0}^x \Big(
           \frac{1}{2}\frac{u-u_{x^\prime x^\prime}}{u_{x^\prime}-u}
           \frac{y-z^{1/2}}{z^{1/2}}\Big) dx^\prime -\frac{1}{2}(x-x_0)
           \Big), \\
        & \psi_2(P,x,x_0)=\frac{y-z^{1/2}}{u_x-u}
              \mathrm{exp}\Big( \int_{x_0}^x \Big(
           \frac{1}{2}\frac{u-u_{x^\prime x^\prime}}{u_{x^\prime}-u}
           \frac{y-z^{1/2}}{z^{1/2}}\Big) dx^\prime
           -\frac{1}{2}(x-x_0)
           \Big).
       \end{split}
     \end{eqnarray}
  The general solution of $\mathrm{s-MCH}_0(u)=2u_x-2u_{xx}=0$ is
  given by
      \begin{equation}\label{4.57}
        u(x)=a_1e^x+\widetilde{E}_1^{1/2},
        \qquad a_1 \in \mathbb{C}.
      \end{equation}
  But, according to the condition $\partial_x^k u \in L^\infty
  (\mathbb{C})$, $k\in\mathbb{N}_0$ in $(\ref{2.1})$, we conclude 
  $a_1=0$, and therefore $u=\widetilde{E}_1^{1/2}$, which is
  the same as the expression of $u$ in $(\ref{4.56})$ from
  trace formula $(\ref{3.51})$ or $(\ref{3.52})$ in the special case
  $n=0$.
  \end{exa4.10}

At the end of this section, we turn to the initial value problem in
the stationary case. We show that the solvability of the Dubrovin
equations (\ref{3.43}) on $\Omega_\mu \subseteq \mathbb{C}$ in fact
implies equation (\ref{2.28}) on $\Omega_\mu$, which amounts to
solving the 
initial value problem in the
stationary case.

\newtheorem{the4.11}[lem4.1]{Theorem}
  \begin{the4.11}
  Assume that $(\ref{2.1})$ holds and
    $\{\hat{\mu}_j\}_{j=1,\ldots,n}$ satisfies the stationary
    Dubrovin equations $(\ref{3.43})$ on $\Omega_\mu$ and
    remain distinct and nonzero for $ x \in
     \Omega_\mu,$ where $\Omega_\mu \subseteq \mathbb{C}$ is open
     and connected. Then, if $u \in C^\infty(\Omega_\mu)$ satisfies
     \begin{equation}\label{4.58}
        -\frac{1}{2}(u-u_{xx})(u^2-u_x^2)
        +\frac{1}{2}(u-u_{xx})\sum_{m=1}^{2n+1}\widetilde{E}_m=
        (u-u_x)\sum_{j=1}^n \mu_j,
     \end{equation}
   $u$ is a solution of 
   the $n$th stationary MCH equation
   $(\ref{2.28}),$ that is
     \begin{equation}\label{4.59}
        \mathrm{s-MCH}_n(u)=0,
        \quad \textrm{on $\Omega_\mu.$}
     \end{equation}
  \end{the4.11}
\textbf{Proof.}~~Given the solutions
   $\hat{\mu}_j=(\mu_j,y(\hat{\mu}_j))\in
   C^\infty(\Omega_\mu,\mathcal {K}_n),j=1,\cdots,n$ of (\ref{3.43}), let us
   introduce
   \begin{equation}\label{4.60}
     F_n(z)=(u_x-u)\prod_{j=1}^n(z-\mu_j)
     \quad \textrm{on $\mathbb{C}\times\Omega_\mu$},
    \end{equation}
 where $u$ is the solution of (\ref{4.58}) up to multiplicative
 constant. Given $F_n$ and $u$, let us denote the polynomial $G_n$ by
      \begin{equation}\label{4.61}
        G_n(z)=(u-u_{xx})^{-1}(F_n(z)+F_{n,x}(z)),
        \quad
         \textrm{on $\mathbb{C}\times\Omega_\mu$},
     \end{equation}
 and from ({\ref{4.60}), one can see that the degree of $G_n$ is $n$
 with respect to $z$. Taking account into (\ref{4.61}),
 the Dubrovin equations (\ref{3.43}) imply
      \begin{eqnarray}\label{4.62}
       y(\hat{\mu}_j)&=&\mu_{j,x}\frac{(u_x-u)}{u-u_{xx}}\sqrt{\mu_j}
         \prod_{ \scriptstyle k=1 \atop \scriptstyle k \neq j}^n
         (\mu_j-\mu_k) \nonumber \\
       &=& -\frac{\sqrt{\mu_j}F_{n,x}(\mu_j)}{u-u_{xx}}
       =-\sqrt{\mu_j}G_n(\mu_j).
      \end{eqnarray}
 Hence
     \begin{equation}\label{4.63}
        R_{2n+1}(\mu_j)^2-\mu_jG_n(\mu_j)^2=
        y(\hat{\mu}_j)^2-\mu_jG_n(\mu_j)^2=0,
        \quad j=1,\ldots,n.
     \end{equation}
 Next, let us define a polynomial $H_n$ on $\mathbb{C}\times\Omega_\mu$
 such that
    \begin{equation}\label{4.64}
      R_{2n+1}(z)-zG_n(z)^2=F_n(z)H_n(z)
    \end{equation}
 holds. Such a polynomial $H_n$ exists since the left-hand side of (\ref{4.64})
 vanishes at $z=\mu_j,~j=1,\cdots,n$ by (\ref{4.63}). We need to
 determine the degree of $H_n$. By (\ref{4.61}), we compute
   \begin{equation}\label{4.65}
     R_{2n+1}(z)-zG_n(z)^2 \underset{|z| \rightarrow \infty}{=}
     (u_x+u)(u_x-u)z^{2n}+O(z^{2n-1}),
   \end{equation}
 with $O(z^{2n-1})$ depending on $x$ by
 inspection. Therefore, combining (\ref{4.60}), (\ref{4.64}) and
 (\ref{4.65}), we conclude that $H_n$ has degree $n$ with respect to
 $z$, with the coefficient $(u_x+u)$ of powers $z^{2n}$. Hence, we
 may write $H_n$ as
    \begin{equation}\label{4.66}
        H_n(z)=(u_x+u) \prod_{j=1}^n
        (z-\nu_j),
         \quad
         \textrm{on $\mathbb{C}\times\Omega_\mu$},
    \end{equation}
 where $u$ is the same as that in (\ref{4.60}).
 Next, let us consider the polynomial $P_{n-1}$ by
    \begin{equation}\label{4.67}
        P_{n-1}(z)=\frac{1}{2}(u-u_{xx})H_n(z)+\frac{1}{2}(u-u_{xx})F_n(z)
        +zG_{n,x}(z).
    \end{equation}
 Using (\ref{4.60}), (\ref{4.61}) and (\ref{4.66}), we see that
 $P_{n-1}$ is a polynomial of degree at most $n-1$. Differentiating
 on both sides of (\ref{4.64}) with respect to $x$ yields
    \begin{equation}\label{4.68}
        2zG_n(z)G_{n,x}(z)+F_{n,x}(z)H_n(z)+F_n(z)H_{n,x}(z)=0
        \quad \textrm{on $\mathbb{C}\times\Omega_\mu$}.
    \end{equation}
 Multiplying (\ref{4.67}) by $G_n$ and using (\ref{4.68}), we have
  \begin{eqnarray}\label{4.69}
     G_n(z)P_{n-1}(z)&=&\frac{1}{2}F_n(z)[(u-u_{xx})G_n(z)-H_{n,x}(z)]
       \nonumber \\
     &&+~
     [(u-u_{xx})G_n(z)-F_{n,x}(z)]\frac{1}{2}H_n(z),
  \end{eqnarray}
 and therefore on $\Omega_\mu$, we have
    \begin{equation}\label{4.70}
        G_n(\mu_j)P_{n-1}(\mu_j)=0,
        \qquad j=1,\ldots,n.
    \end{equation}
 Next, let 
   $x\in\widetilde{\Omega}_\mu \subseteq \Omega_\mu$,
 where $\widetilde{\Omega}_\mu$ is given by
      \begin{eqnarray}\label{4.71}
        \widetilde{\Omega}_\mu&=&\{x\in\Omega_\mu \mid
        G(\mu_j(x),x)=-\frac{y(\hat{\mu}_j(x))}{\sqrt{\mu_j(x)}}\neq0,\,j=1,\cdots,n\}
        \nonumber \\
        &=&\{x\in\Omega_\mu \mid \mu_j(x)
        \notin \{\widetilde{E}_m \}_{m=1,\cdots,2n+1},j=1,\cdots,n\},
      \end{eqnarray}
 Thus, we have
      \begin{equation}\label{4.72}
        P_{n-1}(\mu_j(x),x)=0,\quad
        j=1,\cdots,n,\, ~ x\in\widetilde{\Omega}_\mu.
      \end{equation}
 Since $P_{n-1}$ is a polynomial of degree at most $n-1$, (\ref{4.72})
 implies
       \begin{equation}\label{4.73}
        P_{n-1}=0
        \quad \textrm{on $\mathbb{C}\times\widetilde{\Omega}_\mu$},
       \end{equation}
 So, (2.15) holds, that is,
       \begin{equation}\label{4.74}
        zG_{n,x}(z)=-\frac{1}{2}(u-u_{xx})H_n(z)-\frac{1}{2}(u-u_{xx})F_n(z)
         \quad \textrm{on $\mathbb{C}\times\widetilde{\Omega}_\mu$}.
       \end{equation}
 Inserting (\ref{4.74}) and (\ref{4.61}) into (\ref{4.68}) yields
         \begin{equation}\label{4.75}
            [-(u-u_{xx})G_n(z)-H_n(z)+H_{n,x}(z)]F_n(z)=0,
         \end{equation}
 namely,
       \begin{equation}\label{4.76}
        H_{n,x}(z)=(u-u_{xx})G_n(z)+H_n(z),
         \quad \textrm{on $\mathbb{C}\times\widetilde{\Omega}_\mu$}.
       \end{equation}
 Thus, we obtain the fundamental equations (2.13)-(2.15), and (\ref{2.17})
 on $\mathbb{C}\times\widetilde{\Omega}_\mu$.

 In order to extend these results to all $x\in\Omega_\mu$, let us 
 consider the case where $\hat{\mu}_j$ admits one of the
 branch points $(\widetilde{E}_{m_0},0)$. Hence, we suppose
 \begin{equation}\label{4.77}
      \mu_{j_1}(x)\rightarrow\widetilde{E}_{m_0}
      \quad \textrm{as $x\rightarrow x_0\in\Omega_\mu$},
 \end{equation}
 for some $j_1\in\{1,\cdots,n\},\,m_0\in\{1,\cdots,2n+1\}$.
 Introducing
 \begin{equation}\label{4.78}
   \begin{split}
 & \zeta_{j_1}(x)=\sigma(\mu_{j_1}(x)-\widetilde{E}_{m_0})^{1/2},
   \quad \sigma=\pm1,\quad \\
  & \mu_{j_1}(x)=\widetilde{E}_{m_0}+\zeta_{j_1}(x)^2
    \end {split}
 \end{equation}
 for $x$ near $x_0$, then the Dubrovin
 equation (\ref{3.43}) for $\mu_{j_1}$ becomes
 \begin{eqnarray}\label{4.79}
 \zeta_{j_1,x}(x)&=&
 \frac{c(\sigma)(u-u_{xx})}{2(u_x-u)\sqrt{\widetilde{E}_{m_0}}}
 \Big(\prod_{\scriptstyle m=1, \atop \scriptstyle m\neq
  m_0}^{2n+1}(\widetilde{E}_{m_0}-\widetilde{E}_m)\Big)^{1/2} \nonumber \\
   &&\times \prod_{\scriptstyle k=1, \atop \scriptstyle k\neq
   j_1}^{n}(\widetilde{E}_{m_0}-\mu_k(x))^{-1}(1+O(
   \zeta_{j_1}(x)^2))
 \end{eqnarray}
  for some $|c(\sigma)|=1$. Hence, (\ref{4.73})-(\ref{4.76}) are extended to
  $\Omega_\mu$ by continuity. Consequently, we obtain relations
  (2.13)-(2.15) on $\mathbb{C}\times \Omega_\mu$, and can proceed
  as in Section 2 to see that $u$ satisfies the stationary MCH hierarchy
  (\ref{4.59}). \quad $\square$

   \newtheorem{rem4.12}[lem4.1]{Remark}
     \begin{rem4.12}
       The result in Theorem $4.11$ is derived in terms of
       $u$ and $\{\mu_j\}_{j=1,\cdots,n}$, but we
         can prove the analogous result
       in terms of $u$ and $\{\nu_j\}_{j=1,\cdots,n}$.
     \end{rem4.12}

  \newtheorem{rem4.13}[lem4.1]{Remark}
   \begin{rem4.13}
    Theorem $4.11$ reveals that given $\mathcal{K}_n$ and the initial
    condition $\underline{\hat{\mu}}(x_0)=(\hat{\mu}_1(x_0),\ldots,\hat{\mu}_n(x_0))
    \in \mathrm{Sym}^n(\mathcal{K}_n)$, or equivalently, the
    auxiliary divisor $\mathcal{D}_{\underline{\hat{\mu}}(x_0)}
    \in \mathrm{Sym}^n(\mathcal{K}_n)$ at $x=x_0$, $u$ is uniquely
    determined in an open neighborhood $\Omega$ of $x_0$
    by $(\ref{4.58})$ and satisfies the $n$th stationary MCH equation
    $(\ref{2.28})$. 
    Conversely, given $\mathcal{K}_n$
    and $u$ in an open neighborhood $\Omega$ of $x_0$, we can
    construct the corresponding polynomial $F_n(z,x)$, $G_n(z,x)$
    and $H_n(z,x)$ for $x\in\Omega$, and then obtain the auxiliary
    divisor $\mathcal{D}_{\underline{\hat{\mu}}(x)}$ for
    $x\in\Omega$ from the zeros of $F_n(z,x)$ and $(\ref{3.13})$.
    In that sense, once the curve $\mathcal{K}_n$ is fixed,
    elements of the isospectral class of the MCH potentials $u$ can be characterized by
    nonspecial auxiliary divisor
    $\mathcal{D}_{\underline{\hat{\mu}}(x)}$.
   \end{rem4.13}

\section{The time-dependent MCH formalism}
 In this section, let us go back to the recursive approach detailed in
 Section 2 and
 extend the algebro-geometric analysis in Section 3 to the
 time-dependent MCH hierarchy.

 Throughout this section, we assume that (\ref{2.2}) holds.

 The time-dependent algebro-geometric initial value problem of the
 MCH hierarchy is to solve the time-dependent $r$th MCH flow with
 a stationary solution of the $n$th equation as the initial data in the
 hierarchy. More precisely, given $n\in\mathbb{N}_0$, based on the
 solution $u^{(0)}$ of the $n$th stationary MCH equation
 $\mathrm{s-MCH}_n(u^{(0)})=0$ associated with $\mathcal{K}_n$ and a
 set of integration constants $\{c_l\}_{l=1,\ldots,n} \subset
 \mathbb{C}$, we want to build up a solution $u$ of the $r$th MCH flow
 $\mathrm{MCH}_r(u)=0$ such that $u(t_{0,r})=u^{(0)}$ for some
 $t_{0,r} \in \mathbb{C},~r\in\mathbb{N}_0$.

 We employ the following notation
 $\widetilde{V}_r,$ $\widetilde{F}_r,$ $\widetilde{G}_r,$
 $\widetilde{H}_r,$ $\tilde{f}_{2s}$, $\tilde{g}_{2s},$ $\tilde{h}_{2s}$
 to stand for the time-dependent quantities, which are obtained in
 $V_n,$ $F_n,$ $G_n,$ $H_n,$ $f_{2l},$ $g_{2l},$ $h_{2l}$ by
 replacing $\{c_l\}_{l=1,\ldots,n}$ with
 $\{\tilde{c}_s\}_{s=1,\ldots,r}$, where the integration constants
 $\{c_l\}_{l=1,\ldots,n}\subset \mathbb{C}$ in the stationary MCH hierarchy and
 $\{\tilde{c}_s\}_{s=1,\ldots,r}\subset \mathbb{C}$ in the time-dependent MCH
 hierarchy are independent of each other. In addition, we mark
 the individual $r$th MCH flow by a separate time variable $t_r \in
 \mathbb{C}$.

 Let us now provide the time-dependent
 algebro-geometric initial value problem as follows
   \begin{eqnarray}\label{5.1}
    &&
     \begin{split}
       & \mathrm{MCH}_r(u)=u_{t_r}-u_{xxt_r}-2\tilde{f}_{2r,x}=0, \\
       & u|_{t_r=t_{0,r}}=u^{(0)},
      \end{split}
          \\
   &&
      \mathrm{s-MCH}_n(u^{(0)})=-2f_{2n,x}(u^{(0)})=0,
   \end{eqnarray}
 where $t_{0,r}\in\mathbb{C},$ $n,r\in\mathbb{N}_0$, $u=u(x,t_r)$
 satisfies the condition (\ref{2.2}), and the curve $\mathrm{K}_n$ is
 associated with the initial data $u^{(0)}$ in (5.2). Noticing that
 the MCH flows are isospectral, we are going to a further step and
 assume that (5.2) holds not only at $t_r=t_{0,r}$, but also at all $t_r \in
 \mathbb{C}$.

 Let us now 
 start from the zero-curvature equations (\ref{2.35})
    \begin{equation}\label{5.3}
        U_{t_r}-\widetilde{V}_{r,x}+[U,\widetilde{V}_r]=0,
    \end{equation}
    \begin{equation}\label{5.4}
        -V_{n,x}+[U,V_n]=0,
    \end{equation}
 where
  \begin{equation}\label{5.5}
    \begin{split}
    & U(z)=
     \left(
       \begin{array}{cc}
         -\frac{1}{2} & \frac{1}{2}z^{-\frac{1}{2}}(u-u_{xx}) \\
         -\frac{1}{2}z^{-\frac{1}{2}}(u-u_{xx}) & \frac{1}{2} \\
       \end{array}
     \right)
     \\
    & V_n(z)=
    \left(
      \begin{array}{cc}
      -G_n(z) & z^{-\frac{1}{2}}F_n(z) \\
      z^{-\frac{1}{2}}H_n(z) & G_n(z) \\
      \end{array}
    \right)
      \\
    & \widetilde{V}_r(z)=
       \left(
         \begin{array}{cc}
           -\widetilde{G}_r(z) & z^{-\frac{1}{2}}\widetilde{F}_r(z) \\
           z^{-\frac{1}{2}}\widetilde{H}_r(z) & \widetilde{G}_r(z) \\
         \end{array}
       \right)
    \end{split}
  \end{equation}
and
  \begin{eqnarray}\label{5.6}
   &&
    F_n(z)=\sum_{l=0}^n f_{2l} z^{(n-l)}
    =f_0 \prod_{j=1}^n (z-\mu_j),
    \quad f_0=u_x-u,
     \\
   &&
    G_n(z)=\sum_{l=0}^n g_{2l} z^{(n-l)},
     \\
   &&
    H_n(z)=\sum_{l=0}^n h_{2l} z^{(n-l)}
    =h_0 \prod_{j=1}^n (z-\nu_j),
    \quad h_0=u_x+u,
    \\
   &&
    \widetilde{F}_r(z)=\sum_{s=0}^r \tilde{f}_{2s} z^{(r-s)},
    \quad \tilde{f}_0=u_x-u,
    \\
   &&
    \widetilde{G}_r(z)=\sum_{s=0}^r \tilde{g}_{2s} z^{(r-s)},
    \\
    &&
    \widetilde{H}_r(z)=\sum_{s=0}^r \tilde{h}_{2s} z^{(r-s)},
    \quad \tilde{h}_0=u_x+u,
   \end{eqnarray}
 for fixed $n,r\in \mathbb{N}_0$. Here $f_{2l},$ $g_{2l}$, $h_{2l}$,
 $l=0,\ldots,n$
 and $\tilde{f}_{2s},$ $\tilde{g}_{2s}$, $\tilde{h}_{2s}$,
 $s=0,\ldots,r,$ satisfy the relations in (\ref{2.3}).

 Moreover, it is more convenient for us to rewrite the
 zero-curvature equations (\ref{5.3}) and (\ref{5.4}) as the
 following forms
    \begin{equation}\label{5.12}
        \frac{1}{2}(u_{t_r}-u_{xxt_r})-\widetilde{F}_{r,x}
        -\widetilde{F}_r+(u-u_{xx})\widetilde{G}_r=0,
    \end{equation}
     \begin{equation}\label{5.13}
        -\frac{1}{2}(u_{t_r}-u_{xxt_r})-\widetilde{H}_{r,x}
        +\widetilde{H}_r+(u-u_{xx})\widetilde{G}_r=0,
    \end{equation}
    \begin{equation}\label{5.14}
        z\widetilde{G}_{r,x}=-\frac{1}{2}(u-u_{xx})\widetilde{H}_r
         -\frac{1}{2}(u-u_{xx})\widetilde{F}_r
    \end{equation}
and
   \begin{equation}\label{5.15}
     F_{n,x}=-F_n+(u-u_{xx})G_n,
   \end{equation}
   \begin{equation}\label{5.16}
     H_{n,x}=H_n+(u-u_{xx})G_n,
   \end{equation}
   \begin{equation}\label{5.17}
        zG_{n,x}=-\frac{1}{2}(u-u_{xx})H_n
         -\frac{1}{2}(u-u_{xx})F_n.
    \end{equation}
From (\ref{5.15})-(\ref{5.17}), we may compute 
  \begin{equation}\label{5.18}
    \frac{d}{dx} \mathrm{det}(V_n(z))=-\frac{1}{z}\frac{d}{dx}
    \Big( zG_n(z)^2+F_n(z)H_n(z) \Big)=0,
  \end{equation}
and meanwhile Lemma 5.2 gives
  \begin{equation}\label{5.19}
    \frac{d}{dt_r} \mathrm{det}(V_n(z))=-\frac{1}{z}\frac{d}{dt_r}
    \Big( zG_n(z)^2+F_n(z)H_n(z) \Big)=0.
  \end{equation}
Hence, $zG_n(z)^2+F_n(z)H_n(z)$ is independent of variables both $x$
and $t_r$, which  implies
   \begin{equation}\label{5.20}
   zG_n(z)^2+F_n(z)H_n(z)=R_{2n+1}(z).
   \end{equation}
This reveals that the fundamental identity (\ref{2.17}) still holds
in the time-dependent context. Consequently the hyperelliptic curve
$\mathcal{K}_n$ is still available by (\ref{2.21}).

Next, let us introduce the time-dependent Baker-Akhiezer function
$\psi(P,x,x_0,$ $t_r,t_{0,r})$  on $\mathcal{K}_{n}\setminus
\{P_{\infty},P_0\}$ by
  \begin{equation}\label{5.21}
         \begin{split}
          & \psi(P,x,x_0,t_r,t_{0,r})=\left(
                            \begin{array}{c}
                              \psi_1(P,x,x_0,t_r,t_{0,r}) \\
                              \psi_2(P,x,x_0,t_r,t_{0,r}) \\
                            \end{array}
                          \right), \\
          & \psi_x(P,x,x_0,t_r,t_{0,r})=U(u(x,t_r),z(P))\psi(P,x,x_0,t_r,t_{0,r}),\\
          &  \psi_{t_r}(P,x,x_0,t_r,t_{0,r})=\widetilde{V}_r(u(x,t_r),z(P))
               \psi(P,x,x_0,t_r,t_{0,r}), \\
           & z^\frac{1}{2}V_n(u(x,t_r),z(P))\psi(P,x,x_0,t_r,t_{0,r})
           =y(P)\psi(P,x,x_0,t_r,t_{0,r}),\\
          & \psi_1(P,x_0,x_0,t_{0,r},t_{0,r})=1; \\
          &
          P=(z,y)\in \mathcal{K}_{n}
           \setminus \{P_{\infty},P_0\},~(x,t_r)\in \mathbb{C}^2.
         \end{split}
   \end{equation}
where
   \begin{eqnarray}\label{5.22}
   \psi_1(P,x,x_0,t_r,t_{0,r})&=&\mathrm{exp}\Big(
     \int_{t_{0,r}}^{t_r} ds
     (z^{-1}\widetilde{F}_r(z,x_0,s)\phi(P,x_0,s)
     -\widetilde{G}_r(z,x_0,s)) \nonumber \\
     &+&
     z^{-1} \int_{x_0}^x dx^\prime
     (\frac{1}{2}q(x^\prime,t_r)\phi(P,x^\prime,t_r))
     -\frac{1}{2}(x-x_0) \Big), \nonumber \\
     && ~~~~~~~~~~~~~~~~~
     P=(z,y)\in \mathcal{K}_{n}
           \setminus \{P_{\infty},P_0\}.
   \end{eqnarray}
 Closely related to $\psi(P,x,x_0,t_r,t_{0,r})$ is the following meromorphic
 function $\phi(P,x,t_r)$ on $\mathcal{K}_{n}$ defined by
        \begin{eqnarray}\label{5.23}
         \phi(P,x,t_r)=\frac{2z}{q(x,t_r)} \Big(
         \frac{ \psi_{1,x}(P,x,x_0,t_r,t_{0,r})}
          {\psi_1(P,x,x_0,t_r,t_{0,r})} +\frac{1}{2} \Big),
            \nonumber \\
           \quad P \in
          \mathcal{K}_{n}\setminus \{P_{\infty},P_0\},
          ~ (x,t_r)\in \mathbb{C}^2.
        \end{eqnarray}
 which implies by (\ref{5.21}) that
   \begin{eqnarray}\label{5.24}
        \phi(P,x,t_r)&=&z^{1/2}\frac{y+z^{1/2}G_n(z,x,t_r)}{F_n(z,x,t_r)}
           \nonumber \\
        &=&
         \frac{z^{1/2}H_n(z,x,t_r)}{y-z^{1/2}G_n(z,x,t_r)},
    \end{eqnarray}
 and
      \begin{equation}\label{5.25}
        \psi_2(P,x,x_0,t_r,t_{0,r})=
        \psi_1(P,x,x_0,t_r,t_{0,r})\phi(P,x,t_r)/z^{1/2}.
      \end{equation}
 In analogy to equations (\ref{3.13}) and (\ref{3.14}), we define
     \begin{eqnarray}\label{5.26}
       \hat{\mu}_j(x,t_r)=\Big(\mu_j(x,t_r),
       -\sqrt{\mu_j(x,t_r)}G_n(\mu_j(x,t_r),x,t_r)\Big)
       \in \mathcal{K}_n,
        \nonumber \\
        j=1,\ldots,n, ~(x,t_r)\in \mathbb{C}^2,
     \end{eqnarray}
     \begin{eqnarray}\label{5.27}
       \hat{\nu}_j(x,t_r)=\Big(\nu_j(x,t_r),
       \sqrt{\nu_j(x,t_r)}G_n(\nu_j(x,t_r),x,t_r)\Big)
       \in \mathcal{K}_n,
        \nonumber \\
        j=1,\ldots,n, ~(x,t_r)\in \mathbb{C}^2.
     \end{eqnarray}
 The regular properties of $F_n$, $H_n$, $\mu_j$ and $\nu_j$
 are analogous to those in Section 3 due to assumptions
 (\ref{2.2}).\\
 From (\ref{5.24}), the divisor $(\phi(P,x,t_r))$
 of $\phi(P,x,t_r)$ reads
 \begin{equation}\label{5.28}
           (\phi(P,x,t_r))=
           \mathcal{D}_{P_0\underline{\hat{\nu}}(x,t_r)}(P)
           -\mathcal{D}_{P_{\infty}\underline{\hat{\mu}}(x,t_r)}(P)
  \end{equation}
 where
  \begin{equation}\label{5.29}
    \underline{\hat{\mu}}=\{\hat{\mu}_1,\ldots,\hat{\mu}_{n}\},
    \quad
    \underline{\hat{\nu}}=\{\hat{\nu}_1,\ldots,\hat{\nu}_{n}\}
    \in \mathrm{Sym}^n (\mathcal{K}_n).
  \end{equation}
 That means $P_0,\hat{\nu}_1(x,t_r),\ldots,\hat{\nu}_{n}(x,t_r)$ are
 the  $n+1$ zeros of $\phi(P,x,t_r)$ and
 $ P_{\infty},\hat{\mu}_1(x,t_r), \ldots,\hat{\mu}_{n}(x,t_r)$
 its $n+1$ poles.\\

 Further properties of $\phi(P,x,t_r)$ are summarized as follows.

 \newtheorem{lem5.1}{Lemma}[section]
  \begin{lem5.1}
    Assume that $(\ref{2.2})$,  $(\ref{5.3})$ and
    $(\ref{5.4})$ hold. Let
    $P=(z,y) \in \mathcal{K}_{n}\setminus \{P_{\infty},P_0\},
     ~ (x,t_r)\in \mathbb{C}^2.$
    Then
      \begin{equation}\label{5.30}
       \phi_x(P)+\frac{1}{2}(u-u_{xx})z^{-1}\phi(P)^2
        -\phi(P)=-\frac{1}{2}(u-u_{xx}),
      \end{equation}
      \begin{eqnarray}\label{5.31}
        (q\phi(P))_{t_r}&=&(-2z\widetilde{G}_r(z)+2\widetilde{F}_r(z)\phi(P))_x
            \\
        &=&(u-u_{xx})\widetilde{H}_r(z)+(u-u_{xx})\widetilde{F}_r(z)
        +2(\widetilde{F}_r(z)\phi(P))_x, \nonumber
      \end{eqnarray}
    \begin{equation}\label{5.32}
        \phi_{t_r}(P)=\widetilde{H}_r(z)+2\widetilde{G}_r(z)\phi(P)
         -z^{-1}\widetilde{F}_r(z)\phi(P)^2,
    \end{equation}
    \begin{equation}\label{5.33}
        \phi(P)\phi(P^\ast)=-\frac{zH_n(z)}{F_n(z)},
     \end{equation}
     \begin{equation}\label{5.34}
        \phi(P)+\phi(P^\ast)=2\frac{zG_n(z)}{F_n(z)},
     \end{equation}
     \begin{equation}\label{5.35}
        \phi(P)-\phi(P^\ast)=z^{1/2}\frac{2y}{F_n(z)}.
     \end{equation}
  \end{lem5.1}
\textbf{Proof.} We just need to prove (\ref{5.31}) and (\ref{5.32}).
Equations (\ref{5.30}) and (\ref{5.33})-(\ref{5.35}) can be proved
as in Lemma 3.1. By using (\ref{5.21}) and (\ref{5.23}), we obtain
    \begin{eqnarray}\label{5.36}
        \frac{1}{2}z^{-1}(q\phi)_{t_r}&=&
        (\mathrm{ln}\psi_1)_{xt_r}=(\mathrm{ln}\psi_1)_{t_rx}
        =\Big(\frac{\psi_{1,t_r}}{\psi_1}\Big)_x
          \nonumber \\
        &=&
 \Big(\frac{-\widetilde{G}_r\psi_1+z^{-1/2}\widetilde{F}_r\psi_2}{\psi_1}\Big)_x
          \nonumber \\
      &=&
      (-\widetilde{G}_r+z^{-1}\widetilde{F}_r\phi)_x,
    \end{eqnarray}
which implies the fist line of (\ref{5.31}). Inserting (\ref{5.14})
into (\ref{5.36}) yields the second line of (\ref{5.31}). Then by
the definition of $\phi$ (\ref{5.23}), one may have 
   \begin{eqnarray}\label{5.37}
     \phi_{t_r} &=& z^{1/2} \Big(\frac{\psi_2}{\psi_1}\Big)_{t_r}
       \nonumber \\
     &=&
      z^{1/2} \Big(\frac{\psi_{2,t_r}}{\psi_1}-
      \frac{\psi_2\psi_{1,t_r}}{\psi_1^2} \Big)
        \nonumber \\
     &=&
     z^{1/2} \Big(\frac{z^{-1/2}\widetilde{H}_r\psi_1+\widetilde{G}_r\psi_2}{\psi_1}
     -z^{-1/2}\phi\frac{-\widetilde{G}_r\psi_1+z^{-1/2}\widetilde{F}_r\psi_2}{\psi_1}\Big)
       \nonumber \\
     &=&
     \widetilde{H}_r+2\widetilde{G}_r\phi-z^{-1}\widetilde{F}_r\phi^2,
   \end{eqnarray}
which is (\ref{5.32}). Alternatively, one can insert (5.12)-(5.14)
into (\ref{5.31}) to obtain (\ref{5.32}). \quad $\square$ \\

Next, we study the time evolution of $F_n$, $G_n$ and $H_n$ by using
zero-curvature equations (5.12)-(5.14) and (5.15)-(5.17).

\newtheorem{lem5.2}[lem5.1]{Lemma}
 \begin{lem5.2}
 Assume that $(\ref{2.2})$,  $(\ref{5.3})$. and
 $(\ref{5.4})$ hold. Then
   \begin{equation}\label{5.38}
    F_{n,t_r}=2(G_n\widetilde{F}_r-\widetilde{G}_rF_n),
   \end{equation}
   \begin{equation}\label{5.39}
    zG_{n,t_r}=\widetilde{H}_rF_n-H_n\widetilde{F}_r,
   \end{equation}
    \begin{equation}\label{5.40}
        H_{n,t_r}=2(H_n\widetilde{G}_r-G_n\widetilde{H}_r).
    \end{equation}
 Equations $(\ref{5.38})-(\ref{5.40})$ imply
    \begin{equation}\label{5.41}
        -V_{n,t_r}+[\widetilde{V}_r,V_n]=0.
    \end{equation}
 \end{lem5.2}
\textbf{Proof.}~~Differentiating both sides of (\ref{5.35}) with respect to $t_r$
leads to
  \begin{equation}\label{5.42}
    (\phi(P)-\phi(P^\ast))_{t_r}=-2z^{1/2}yF_{n,t_r}F_n^{-2}.
  \end{equation}
On the other hand, by (\ref{5.32}), (\ref{5.34}), and
(\ref{5.35}), the left-hand side of (\ref{5.42}) equals to
   \begin{eqnarray}\label{5.43}
    \phi(P)_{t_r}-\phi(P^\ast)_{t_r}&=&
    2\widetilde{G}_r(\phi(P)-\phi(P^\ast))-z^{-1}\widetilde{F}_r
    (\phi(P)^2-\phi(P^\ast)^2)
     \nonumber \\
    &=&
    4z^{1/2}y(\widetilde{G}_rF_n-\widetilde{F}_rG_n)F_n^{-2}.
   \end{eqnarray}
Combining (\ref{5.42}) with (\ref{5.43}) yields (\ref{5.38}).
Similarly, Differentiating both sides of (\ref{5.34}) with respect to $t_r$ gives
  \begin{equation}\label{5.44}
    (\phi(P)+\phi(P^\ast))_{t_r}=2z(G_{n,t_r}F_n-G_nF_{n,t_r})F_n^{-2}.
  \end{equation}
Meanwhile, by (\ref{5.32}), (\ref{5.33}), and (\ref{5.34}),
the left-hand side of (\ref{5.44}) equals to
  \begin{eqnarray}\label{5.45}
   \phi(P)_{t_r}+\phi(P^\ast)_{t_r}&=&
   2\widetilde{G}_r(\phi(P)+\phi(P^\ast))
   -z^{-1}\widetilde{F}_r(\phi(P)^2+\phi(P^\ast)^2)+2\widetilde{H}_r
    \nonumber \\
   &=&
   -2zG_nF_n^{-2}F_{n,t_r}+2F_n^{-1}(\widetilde{H}_rF_n-\widetilde{F}_rH_n).
  \end{eqnarray}
Thus, (\ref{5.39}) clearly follows by (\ref{5.44}) and (\ref{5.45}). Hence,
insertion of (\ref{5.38}) and (\ref{5.39}) into the differentiation
of  $zG_n^2+F_nH_n=R_{2n+1}(z)$ can derive (\ref{5.40}). Finally, a
direct calculation shows that (\ref{5.38})-(\ref{5.40}) are
equivalent to (\ref{5.41}). \quad $\square$ \\

Further properties of $\psi$ is summarized as follows.

\newtheorem{lem5.3}[lem5.1]{Lemma}
 \begin{lem5.3}
  Assume that $(\ref{2.2})$,  $(\ref{5.3})$, and
  $(\ref{5.4})$ hold. Let
  $P=(z,y) \in \mathcal{K}_{n}\setminus \{P_{\infty},P_0\},
   ~ (x,x_0,t_r.t_{0,r})\in \mathbb{C}^4.$
  Then, we have
  \begin{eqnarray}\label{5.46}
    \psi_1(P,x,x_0,t_r,t_{0,r}) &=&
    \Big(\frac{F_n(z,x,t_r)}{F_n(z,x_0,t_{0,r})}\Big)^{1/2}
     \nonumber \\
     &\times&
    \mathrm{exp} \Big(
    \frac{y}{\sqrt{z}}\int_{t_{0,r}}^{t_r} ds
    \widetilde{F}_r(z,x_0,s)F_n(z,x_0,s)^{-1}
    \nonumber \\
    &&
    +\frac{y}{2\sqrt{z}} \int_{x_0}^x dx^\prime
    q(x^\prime,t_r)F_n(z,x^\prime,t_r)^{-1}
    \Big),
  \end{eqnarray}
  \begin{equation}\label{5.47}
    \psi_1(P,x,x_0,t_r,t_{0,r}) \psi_1(P^\ast,x,x_0,t_r,t_{0,r})
    =\frac{F_n(z,x,t_r)}{F_n(z,x_0,t_{0,r})},
  \end{equation}
  \begin{equation}\label{5.48}
    \psi_2(P,x,x_0,t_r,t_{0,r}) \psi_2(P^\ast,x,x_0,t_r,t_{0,r})
    =-\frac{H_n(z,x,t_r)}{F_n(z,x_0,t_{0,r})},
  \end{equation}
  \begin{eqnarray}\label{5.49}
  &&
    \psi_1(P,x,x_0,t_r,t_{0,r}) \psi_2(P^\ast,x,x_0,t_r,t_{0,r})
      \\
  && ~~~~
    + \psi_1(P^\ast,x,x_0,t_r,t_{0,r}) \psi_2(P,x,x_0,t_r,t_{0,r})
    =2\frac{\sqrt{z}G_n(z,x,t_r)}{F_n(z,x_0,t_{0,r})},
    \nonumber
  \end{eqnarray}
   \begin{eqnarray}\label{5.50}
  &&
    \psi_1(P,x,x_0,t_r,t_{0,r}) \psi_2(P^\ast,x,x_0,t_r,t_{0,r})
      \\
  && ~~~~
    - \psi_1(P^\ast,x,x_0,t_r,t_{0,r}) \psi_2(P,x,x_0,t_r,t_{0,r})
    =-\frac{2y}{F_n(z,x_0,t_{0,r})}.
    \nonumber
  \end{eqnarray}
 \end{lem5.3}
\textbf{Proof.}~~In order to prove (\ref{5.46}), let us first consider
the part of time variable in the definition (\ref{5.22}), that is,
   \begin{equation}\label{5.51}
    \mathrm{exp} \Big( \int_{t_{0,r}}^{t_r} ds
    [z^{-1}\widetilde{F}_r(z,x_0,s)\phi(z,x_0,s)
    -\widetilde{G}_r(z,x_0,s)]
    \Big).
   \end{equation}
The integrand in the above integral equals to
   \begin{eqnarray}\label{5.52}
   &&
     z^{-1}\widetilde{F}_r(z,x_0,s)\phi(z,x_0,s)
     -\widetilde{G}_r(z,x_0,s)
      \nonumber \\
     &&
     =z^{-1}\widetilde{F}_r(z,x_0,s)
     z^{1/2}\frac{y+z^{1/2}G_n(z,x_0,s)}{F_n(z,x_0,s)}
     -\widetilde{G}_r(z,x_0,s)
       \nonumber \\
     &&
     =\frac{y}{\sqrt{z}}\widetilde{F}_r(z,x_0,s)F_n(z,x_0,s)^{-1}
      +(\widetilde{F}_r(z,x_0,s)G_n(z,x_0,s)
      \nonumber \\
     && ~~~~
     -\widetilde{G}_r(z,x_0,s)F_n(z,x_0,s))F_n(z,x_0,s)^{-1}
       \nonumber \\
      &&
      =\frac{y}{\sqrt{z}}\widetilde{F}_r(z,x_0,s)F_n(z,x_0,s)^{-1}
      +\frac{1}{2}\frac{F_{n,s}(z,x_0,s)}{F_n(z,x_0,s)},
   \end{eqnarray}
By (\ref{5.24}), (\ref{5.38}), and (\ref{5.52}), (\ref{5.51}) reads
   \begin{equation}\label{5.53}
   \Big(\frac{F_n(z,x_0,t_r)}{F_n(z,x_0,t_{0,r})}\Big)^{1/2}
     \mathrm{exp} \Big( \frac{y}{\sqrt{z}} \int_{t_{0,r}}^{t_r} ds
     \widetilde{F}_r(z,x_0,s)F_n(z,x_0,s)^{-1}
     \Big).
   \end{equation}
On the other hand, the
part of space variable in (\ref{5.22}) can be written as
 \begin{equation}\label{5.54}
    \Big(\frac{F_n(z,x,t_r)}{F_n(z,x_0,t_{r})}\Big)^{1/2}
    \mathrm{exp}\Big(\frac{y}{2\sqrt{z}} \int_{x_0}^x dx^\prime
    q(x^\prime,t_r)F_n(z,x^\prime,t_r)^{-1}  \Big),
 \end{equation}
which can be proved using the similar procedure to Lemma 3.2. Combining (\ref{5.53}) with
(\ref{5.54}) yields (\ref{5.46}). Evaluating (\ref{5.46}) at the
points $P$ and $P^\ast$ with 
noticing 
    \begin{equation}\label{5.55}
        y(P)+y(P^\ast)=0,
    \end{equation}
leads to (\ref{5.47}). Hence, we have
   \begin{eqnarray}\label{5.56}
   &&
     \psi_2(P,x,x_0,t_r,t_{0,r})\psi_2(P^\ast,x,x_0,t_r,t_{0,r})
       \nonumber \\
   &&
      =z^{-1} \psi_1(P,x,x_0,t_r,t_{0,r})\phi(P)
     \psi_1(P^\ast,x,x_0,t_r,t_{0,r})\phi(P^\ast)
       \nonumber \\
     &&
     =z^{-1}\frac{F_n(z,x,t_r)}{F_n(z,x_0,t_{0,r})}
      \frac{-zH_n(z,x,t_r)}{F_n(z,x,t_r)}
       \nonumber \\
     &&
     =-\frac{H_n(z,x,t_r)}{F_n(z,x_0,t_{0,r})},
   \end{eqnarray}
   \begin{eqnarray}\label{5.57}
   &&
    \psi_1(P,x,x_0,t_r,t_{0,r})\psi_2(P^\ast,x,x_0,t_r,t_{0,r})
      \nonumber \\
    && ~~~~~~~~~~~~~~+
    \psi_1(P^\ast,x,x_0,t_r,t_{0,r})\psi_2(P,x,x_0,t_r,t_{0,r})
       \nonumber \\
    &&
    =(\phi(P)+\phi(P^\ast)) \psi_1(P)\psi_1(P^\ast)/z^{1/2}
      \nonumber \\
    &&
    =\frac{F_n(z,x,t_r)}{F_n(z,x_0,t_{0,r})}
     \frac{2zG_n(z,x,t_r)}{F_n(z,x,t_r)}/z^{1/2}
       \nonumber \\
     &&
     =2\frac{\sqrt{z}G_n(z,x,t_r)}{F_n(z,x_0,t_{0,r})},
   \end{eqnarray}
   \begin{eqnarray}\label{5.58}
   &&
    \psi_1(P,x,x_0,t_r,t_{0,r})\psi_2(P^\ast,x,x_0,t_r,t_{0,r})
      \nonumber \\
    && ~~~~~~~~~~~~~~-
    \psi_1(P^\ast,x,x_0,t_r,t_{0,r})\psi_2(P,x,x_0,t_r,t_{0,r})
       \nonumber \\
    &&
    =(-\phi(P)+\phi(P^\ast)) \psi_1(P)\psi_1(P^\ast)/z^{1/2}
      \nonumber \\
    &&
    =\frac{F_n(z,x,t_r)}{F_n(z,x_0,t_{0,r})}
     \frac{-2z^{1/2}y}{F_n(z,x,t_r)}/z^{1/2}
       \nonumber \\
     &&
     =-\frac{2y}{F_n(z,x_0,t_{0,r})},
   \end{eqnarray}
which are (\ref{5.48})-(\ref{5.50}). \quad $\square$ \\

In analogy to Lemma 3.4, the dynamics of the zeros
$\{\mu_j(x,t_r)\}_{j=1,\ldots,n}$ and
$\{\nu_j(x,t_r)\}_{j=1,\ldots,n}$ of $F_n(z,x,t_r)$ and
$H_n(z,x,t_r)$ with respect to $x$ and $t_r$ are described in
terms of Dubrovin-type equations (see the following Lemmas). We assume
that the affine part of $\mathcal{K}_n$ is nonsingular 
    \begin{equation}\label{5.59}
        \{\widetilde{E}_m\}_{m=1,\ldots,2n+1} \subset \mathbb{C},
        \quad \widetilde{E}_m \neq \widetilde{E}_{m^\prime}
        \quad \textrm{for $m \neq m^\prime,~
        m,m^\prime=1,\ldots,2n+1$}.
     \end{equation}

\newtheorem{lem5.4}[lem5.1]{Lemma}
 \begin{lem5.4}
   Assume that $(\ref{2.2})$, $(\ref{5.3})$, and
   $(\ref{5.4})$ hold.

 $(\mathrm{i})$
 Suppose that the zeros $\{\mu_j(x,t_r)\}_{j=1,\ldots,n}$
 of $F_n(z,x,t_r)$ remain distinct for $(x,t_r) \in
 \Omega_\mu,$ where $\Omega_\mu \subseteq \mathbb{C}^2$ is open
 and connected. Then, $\{\mu_j(x,t_r)\}_{j=1,\ldots,n}$ satisfy
 the system of differential equations
    \begin{equation}\label{5.60}
        \mu_{j,x}=\frac{(u-u_{xx})y(\hat{\mu}_j)}{(u_x-u)\sqrt{\mu_j}}
        \prod_{ \scriptstyle k=1 \atop \scriptstyle k \neq j}^n
        (\mu_j-\mu_k)^{-1},
        \qquad j=1,\ldots,n,
    \end{equation}
    \begin{equation}\label{5.61}
        \mu_{j,t_r}=\frac{2\widetilde{F}_r(\mu_j) y(\hat{\mu}_j) }
                    {(u_x-u)\sqrt{\mu_j}}
         \prod_{ \scriptstyle k=1 \atop \scriptstyle k \neq j}^n
        (\mu_j-\mu_k)^{-1},
        \qquad j=1,\ldots,n,
    \end{equation}
with initial conditions
       \begin{equation}\label{5.62}
         \{\hat{\mu}_j(x_0,t_{0,r})\}_{j=1,\ldots,n}
         \in \mathcal{K}_{n},
       \end{equation}
for some fixed $(x_0,t_{0,r}) \in \Omega_\mu$. The initial value
problem $(\ref{5.61})$, $(\ref{5.62})$ has a unique solution
satisfying
        \begin{equation}\label{5.63}
         \hat{\mu}_j \in C^\infty(\Omega_\mu,\mathcal{K}_{n}),
         \quad j=1,\ldots,n.
        \end{equation}

$\mathrm{(ii)}$
 Suppose that the zeros $\{\nu_j(x,t_r)\}_{j=1,\ldots,n}$
 of $H_n(z,x,t_r)$ remain distinct for $(x,t_r) \in
 \Omega_\nu,$ where $\Omega_\nu \subseteq \mathbb{C}^2$ is open
 and connected. Then
 $\{\nu_j(x,t_r)\}_{j=1,\ldots,n}$ satisfy the system of
 differential equations
    \begin{equation}\label{5.64}
        \nu_{j,x}=-\frac{(u-u_{xx})y(\hat{\nu}_j)}{(u_x+u)\sqrt{\nu_j}}
        \prod_{ \scriptstyle k=1 \atop \scriptstyle k \neq j}^n
        (\nu_j-\nu_k)^{-1},
        \qquad j=1,\ldots,n,
    \end{equation}
    \begin{equation}\label{5.65}
        \nu_{j,t_r}=\frac{2\widetilde{H}_r(\nu_j) y(\hat{\nu}_j) }
                     {(u_x+u)\sqrt{\nu_j}}
        \prod_{ \scriptstyle k=1 \atop \scriptstyle k \neq j}^n
        (\nu_j-\nu_k)^{-1},
        \qquad j=1,\ldots,n,
    \end{equation}
with initial conditions
       \begin{equation}\label{5.66}
         \{\hat{\nu}_j(x_0,t_{0,r})\}_{j=1,\ldots,n}
         \in \mathcal{K}_{n},
       \end{equation}
for some fixed $(x_0,t_{0,r}) \in \Omega_\nu$. The initial value
problem $(\ref{5.65})$, $(\ref{5.66})$ has a unique solution
satisfying
        \begin{equation}\label{5.67}
         \hat{\nu}_j \in C^\infty(\Omega_\nu,\mathcal{K}_{n}),
         \quad j=1,\ldots,n.
        \end{equation}
\end{lem5.4}
\textbf{Proof.}~~ It suffices to focus on (\ref{5.60}), (\ref{5.61})
and (\ref{5.63}), since the proof procedure for (\ref{5.64}), (\ref{5.65})
and (\ref{5.67}) is similar.

The proof of (\ref{5.60}) has
been given in Lemma 3.4. We just derive (\ref{5.61}). Differentiating
on both sides of (\ref{5.6}) with respect to $t_r$ yields
     \begin{equation}\label{5.68}
        F_{n,t_r}(\mu_j)=-(u_x-u)\mu_{j,t_r}
         \prod_{ \scriptstyle k=1 \atop \scriptstyle k \neq j}^n
        (\mu_j-\mu_k).
     \end{equation}
On the other hand, inserting $z=\mu_j$ into (\ref{5.38}) and
considering (\ref{5.26}), we arrive at
     \begin{equation}\label{5.69}
        F_{n,t_r}(\mu_j)=2G_n(\mu_j)\widetilde{F}_r(\mu_j)
         =2\frac{y(\hat{\mu}_j)}{-\sqrt{\mu_j}}\widetilde{F}_r(\mu_j).
     \end{equation}
Combining (\ref{5.68}) with (\ref{5.69}) leads to (\ref{5.61}). The
smoothness assertion of (\ref{5.63}) is clear as long as $\hat{\mu}_j$
stays away from the branch points $(\widetilde{E}_m,0)$. In case
$\hat{\mu}_j$ admits such a branch point, one can use the local chart
around $(\widetilde{E}_m,0)$
($\zeta=\sigma(z-\widetilde{E}_m)^{1/2}$,
$\sigma=\pm 1$) to verify (\ref{5.63}). \quad $\square$ \\

Let us now present the $t_r$-dependent trace formulas of the MCH hierarchy,
which are used to construct the algebro-geometric solutions $u$
in section 6. For simplicity, we just take the simplest case.

\newtheorem{lem5.5}[lem5.1]{Lemma}
 \begin{lem5.5}
  Assume that $(\ref{2.2})$,  $(\ref{5.3})$, and $(\ref{5.4})$ hold.
  Then, we have
      \begin{equation}\label{5.70}
        -\frac{1}{2}(u-u_{xx})(u^2-u_x^2)
        +\frac{1}{2}(u-u_{xx})\sum_{m=1}^{2n+1}\widetilde{E}_m=
        (u-u_x)\sum_{j=1}^n \mu_j,
      \end{equation}
      \begin{equation}\label{5.71}
         \frac{1}{2}(u-u_{xx})(u^2-u_x^2)
        -\frac{1}{2}(u-u_{xx})\sum_{m=1}^{2n+1}\widetilde{E}_m=
        -(u+u_x)\sum_{j=1}^n \nu_j,
      \end{equation}
    \end{lem5.5}
\textbf{Proof.}~~ The proof is similar to the corresponding
stationary case in Lemma 3.5. \quad $\square$

\section{Time-dependent algebro-geometric solutions}
  In this final section, we extend the results in section 4
  from the stationary MCH hierarchy to the time-dependent case.
  In particular,
  we obtain Riemann theta function representations for the
  Baker-Akhiezer function, the meromorphic function $\phi$, and the
  algebro-geometric solutions for the MCH hierarchy.

  Let us first consider the asymptotic properties of $\phi$ in the
  time-dependent case.

  \newtheorem{lem6.1}{Lemma}[section]
   \begin{lem6.1}
     Assume $(\ref{2.2})$, $(\ref{5.3})$, and
     $(\ref{5.4})$ hold. Let $P=(z,y)
    \in \mathcal{K}_n \setminus \{P_\infty,P_0\},$ $(x,t_r) \in
    \mathbb{C}^2$. Then, we have
     \begin{equation}\label{6.1}
        \phi(P)\underset{\zeta \rightarrow 0}{=}
        \frac{2}{u-u_x}\zeta^{-1}+O(1),
        \qquad P \rightarrow P_\infty, \qquad \zeta=z^{-1},
     \end{equation}
     \begin{equation}\label{6.2}
        \phi(P)\underset{\zeta \rightarrow 0}{=}
         i\zeta+O(\zeta^2),
        \qquad P \rightarrow P_0, \qquad \zeta=z^{1/2}.
     \end{equation}
   \end{lem6.1}
\textbf{Proof.}~~The proof is analogous to the corresponding
stationary case in Lemma 4.1. \quad $\square$ \\

Next, we study the properties of Abel map, which dose not linearize
the divisor $\mathcal{D}_{\underline{\hat{\mu}}(x,t_r)}$ and
$\mathcal{D}_{\underline{\hat{\nu}}(x,t_r)}$ in the time-dependent
MCH hierarchy. This is the remarkable difference between 
the MCH hierarchy and other integrable systems such as KdV and
AKNS hierarchies. For that purpose, we introduce some notations of
symmetric functions.

Let us  define
  \begin{equation}\label{6.3}
     \begin{split}
       \mathcal{S}_k&=\{\underline{l}=(l_1,\ldots,l_n) \in \mathbb{N}^k
        |~l_1 < \cdots < l_k \leq n\},
        \quad k=1,\ldots, n, \\
       \mathcal{T}_k^{(j)}&=\{\underline{l}=(l_1,\ldots,l_n) \in \mathcal{S}_k
      |~ l_m \neq j\}, \quad k=1,\ldots, n-1, ~ j=1, \ldots, n.
     \end{split}
  \end{equation}
The symmetric functions are defined by
    \begin{equation}\label{6.4}
        \Psi_0(\underline{\mu})=1, \quad
        \Psi_k(\underline{\mu})=(-1)^k \sum_{\underline{l}\in \mathcal{S}_k}
        \mu_{l_1}\cdots \mu_{l_k}, \quad
        k=1,\ldots,n,
    \end{equation}
and
    \begin{equation}\label{6.5}
        \begin{split}
         & \Phi_0^{(j)}(\underline{\mu})=1, \\
         & \Phi_k^{(j)}(\underline{\mu})= \sum_{\underline{l}\in \mathcal{T}_k^{(j)}}
           \mu_{l_1}\cdots \mu_{l_k}, \quad
        k=1,\ldots,n-1, \quad j=1,\ldots,n,
        \end{split}
    \end{equation}
where $\underline{\mu}=(\mu_1,\ldots,\mu_n) \in \mathbb{C}^n$. The
properties of $\Psi_k(\underline{\mu})$ and
$\Phi_k^{(j)}(\underline{\mu})$ can be found in Appendix E \cite{12}.
Here we freely use those relations.

Moreover, for the MCH hierarchy we have
 \footnote{$m \wedge n = \mathrm{min}\{m,n\}$,
 $m \vee n =\mathrm{max} \{m,n\}$}
    \begin{equation}\label{6.6}
      \begin{split}
     & \widehat{F}_r(\mu_j)=(u_x-u) \sum_{s=(r-n)\vee 0}^r
      \hat{c}_s(\underline{\widetilde{E}})
      \Phi_{r-s}^{(j)}(\underline{\mu}),
       \\
  & \widetilde{F}_r(\mu_j)= \sum_{s=0}^r \tilde{c}_{r-s}\widehat{F}_r(\mu_j)
  =(u_x-u)
  \sum_{k=0}^{r\wedge n} \tilde{d}_{r,k}(\underline{\widetilde{E}})
  \Phi_k^{(j)}(\underline{\mu}), \quad
    r\in \mathbb{N}_0, ~\tilde{c}_0=1,
     \end{split}
    \end{equation}
where
   \begin{equation}\label{6.7}
     \tilde{d}_{r,k}(\underline{\widetilde{E}})
     =\sum_{s=0}^{r-k} \tilde{c}_{r-k-s} \hat{c}_s(\underline{\widetilde{E}})
     \qquad k=0,\ldots, r\wedge n.
   \end{equation}

\newtheorem{the6.2}[lem6.1]{Theorem}
 \begin{the6.2}
   Assume that $\mathcal{K}_n$ is nonsingular and $(\ref{2.2})$
   holds.

   $\mathrm{(i)}$ Suppose $\{\hat{\mu}_j\}_{j=1,\ldots,n}$
   satisfies the Dubrovin equations $(\ref{5.60})$, $(\ref{5.61})$
   on $\Omega_\mu$ and remain distinct and $\widetilde{F}_r(\mu_j)
   \neq 0$ for $(x,t_r)\in \Omega_\mu$, where $\Omega_\mu \subseteq \mathbb{C}^2$
   is open and connected, and the associated divisor is defined by
     \begin{equation}\label{6.8}
        \mathcal{D}_{\underline{\hat{\mu}}(x,t_r)} \in \mathrm{Sym}^n
        (\mathcal{K}_n),
        \qquad
        \underline{\hat{\mu}}=\{\hat{\mu}_1,\ldots,\hat{\mu}_n\}
        \in  \mathrm{Sym}^n (\mathcal{K}_n).
       \end{equation}
 Then, we have
        \begin{equation}\label{6.9}
            \partial_x
            \underline{\alpha}_{Q_0}(\mathcal{D}_{\underline{\hat{\mu}}(x,t_r)})
            =-\frac{u-u_{xx}}{u_x-u}
             \frac{1}{ \Psi_n(\underline{\mu}(x,t_r))}
             \underline{c}(1),
            \qquad    (x,t_r) \in \Omega_\mu,
        \end{equation}
        \begin{eqnarray}\label{6.10}
         &&
         \partial_{t_r}
         \underline{\alpha}_{Q_0}(\mathcal{D}_{\underline{\hat{\mu}}(x,t_r)})
         =-\frac{2}{ \Psi_n(\underline{\mu}(x,t_r))}
         \Big( \sum_{k=0}^{r \wedge n} \tilde{d}_{r,k}(\underline{\widetilde{E}})
          \Psi_k(\underline{\mu}(x,t_r)) \Big) \underline{c}(1)
          \nonumber \\
         && ~~~~~
         +2 \Big( \sum_{l=1 \vee (n+1-r)}^n
         \tilde{d}_{r,n+1-l}(\underline{\widetilde{E}})
         \underline{c}(l)
         \Big),
         \qquad
         (x,t_r) \in \Omega_\mu.
        \end{eqnarray}
 In particular, the Abel map dose not linearize the divisor
 $\mathcal{D}_{\underline{\hat{\mu}}(x,t_r)}$ on $\Omega_\mu$.

 $\mathrm{(ii)}$ Suppose that $\{\hat{\nu}_j\}_{j=1,\ldots,n}$
   satisfies the Dubrovin equations $(\ref{5.64})$, $(\ref{5.65})$
   on $\Omega_\nu$ and remain distinct and $\widetilde{H}_r(\nu_j)
   \neq 0$ for $(x,t_r)\in \Omega_\nu$, where $\Omega_\nu \subseteq \mathbb{C}^2$
   is open and connected, and the associated divisor is defined by
     \begin{equation}\label{6.11}
        \mathcal{D}_{\underline{\hat{\nu}}(x,t_r)} \in \mathrm{Sym}^n
        (\mathcal{K}_n),
        \qquad
        \underline{\hat{\nu}}=\{\hat{\nu}_1,\ldots,\hat{\nu}_n\}
        \in  \mathrm{Sym}^n (\mathcal{K}_n).
       \end{equation}
 Then, we have
        \begin{equation}\label{6.12}
            \partial_x
            \underline{\alpha}_{Q_0}(\mathcal{D}_{\underline{\hat{\nu}}(x,t_r)})
            =\frac{u-u_{xx}}{u_x+u}
             \frac{1}{ \Psi_n(\underline{\nu}(x,t_r))}
             \underline{c}(1),
            \qquad    (x,t_r) \in \Omega_\nu,
        \end{equation}
        \begin{eqnarray}\label{6.13}
         &&
         \partial_{t_r}
         \underline{\alpha}_{Q_0}(\mathcal{D}_{\underline{\hat{\nu}}(x,t_r)})
         =-\frac{2}{ \Psi_n(\underline{\nu}(x,t_r))}
         \Big( \sum_{k=0}^{r \wedge n} \tilde{d}_{r,k}(\underline{\widetilde{E}})
          \Psi_k(\underline{\nu}(x,t_r)) \Big) \underline{c}(1)
          \nonumber \\
         && ~~~~~
         +2 \Big( \sum_{l=1 \vee (n+1-r)}^n
         \tilde{d}_{r,n+1-l}(\underline{\widetilde{E}})
         \underline{c}(l)
         \Big),
         \qquad
         (x,t_r) \in \Omega_\nu.
        \end{eqnarray}
 In particular, the Abel map dose not linearize the divisor
 $\mathcal{D}_{\underline{\hat{\nu}}(x,t_r)}$ on $\Omega_\nu$.
 \end{the6.2}
 \textbf{Proof.}~~It suffices to prove (\ref{6.10}),  since the proof procedure
  is similar for (\ref{6.13}).  The proof for (\ref{6.9}) and
 (\ref{6.12}) has been given in the stationary context of Theorem
 4.3. Let us first give a fundamental identity (E.17) \cite{12}, that is,
   \begin{equation}\label{6.14}
     \Phi_{k+1}^{(j)}(\underline{\mu})=\mu_j \Phi_{k}^{(j)}(\underline{\mu})
     +\Psi_{k+1}(\underline{\mu}),
     \quad k=0,\ldots,n-1, ~
     j=1,\ldots,n.
   \end{equation}
 Then, together with (\ref{6.6}) and (\ref{4.18}), we have
    \begin{eqnarray}\label{6.15}
     \frac{\widetilde{F}_r(\mu_j)}{(u_x-u)\mu_j}
     &=&\mu_j^{-1}
     \sum_{m=0}^{r\wedge n} \tilde{d}_{r,m}(\underline{\widetilde{E}})
     \Phi_m^{(j)}(\underline{\mu})
       \\
     &=&
     \mu_j^{-1}
     \sum_{m=0}^{r\wedge n} \tilde{d}_{r,m}(\underline{\widetilde{E}})
     \Big(\mu_j\Phi_{m-1}^{(j)}(\underline{\mu})
     +\Psi_{m}(\underline{\mu})\Big)
       \nonumber \\
     &=&
    \sum_{m=1}^{r\wedge n} \tilde{d}_{r,m}(\underline{\widetilde{E}})
    \Phi_{m-1}^{(j)}(\underline{\mu})
    -\sum_{m=0}^{r\wedge n} \tilde{d}_{r,m}(\underline{\widetilde{E}})
    \Psi_{m}(\underline{\mu})
    \frac{\Phi_{n-1}^{(j)}(\underline{\mu})}{\Psi_{n}(\underline{\mu})}.
     \nonumber
    \end{eqnarray}
So, using (\ref{6.15}), (\ref{5.61}), (E.9), (E.25) and (E.26)
\cite{12}, we obtain
   \begin{eqnarray}
    \partial_{t_r} \Big(\sum_{j=1}^n \int_{Q_0}^{\hat{\mu}_j}\underline{\omega}\Big)
    &=&\sum_{j=1}^n \mu_{j,t_r} \sum_{k=1}^n
    \underline{c}(k)\frac{\mu_j^{k-1}}{\sqrt{\mu_j}y(\hat{\mu}_j)}
      \nonumber \\
    &=&
    2\sum_{j=1}^n \sum_{k=1}^n \underline{c}(k)
    \frac{\mu_j^{k-1}}{\prod_{\scriptstyle l=1 \atop \scriptstyle l \neq j}^n (\mu_j-\mu_l)}
    \frac{\widetilde{F}_r(\mu_j)}{\mu_j(u_x-u)}
      \nonumber \\
    &=&
     2\sum_{j=1}^n \sum_{k=1}^n \underline{c}(k)
    \frac{\mu_j^{k-1}}{\prod_{\scriptstyle l=1 \atop \scriptstyle l \neq j}^n (\mu_j-\mu_l)}
    \Big(
      -\sum_{m=0}^{r\wedge n} \tilde{d}_{r,m}(\underline{\widetilde{E}})
    \Psi_{m}(\underline{\mu})
    \frac{\Phi_{n-1}^{(j)}(\underline{\mu})}{\Psi_{n}(\underline{\mu})}
      \nonumber \\
    &&
    +\sum_{m=1}^{r\wedge n} \tilde{d}_{r,m}(\underline{\widetilde{E}})
    \Phi_{m-1}^{(j)}(\underline{\mu}) \Big)
      \nonumber
   \end{eqnarray}
   \begin{eqnarray}\label{6.16}
     &=&
     -2\sum_{m=0}^{r\wedge n} \tilde{d}_{r,m}(\underline{\widetilde{E}})
     \frac{\Psi_{m}(\underline{\mu})}{\Psi_{n}(\underline{\mu})}
     \sum_{k=1}^n \sum_{j=1}^n \underline{c}(k)
     (U_n(\underline{\mu}))_{k,j}(U_n(\underline{\mu}))_{j,1}^{-1}
       \nonumber \\
     &&
     +2\sum_{m=1}^{r\wedge n} \tilde{d}_{r,m}(\underline{\widetilde{E}})
     \sum_{k=1}^n \sum_{j=1}^n \underline{c}(k)
     (U_n(\underline{\mu}))_{k,j}(U_n(\underline{\mu}))_{j,n-m+1}^{-1}
       \nonumber \\
     &=&
     -\frac{2}{\Psi_n(\underline{\mu})}
     \sum_{m=0}^{r\wedge n} \tilde{d}_{r,m}(\underline{\widetilde{E}})
     \Psi_{m}(\underline{\mu}) \underline{c}(1)
     +2\sum_{m=1}^{r\wedge n} \tilde{d}_{r,m}(\underline{\widetilde{E}})
     \underline{c}(n-m+1)
        \nonumber \\
     &=&
     -\frac{2}{\Psi_n(\underline{\mu})}
     \sum_{m=0}^{r\wedge n} \tilde{d}_{r,m}(\underline{\widetilde{E}})
     \Psi_{m}(\underline{\mu}) \underline{c}(1)
     +2\sum_{m=1\vee (n+1-r) }^{ n} \tilde{d}_{r,n+1-m}(\underline{\widetilde{E}})
     \underline{c}(m).
      \nonumber \\
   \end{eqnarray}
Therefore,  we complete the proof of (\ref{6.10}). \quad $\square$ \\

The following result is a special form of Theorem 6.2, which
provides the constraint condition to linearize the divisor
$\mathcal{D}_{\underline{\hat{\mu}}(x,t_r)}$ and
$\mathcal{D}_{\underline{\hat{\nu}}(x,t_r)}$ associated with
$\phi(P,x,t_r)$. We recall the definitions of
$\underline{\widehat{B}}_{Q_0}$ and $\underline{\hat{\beta}}_{Q_0}$
in (\ref{4.22}) and (\ref{4.23}).

\newtheorem{the6.3}[lem6.1]{Theorem}
  \begin{the6.3}
   Assume that $(\ref{2.2})$ and the statements of $\{\mu_j\}_{j=1,\ldots,n}$
   and $\{\nu_j\}_{j=1,\ldots,n}$ in Theorem $6.2$ hold.
   Then,

        $\mathrm{(i)}$
       for $\{\mu_j\}_{j=1,\ldots,n}$, we have
   \begin{equation}\label{6.17}
        \partial_x \sum_{j=1}^n \int_{Q_0}^{\hat{\mu}_j(x,t_r)} \eta_1
        = -\frac{1}{\Psi_n(\underline{\mu}(x,t_r))} \frac{u-u_{xx}}{u_x-u},
        \qquad (x,t_r) \in \Omega_\mu,
    \end{equation}
    \begin{equation}\label{6.18}
        \partial_x \underline{\hat{\beta}}
        (\mathcal{D}_{\underline{\hat{\mu}}(x,t_r)})
        =
        \begin{cases}
         \frac{u-u_{xx}}{u_x-u},
         ~~~~~~~~n=1,\\
          \frac{u-u_{xx}}{u_x-u}(0,\ldots,0,1),
          ~~~~~~~~n\geq 2,
        \end{cases}
        \quad  (x,t_r) \in \Omega_\mu,
    \end{equation}
    \begin{eqnarray}\label{6.19}
     \partial_{t_r} \sum_{j=1}^n \int_{Q_0}^{\hat{\mu}_j(x,t_r)} \eta_1
     &=&-\frac{2}{\Psi_n(\underline{\mu}(x,t_r))}
     \sum_{k=0}^{r\wedge n} \tilde{d}_{r,k}(\underline{\widetilde{E}})
     \Psi_{k}(\underline{\mu}(x,t_r))
      \nonumber \\
      &&
     +~2 \tilde{d}_{r,n}(\underline{\widetilde{E}}) \delta_{n,r\wedge n},
     \quad  (x,t_r) \in \Omega_\mu,
    \end{eqnarray}
    \begin{eqnarray}\label{6.20}
    &&
    \partial_{t_r} \underline{\hat{\beta}}
        (\mathcal{D}_{\underline{\hat{\mu}}(x,t_r)})
        \nonumber \\
        && ~~~~~
        =2\Big(
     \sum_{s=0}^r \tilde{c}_{r-s}\hat{c}_{s+1-n}(\underline{\widetilde{E}}),
     \ldots,
     \sum_{s=0}^r \tilde{c}_{r-s}\hat{c}_{s+1}(\underline{\widetilde{E}}),
           \sum_{s=0}^r \tilde{c}_{r-s}\hat{c}_{s}(\underline{\widetilde{E}}),
        \Big),
         \nonumber \\
     && ~~~~~~~~~~~~~~~
     \quad \hat{c}_{-l}(\underline{\widetilde{E}})=0,~ l\in
     \mathbb{N},
     \quad  (x,t_r) \in \Omega_\mu.
    \end{eqnarray}

      $\mathrm{(ii)}$ for $\{\nu_j\}_{j=1,\ldots,n}$, we have
        \begin{equation}\label{6.21}
        \partial_x \sum_{j=1}^n \int_{Q_0}^{\hat{\nu}_j(x,t_r)} \eta_1
        = \frac{1}{\Psi_n(\underline{\nu}(x,t_r))} \frac{u-u_{xx}}{u_x+u},
        \qquad (x,t_r) \in \Omega_\nu,
    \end{equation}
    \begin{equation}\label{6.22}
        \partial_x \underline{\hat{\beta}}
        (\mathcal{D}_{\underline{\hat{\nu}}(x,t_r)})
        =
        \begin{cases}
         \frac{-u+u_{xx}}{u_x+u},
         ~~~~~~~~n=1,\\
          \frac{-u+u_{xx}}{u_x+u}(0,\ldots,0,1),
          ~~~~~~~~n\geq 2,
        \end{cases}
        \quad  (x,t_r) \in \Omega_\nu,
    \end{equation}
    \begin{eqnarray}\label{6.23}
     \partial_{t_r} \sum_{j=1}^n \int_{Q_0}^{\hat{\nu}_j(x,t_r)} \eta_1
     &=&-\frac{2}{\Psi_n(\underline{\nu}(x,t_r))}
     \sum_{k=0}^{r\wedge n} \tilde{d}_{r,k}(\underline{\widetilde{E}})
     \Psi_{k}(\underline{\nu}(x,t_r))
      \nonumber \\
      &&
     +~2 \tilde{d}_{r,n}(\underline{\widetilde{E}}) \delta_{n,r\wedge n},
     \quad  (x,t_r) \in \Omega_\nu,
    \end{eqnarray}
    \begin{eqnarray}\label{6.24}
    &&
    \partial_{t_r} \underline{\hat{\beta}}
        (\mathcal{D}_{\underline{\hat{\nu}}(x,t_r)})
        \nonumber \\
        && ~~~~~
        =2\Big(
     \sum_{s=0}^r \tilde{c}_{r-s}\hat{c}_{s+1-n}(\underline{\widetilde{E}}),
     \ldots,
     \sum_{s=0}^r \tilde{c}_{r-s}\hat{c}_{s+1}(\underline{\widetilde{E}}),
           \sum_{s=0}^r \tilde{c}_{r-s}\hat{c}_{s}(\underline{\widetilde{E}}),
        \Big),
         \nonumber \\
     && ~~~~~~~~~~~~~~~
     \quad \hat{c}_{-l}(\underline{\widetilde{E}})=0,~ l\in
     \mathbb{N},
     \quad  (x,t_r) \in \Omega_\nu.
    \end{eqnarray}
  \end{the6.3}
\textbf{Proof.}~~Equations (\ref{6.17}), (\ref{6.18}),
(\ref{6.21}), and (\ref{6.22}) have been proved in the stationary
case in Theorem 4.4. Equations (\ref{6.19}) and (\ref{6.20}) follows
by (\ref{6.16}) through taking account into (E.9) \cite{12}. Similarly, one
can obtain (\ref{6.23}) and (\ref{6.24}). \quad $\square$.\\

Motivated by Theorems 6.2 and 6.3, the change of variables
    \begin{equation}\label{6.25}
        x \mapsto \tilde{x}= \int^x dx^\prime
        \Big(\frac{1}{\Psi_n(\underline{\mu}(x^\prime))}
        \frac{u-u_{x^\prime x^\prime}}{u_{x^\prime}-u}\Big)
    \end{equation}
and
    \begin{eqnarray}\label{6.26}
     t_r \mapsto  \tilde{t}_r &=&\int^{t_r} ds
     \Big(
     \frac{2}{ \Psi_n(\underline{\mu}(x,t_r))}
          \sum_{k=0}^{r \wedge n} \tilde{d}_{r,k}(\underline{\widetilde{E}})
          \Psi_k(\underline{\mu}(x,t_r))
          \nonumber \\
         && ~~~~~
         -2 \sum_{l=1 \vee (n+1-r)}^n
         \tilde{d}_{r,n+1-l}(\underline{\widetilde{E}})
         \frac{\underline{c}(l)}{\underline{c}(1)}
              \Big)
    \end{eqnarray}
linearizes the Abel map
$\underline{A}_{Q_0}(\mathcal{D}_{\underline{\hat{\tilde{\mu}}}(\tilde{x},\tilde{t}_r)})$,
$\tilde{\mu}_j(\tilde{x},\tilde{t}_r)=\mu_j(x,t_r)$, $j=1,\ldots,n$.
The intricate relation between the variables $(x,t_r)$ and
$(\tilde{x},\tilde{t}_r)$ is detailedly studied in Theorem 6.4.

Next, we shall provide an explicit representation of $\phi$ and $u$
in terms of the Riemann theta function associated with
$\mathcal{K}_n$ under the assumption of the affine part of $\mathcal{K}_n$ being
nonsingular. We still use $n\in\mathbb{N}$ for the remainder of
this argument to avoid the trivial case $n=0$. By 
(\ref{4.26})-(\ref{4.35}), one of the principal results 
reads as follows.

\newtheorem{the6.4}[lem6.1]{Theorem}
 \begin{the6.4}
   Suppose that the curve $\mathcal{K}_n$ is nonsingular,
   $(\ref{2.2})$, $(\ref{5.3})$, and
   $(\ref{5.4})$ hold on $\Omega$. Let $P=(z,y) \in \mathcal{K}_n \setminus \{P_\infty,P_0\}$,
    $(x,t_r),(x_0,t_{0,r}) \in \Omega$, and $\mathcal{D}_{\underline{\hat{\mu}}(x,t_r)}$, or
   $\mathcal{D}_{\underline{\hat{\nu}}(x,t_r)}$ is nonspecial for
   $(x,t_r) \in \Omega$, where $\Omega \subseteq
   \mathbb{C}^2$ is open and connected. 
   Then, $\phi$ and $u$ have the following
   representations
   \begin{equation}\label{6.27}
    \phi(P,x,t_r)=i
    \frac{\theta(\underline{z}(P,\underline{\hat{\nu}}(x,t_r) ))
         \theta(\underline{z}(P_0,\underline{\hat{\mu}}(x,t_r) ))}
     {\theta(\underline{z}(P_0,\underline{\hat{\nu}}(x,t_r) ))
     \theta(\underline{z}(P,\underline{\hat{\mu}}(x,t_r) ))}
     \mathrm{exp} \Big(d_0-\int_{Q_0}^P \omega_{P_\infty P_0}^{(3)}
     \Big),
   \end{equation}
   \begin{eqnarray}\label{6.28}
    u(x,t_r)&=&i
    \frac{\theta(\underline{z}(P_0,\underline{\hat{\nu}}(x,t_r) ))
         \theta(\underline{z}(P_\infty,\underline{\hat{\mu}}(x,t_r) ))}
     {\theta(\underline{z}(P_\infty,\underline{\hat{\nu}}(x,t_r) ))
     \theta(\underline{z}(P_0,\underline{\hat{\mu}}(x,t_r) ))}
        \\
    &\times&
    \left(
         \frac{\sum_{j=1}^n \lambda_j - \sum_{j=1}^n U_j \partial_{\omega_j}
              \mathrm{ln}
              \Big(
    \frac{\theta(\underline{z}(P_\infty,\underline{\hat{\mu}}(x,t_r) )+\underline{\omega})}
         {\theta(\underline{z}(P_0,\underline{\hat{\mu}}(x,t_r) )+\underline{\omega})}
              \Big)
              \Big|_{\underline{\omega}=0} }
    {\sum_{j=1}^n \lambda_j - \sum_{j=1}^n U_j \partial_{\omega_j}
              \mathrm{ln}
              \Big(
    \frac{\theta(\underline{z}(P_\infty,\underline{\hat{\nu}}(x,t_r) )+\underline{\omega})}
         {\theta(\underline{z}(P_0,\underline{\hat{\nu}}(x,t_r) )+\underline{\omega})}
              \Big)
              \Big|_{\underline{\omega}=0} }
    +1 \right).
    \nonumber
   \end{eqnarray}
Moreover, let $\mu_j$, $j=1,\ldots,n,$ be non-vanishing on $\Omega$.
Then, we have the following constraint
    \begin{eqnarray}\label{6.29}
     &&
     \int_{x_0}^x dx^\prime \frac{u-u_{x^\prime x^\prime }}{u_{x^\prime}-u}
     +2(t_r-t_{0,r}) \sum_{s=0}^r \tilde{c}_{r-s} \hat{c}_s(\underline{\widetilde{E}})
      \nonumber \\
     &&=
     \Big(-\int_{x_0}^x \frac{u-u_{x^\prime x^\prime }}{u_{x^\prime}-u}
         \frac{dx^\prime}{\prod_{k=1}^n \mu_k(x^\prime, t_r)}
     -2 \sum_{k=0}^{r \wedge n} \tilde{d}_{r,k}(\underline{\widetilde{E}})
     \int_{t_{0,r}}^{t_r} \frac{\Psi_k(\underline{\mu}(x_0,s))}{\Psi_n(\underline{\mu}(x_0,s))}
     ds
     \Big)
     \nonumber \\
     &&~~~~~ \times ~
     \sum_{j=1}^n \Big( \int_{a_j} \tilde{\omega}_{P_\infty P_0}^{(3)}
     \Big) c_j(1)
      \nonumber \\
     &&~~~~~+~
     2(t_r-t_{0,r}) \sum_{l=1 \vee (n+1-l) }^n
     \tilde{d}_{r,n+1-l}(\underline{\widetilde{E}})
     \sum_{j=1}^n \Big( \int_{a_j} \tilde{\omega}_{P_\infty P_0}^{(3)}
     \Big) c_j(l)
      \nonumber \\
     && ~~~~~+~
     \mathrm{ln} \left(
     \frac{\theta(\underline{z}(P_\infty,\underline{\hat{\mu}}(x,t_r) ))
           \theta(\underline{z}(P_0,\underline{\hat{\mu}}(x_0,t_{0,r}) ))}
          {\theta(\underline{z}(P_0,\underline{\hat{\mu}}(x,t_{r})))
          \theta(\underline{z}(P_\infty,\underline{\hat{\mu}}(x_0,t_{0,r}) ))}
          \right),
    \end{eqnarray}
     $$(x,t_r), (x_0,t_{0,r}) \in \Omega$$
with
    \begin{eqnarray}\label{6.30}
      &&
       \underline{\hat{z}}(P_\infty,\underline{\hat{\mu}}(x,t_r))=
       \underline{\widehat{\Xi}}_{Q_0}
       -\underline{\widehat{A}}_{Q_0}(P_\infty)
       +\underline{\hat{\alpha}}_{Q_0}(\mathcal{D}_{\underline{\hat{\mu}}(x,t_r)})
        \nonumber \\
      && =
      \underline{\widehat{\Xi}}_{Q_0}
       -\underline{\widehat{A}}_{Q_0}(P_\infty)
       +\underline{\hat{\alpha}}_{Q_0}(\mathcal{D}_{\underline{\hat{\mu}}(x_0,t_r)})
       -  \Big(   \int_{x_0}^x
       \frac{u-u_{x^\prime x^\prime}}{u_{x^\prime}-u}
             \frac{dx^\prime}{ \Psi_n(\underline{\mu}(x^\prime,t_r))} \Big)
             \underline{c}(1)
          \nonumber \\ \\
        &&=
        \underline{\widehat{\Xi}}_{Q_0}
       -\underline{\widehat{A}}_{Q_0}(P_\infty)
       +\underline{\hat{\alpha}}_{Q_0}(\mathcal{D}_{\underline{\hat{\mu}}(x,t_{0,r})})
            \nonumber \\
        &&~~~~~
       -2 \Big(
       \sum_{k=0}^{r \wedge n} \tilde{d}_{r,k}(\underline{\widetilde{E}})
       \int_{t_{0,r}}^{t_r}
       \frac{\Psi_k(\underline{\mu}(x,s))}{\Psi_n(\underline{\mu}(x,s))}
       ds \Big)\underline{c}(1)
            \nonumber \\
       &&~~~~~+
       2(t_r-t_{0,r}) \Big(
       \sum_{l=1 \vee (n+1-l) }^n
       \tilde{d}_{r,n+1-l}(\underline{\widetilde{E}})
       \underline{c}(l)
       \Big),
       \end{eqnarray}
and
       \begin{eqnarray}\label{6.32}
      &&
       \underline{\hat{z}}(P_0,\underline{\hat{\mu}}(x,t_r))=
       \underline{\widehat{\Xi}}_{Q_0}
       -\underline{\widehat{A}}_{Q_0}(P_0)
       +\underline{\hat{\alpha}}_{Q_0}(\mathcal{D}_{\underline{\hat{\mu}}(x,t_r)})
        \nonumber \\
      && =
      \underline{\widehat{\Xi}}_{Q_0}
       -\underline{\widehat{A}}_{Q_0}(P_0)
       +\underline{\hat{\alpha}}_{Q_0}(\mathcal{D}_{\underline{\hat{\mu}}(x_0,t_r)})
       -  \Big(   \int_{x_0}^x
       \frac{u-u_{x^\prime x^\prime}}{u_{x^\prime}-u}
             \frac{dx^\prime}{ \Psi_n(\underline{\mu}(x^\prime,t_r))} \Big)
             \underline{c}(1)
          \nonumber \\ \\
       &&=
        \underline{\widehat{\Xi}}_{Q_0}
       -\underline{\widehat{A}}_{Q_0}(P_0)
       +\underline{\hat{\alpha}}_{Q_0}(\mathcal{D}_{\underline{\hat{\mu}}(x,t_{0,r})})
            \nonumber \\
        &&~~~~~
       -2 \Big(
       \sum_{k=0}^{r \wedge n} \tilde{d}_{r,k}(\underline{\widetilde{E}})
       \int_{t_{0,r}}^{t_r}
       \frac{\Psi_k(\underline{\mu}(x,s))}{\Psi_n(\underline{\mu}(x,s))}
       ds \Big)\underline{c}(1)
            \nonumber \\
       &&~~~~~+
       2(t_r-t_{0,r}) \Big(
       \sum_{l=1 \vee (n+1-l) }^n
       \tilde{d}_{r,n+1-l}(\underline{\widetilde{E}})
       \underline{c}(l)
       \Big).
       \end{eqnarray}
    \end{the6.4}
\textbf{Proof.}~~Let us first 
assume that $\mu_j$,
$j=1,\ldots,n$, are distinct and non-vanishing on
$\widetilde{\Omega}$ and $\widetilde{F}_r(\mu_j) \neq 0$ on
$\widetilde{\Omega}$, $j=1,\ldots,n,$ where $\widetilde{\Omega}
\subseteq \Omega$. Then, the representation (\ref{6.27}) for $\phi$
on $\widetilde{\Omega}$ follows by combining (\ref{6.1}) with
(\ref{6.2}). The representation (\ref{6.28}) for $u$ on
$\widetilde{\Omega}$ follows by the trace formulas (\ref{5.70}),
(\ref{5.71}) and (F.89) \cite{12}. In fact, since the proofs of
(\ref{6.27}) and (\ref{6.28}) are identical to the corresponding
stationary results in Theorem 4.5, which can be extended line by
line to the time-dependent setting. Here we skip the 
details.

Let us now turn to the relation (\ref{6.29}). We first consider
the time variation part of (\ref{6.29}). From (\ref{6.20}), it is
easy to see that
    \begin{eqnarray*}
           \sum_{j=1}^n \int_{Q_0}^{\hat{\mu}_j(x_0,t_r)}
           \tilde{\omega}_{P_\infty P_0}^{(3)}
      -\sum_{j=1}^n \int_{Q_0}^{\hat{\mu}_j(x_0,t_{0,r})}
           \tilde{\omega}_{P_\infty P_0}^{(3)}
     \end{eqnarray*}
      \begin{eqnarray}\label{6.34}
      &&
      =\int_{t_{0,r}}^{t_r}
        \partial_s \Big(
        \sum_{j=1}^n \int_{Q_0}^{\hat{\mu}_j(x_0,s)}
           \tilde{\omega}_{P_\infty P_0}^{(3)}
      \Big) ds
         \nonumber \\
      &&
      =\int_{t_{0,r}}^{t_r} \Big(
        2\sum_{s=0}^r \tilde{c}_{r-s} \hat{c}_s(\underline{\widetilde{E}})
      \Big) ds
      =2(t_r-t_{0,r})\sum_{s=0}^r \tilde{c}_{r-s}
      \hat{c}_s(\underline{\widetilde{E}}).
    \end{eqnarray}
On the other hand, combining (\ref{6.10}) with (F.88) \cite{12} yields
    \begin{eqnarray}\label{6.35}
     &&
     \sum_{j=1}^n \int_{Q_0}^{\hat{\mu}_j(x_0,t_r)}
           \tilde{\omega}_{P_\infty P_0}^{(3)}
      -\sum_{j=1}^n \int_{Q_0}^{\hat{\mu}_j(x_0,t_{0,r})}
           \tilde{\omega}_{P_\infty P_0}^{(3)}
             \nonumber \\
     &&
     =\sum_{j=1}^n \Big(
        \int_{a_j}  \tilde{\omega}_{P_\infty P_0}^{(3)}
      \Big)
     \Big(
     \sum_{k=1}^n \int_{Q_0}^{\hat{\mu}_k(x_0,t_r)}
           \omega_j
      -\sum_{k=1}^n \int_{Q_0}^{\hat{\mu}_k(x_0,t_{0,r})}
           \omega_j
     \Big)
      \nonumber \\
     && ~~~~~+ ~
     \mathrm{ln}~\Big(
        \frac{\theta(\underline{z}(P_\infty, \underline{\hat{\mu}}(x_0,t_r) ))}
             {\theta(\underline{z}(P_0, \underline{\hat{\mu}}(x_0,t_r)))} \Big)
     -\mathrm{ln}~\Big(
        \frac{\theta(\underline{z}(P_\infty, \underline{\hat{\mu}}(x_0,t_{0,r}) ))}
             {\theta(\underline{z}(P_0, \underline{\hat{\mu}}(x_0,t_{0,r})))} \Big)
        \nonumber \\
     &&
     =\sum_{j=1}^n \Big(
        \int_{a_j}  \tilde{\omega}_{P_\infty P_0}^{(3)}
      \Big)
      \int_{t_{0,r}}^{t_r}
      \partial_s \Big(
      \sum_{k=1}^n \int_{Q_0}^{\hat{\mu}_k(x_0,s)} \omega_j
      \Big)ds
         \nonumber \\
      &&~~~~~+~
      \mathrm{ln}~\Big(
        \frac{\theta(\underline{z}(P_\infty, \underline{\hat{\mu}}(x_0,t_r) ))
              \theta(\underline{z}(P_0, \underline{\hat{\mu}}(x_0,t_{0,r})))}
             {\theta(\underline{z}(P_0, \underline{\hat{\mu}}(x_0,t_r)))
              \theta(\underline{z}(P_\infty, \underline{\hat{\mu}}(x_0,t_{0,r}) ))}
               \Big)
         \nonumber \\
      &&
      =\sum_{j=1}^n \Big(
        \int_{a_j}  \tilde{\omega}_{P_\infty P_0}^{(3)}
      \Big)
       \int_{t_{0,r}}^{t_r}
      \Big(
       -\frac{2 \Psi_k(\underline{\mu}(x_0,s))}
       {\Psi_n(\underline{\mu}(x_0,s))} \sum_{k=0}^{r \wedge n}
       \tilde{d}_{r,k}(\underline{\widetilde{E}})
       c_j(1)
         \nonumber \\
      &&~~~~~+~
       2\sum_{l=1 \vee (n+1-l)}^n \tilde{d}_{r,n+1-l}(\underline{\widetilde{E}})
       c_j(l)
      \Big) ds
          \nonumber \\
      &&~~~~~+~
      \mathrm{ln}~\Big(
        \frac{\theta(\underline{z}(P_\infty, \underline{\hat{\mu}}(x_0,t_r) ))
              \theta(\underline{z}(P_0, \underline{\hat{\mu}}(x_0,t_{0,r})))}
             {\theta(\underline{z}(P_0, \underline{\hat{\mu}}(x_0,t_r)))
              \theta(\underline{z}(P_\infty, \underline{\hat{\mu}}(x_0,t_{0,r}) ))}
               \Big)
         \nonumber \\
      &&
      =-2\sum_{j=1}^n \Big(
        \int_{a_j}  \tilde{\omega}_{P_\infty P_0}^{(3)}
      \Big) c_j(1)
      \sum_{k=0}^{r \wedge n} \tilde{d}_{r,k}(\underline{\widetilde{E}})
       \int_{t_{0,r}}^{t_r}
         \frac{\Psi_k(\underline{\mu}(x_0,s))}
       {\Psi_n(\underline{\mu}(x_0,s))} ds
          \nonumber \\
       &&~~~~~+~
    2(t_r-t_{0,r})
    \sum_{l=1 \vee (n+1-l)}^n \tilde{d}_{r,n+1-l}(\underline{\widetilde{E}})
     \sum_{j=1}^n \Big(
        \int_{a_j}  \tilde{\omega}_{P_\infty P_0}^{(3)}
      \Big)  c_j(l)
         \nonumber \\
       &&~~~~~+~
      \mathrm{ln}~\Big(
        \frac{\theta(\underline{z}(P_\infty, \underline{\hat{\mu}}(x_0,t_r) ))
              \theta(\underline{z}(P_0, \underline{\hat{\mu}}(x_0,t_{0,r})))}
             {\theta(\underline{z}(P_0, \underline{\hat{\mu}}(x_0,t_r)))
              \theta(\underline{z}(P_\infty, \underline{\hat{\mu}}(x_0,t_{0,r}) ))}
               \Big).
            \end{eqnarray}
The space variation part of (\ref{6.29}) has been given in the
stationary case in (\ref{4.49}) and (\ref{4.50}), that is,
    \begin{eqnarray}\label{6.36}
    \int_{x_0}^x dx^\prime \frac{u-u_{x^\prime x^\prime}}{u_{x^\prime}-u}
    &=&- \int_{x_0}^x
     \Big(
     \frac{u-u_{x^\prime x^\prime}}{u_{x^\prime}-u}
     \frac{1}{\prod_{k=1}^n \mu_k(x^\prime, t_r)}
     \Big) dx^\prime
      \nonumber \\
      && ~~~~\times ~
     \sum_{j=1}^n \Big(
        \int_{a_j}  \tilde{\omega}_{P_\infty P_0}^{(3)}
      \Big)  c_j(1)
       \nonumber \\
      &+&
      \mathrm{ln}\Big(
        \frac{\theta(\underline{z}(P_\infty, \underline{\hat{\mu}}(x,t_r) ))
              \theta(\underline{z}(P_0, \underline{\hat{\mu}}(x_0,t_{r})))}
             {\theta(\underline{z}(P_0, \underline{\hat{\mu}}(x,t_r)))
              \theta(\underline{z}(P_\infty, \underline{\hat{\mu}}(x_0,t_{r}) ))}
               \Big).
    \end{eqnarray}
Hence, combining all of these three (\ref{6.34}), (\ref{6.35}) and (\ref{6.36}) leads
to (\ref{6.29}). Equations (\ref{6.30})-(6.33) are valid from
(\ref{6.9}) and (\ref{6.10}). The extension of all results from
$x\in\widetilde{\Omega}$ to $x\in\Omega$  simply follows by
the continuity of $\underline{\alpha}_{Q_0}$ and the hypothesis of
$\mathcal{D}_{\underline{\hat{\mu}}(x,t_r)}$ being nonspecial for
$x\in\Omega$. \quad $\square$

\newtheorem{rem6.5}[lem6.1]{Remark}
 \begin{rem6.5}
  A closer look at Theorem $6.4$ shows that
  $(\ref{6.30})$-$(6.33)$ equal to
  \begin{eqnarray}\label{6.37}
       \underline{\hat{z}}(P_\infty,\underline{\hat{\mu}}(x,t_r))&=&
       \underline{\widehat{\Xi}}_{Q_0}
       -\underline{\widehat{A}}_{Q_0}(P_\infty)
       +\underline{\hat{\alpha}}_{Q_0}(\mathcal{D}_{\underline{\hat{\mu}}(x,t_r)})
         \\
      &=&
        \underline{\widehat{\Xi}}_{Q_0}
       -\underline{\widehat{A}}_{Q_0}(P_\infty)
       +\underline{\hat{\alpha}}_{Q_0}(\mathcal{D}_{\underline{\hat{\mu}}(x_0,t_r)})
       -\underline{c}(1)(\tilde{x}-\tilde{x}_0)
         \nonumber \\
        &=&
        \underline{\widehat{\Xi}}_{Q_0}
       -\underline{\widehat{A}}_{Q_0}(P_\infty)
       +\underline{\hat{\alpha}}_{Q_0}(\mathcal{D}_{\underline{\hat{\mu}}(x,t_{0,r})})
        -\underline{c}(1)(\tilde{t}_r-\tilde{t}_{0,r}),
        \nonumber
       \end{eqnarray}
  and
     \begin{eqnarray}\label{6.38}
       \underline{\hat{z}}(P_0,\underline{\hat{\mu}}(x,t_r))&=&
       \underline{\widehat{\Xi}}_{Q_0}
       -\underline{\widehat{A}}_{Q_0}(P_0)
       +\underline{\hat{\alpha}}_{Q_0}(\mathcal{D}_{\underline{\hat{\mu}}(x,t_r)})
         \\
      &=&
        \underline{\widehat{\Xi}}_{Q_0}
       -\underline{\widehat{A}}_{Q_0}(P_0)
       +\underline{\hat{\alpha}}_{Q_0}(\mathcal{D}_{\underline{\hat{\mu}}(x_0,t_r)})
       -\underline{c}(1)(\tilde{x}-\tilde{x}_0)
         \nonumber \\
        &=&
        \underline{\widehat{\Xi}}_{Q_0}
       -\underline{\widehat{A}}_{Q_0}(P_0)
       +\underline{\hat{\alpha}}_{Q_0}(\mathcal{D}_{\underline{\hat{\mu}}(x,t_{0,r})})
        -\underline{c}(1)(\tilde{t}_r-\tilde{t}_{0,r})
        \nonumber
       \end{eqnarray}
  based on the changing of variables $x\mapsto \tilde{x}$ and $t_r \mapsto
  \tilde{t}_r$ in $(\ref{6.25})$ and $(\ref{6.26})$.
  Hence, the Abel map linearizes the divisor
  $\mathcal{D}_{\underline{\hat{\mu}}(x,t_{r})}$ on $\Omega$ with
  respect to $\tilde{x}, \tilde{t}_r$.
  This fact reveals that the Abel map does not effect the
  linearization of the divisor $\mathcal{D}_{\underline{\hat{\mu}}(x,t_{r})}$
  in the time-dependent MCH case.
 \end{rem6.5}

 \newtheorem{rem6.6}[lem6.1]{Remark}
  \begin{rem6.6}
   The Abel map linearizes the divisor
   $\mathcal{D}_{\underline{\hat{\nu}}(x,t_{r})}$ on $\Omega$ with
   respect to $\tilde{x}, \tilde{t}_r$, and the change of variables
   is given by $(\ref{4.51})$ and
   \begin{eqnarray}\label{6.39}
     t_r \mapsto  \tilde{t}_r &=&\int^{t_r} ds
     \Big(
     \frac{2}{ \Psi_n(\underline{\nu}(x,t_r))}
          \sum_{k=0}^{r \wedge n} \tilde{d}_{r,k}(\underline{\widetilde{E}})
          \Psi_k(\underline{\nu}(x,t_r))
          \nonumber \\
         && ~~~~~
         -2 \sum_{l=1 \vee (n+1-r)}^n
         \tilde{d}_{r,n+1-l}(\underline{\widetilde{E}})
         \frac{\underline{c}(l)}{\underline{c}(1)}
              \Big).
    \end{eqnarray}
  \end{rem6.6}

\newtheorem{rem6.7}[lem6.1]{Remark}
  \begin{rem6.7}
    Remark $4.8$ is applicable to the present time-dependent
    context. Moreover, in order to obtain the theta function
    representation of $\psi_j$, $j=1,2,$, one can write $\widetilde{F}_r$
    in terms of $\Psi_k(\underline{\mu})$ 
    in analogy to the stationary case studied in Remark $4.9$. Here
    we skip the corresponding details.
  \end{rem6.7}

  In analogy to Example 4.10, the special case $n=0$ is excluded in
Theorem 6.4. Let us summarize it as follows. For simplicity, we just
consider the elementary case $n=0$, $r=0$.

\newtheorem{exa6.8}[lem6.1]{Example}
  \begin{exa6.8}
    Suppose $n=0$, $r=0$, $P=(z,y) \in \mathcal{K}_0 \setminus
    \{P_\infty,P_0\}$, and let $(x,t_r), (x_0,t_{0,r}) \in \mathbb{C}^2$.
    Then, we have
     \begin{equation}\label{6.40}
       \begin{split}
       & \mathcal{K}_0: \mathcal{F}_0(z,y)=y^2-R_1(z)=y^2-(z-\widetilde{E}_1)
          =0 \quad \widetilde{E}_1 \in \mathbb{C}, \\
       & u(x,t_0)=\widetilde{E}_1 ^{1/2}, \\
       & \mathrm{MCH}_0(u)=u_{t_0}-u_{xxt_0}+2u_x-2u_{xx}=0,\\
       & F_0(z,x)=\widetilde{F}_0(z,x)=f_0=u_x-u, \\
       & G_0(z,x)=\widetilde{G}_0(z,x)=g_0=-1, \\
       & H_0(z,x)=\widetilde{H}_0(z,x)=h_0=u_x+u, \\
       & \phi(P,x,t_0)=\frac{z^{1/2}y-z}{u_x-u}
                      =\frac{u_x+u}{z^{-1/2}y+1}, \\
       & \psi_1(P,x,x_0,t_0,t_{0,0})= \mathrm{exp} \Big(
           \int_{x_0}^x \Big(
          \frac{1}{2}\frac{u-u_{x^\prime x^\prime}}{u_{x^\prime}-u}
          \frac{y-\sqrt{z}}{\sqrt{z}} \Big) dx^\prime \\
       &  ~~~~~~~~~~~~~~~~~~~~~~~
          -\frac{1}{2}(x-x_0)
          +\int_{t_{0,0}}^{t_0} \frac{y}{\sqrt{z}} ds \Big),\\
       &  \psi_2(P,x,x_0,t_0,t_{0,0})=\frac{y-\sqrt{z}}{u_x-u}
          \mathrm{exp} \Big(
           \int_{x_0}^x \Big(
          \frac{1}{2}\frac{u-u_{x^\prime x^\prime}}{u_{x^\prime}-u}
          \frac{y-\sqrt{z}}{\sqrt{z}} \Big) dx^\prime \\
       &  ~~~~~~~~~~~~~~~~~~~~~~~
          -\frac{1}{2}(x-x_0)
          +\int_{t_{0,0}}^{t_0} \frac{y}{\sqrt{z}} ds \Big).
       \end{split}
     \end{equation}
  The general solution of $\mathrm{MCH}_0(u)=u_{t_0}-u_{xxt_0}+2u_x-2u_{xx}=0$
  is given by
      \begin{equation}\label{6.41}
        u(x,t_0)=a_1e^{x+a_2t_0} + \widetilde{E}_1^{1/2},
        \qquad a_1,a_2 \in \mathbb{C}.
      \end{equation}
  But, according to the condition $\partial_x^k u(x,t) \in
  L^\infty(\mathbb{C})$, $k\in \mathbb{N}_0$, $t\in \mathbb{C}$ in
  $(\ref{2.2})$, we may conclude  $a_1=0$, and therefore derive
  $u(x,t_0)=\widetilde{E}_1^{1/2}$, which equals to the expression of
  $u(x,t_0)$ in $(\ref{6.40})$ obtained from trace formula
  $(\ref{5.70})$ or $(\ref{5.71})$ in the special case $n=0$.
  \end{exa6.8}

  Let us end this section by providing another principle result about
  algebro-geometric initial value problem of the MCH hierarchy. We will
  show that the solvability of the Dubrovin equations (\ref{5.60})
  and (\ref{5.61}) on $\Omega_\mu \subseteq \mathbb{C}^2 $ in fact
  implies (\ref{5.3}) and (\ref{5.4}) on $\Omega_\mu$. As pointed
  out in Remark 4.13, this amounts to solving the
  time-dependent algebro-geometric initial value problem (\ref{5.1})
  and (5.2) on $\Omega_\mu$. Recalling definition of
  $\widetilde{F}_r (\mu_j)$ introduced in (\ref{6.6}), then we
  may present the following result.

  \newtheorem{the6.9}[lem6.1]{Theorem}
   \begin{the6.9}
     Assume that $(\ref{2.2})$ holds and
     $\{\hat{\mu}_j\}_{j=1,\ldots,n}$ satisfies the Dubrovin
     equations $(\ref{5.60})$ and $(\ref{5.61})$ on $\Omega_\mu$ and
     remain distinct and nonzero
     for $(x,t_r) \in \Omega_\mu$, where
     $\Omega_\mu \subseteq \mathbb{C}^2 $ is open and connected.
     If $\widetilde{F}_r (\mu_j)$ in
     $(\ref{5.61})$ are expressed in terms of $\mu_k$, $k=1,\ldots,n$ by
     $(\ref{6.6})$, then $u \in C^\infty(\Omega_\mu)$ satisfying
       \begin{equation}\label{6.42}
        -\frac{1}{2}(u-u_{xx})(u^2-u_x^2) + \frac{1}{2}(u-u_{xx})
          \sum_{m=1}^{2n+1}\widetilde{E}_m
          =(u-u_x)\sum_{j=1}^n \mu_j
       \end{equation}
     will also satisfy the $r$th MCH equation $(\ref{5.1})$, that
     is,
       \begin{equation}\label{6.43}
        \mathrm{MCH}_r(u)=0
        \quad \textrm{on $\Omega_\mu$},
       \end{equation}
     with initial values satisfying the $n$th stationary MCH
     equation $(5.2)$.
   \end{the6.9}
 \textbf{Proof.}~~Given the solutions
  $\hat{\mu}_j=(\mu_j,y(\hat{\mu}_j))\in
  C^\infty(\Omega_\mu,\mathcal {K}_n),$ $j=1,\cdots,n$ of (\ref{5.60}) and (\ref{5.61}), we
  introduce polynomials $F_n, G_n,$ and $H_n$ on $\Omega_\mu$, which are exactly the same as in Theorem 4.11 in the
  stationary case
     \begin{eqnarray}
       &&
        F_n(z)=(u_x-u) \prod_{j=1}^n (z-\mu_j), \\
       &&
        (u-u_{xx})G_n(z)=F_n(z)+F_{n,x}(z), \\
       &&
       H_n(z)=(u_x+u) \prod_{j=1}^n (z-\nu_j), \\
       &&
       zG_{n,x}(z)=-\frac{1}{2}(u-u_{xx})H_n(z)-\frac{1}{2}(u-u_{xx})F_n(z),
          \\
       &&
       H_{n,x}(z)=(u-u_{xx})G_n(z)+H_n(z), \\
       &&
       R_{2n+1}(z)=zG_n^2(z)+F_n(z)H_n(z),
     \end{eqnarray}
  where  $t_r$ is treated as a parameter. Hence, let us focus on the
  proof of (\ref{5.1}).\\
  Let us denote the polynomial $\widetilde{G}_r$ of degree $r$ by
     \begin{equation}\label{6.50}
        \frac{1}{2}(u_{t_r}-u_{xxt_r})=\widetilde{F}_{r,x}(z)+\widetilde{F}_r(z)
        -(u-u_{xx})\widetilde{G}_r(z)
        \quad \textrm{on $ \mathbb{C} \times \Omega_\mu$}.
     \end{equation}
  Next, we want to establish
    \begin{equation}\label{6.51}
        F_{n,t_r}(z)=2(G_n(z)\widetilde{F}_r(z)-F_n(z)\widetilde{G}_r(z))
        \quad \textrm{on $ \mathbb{C} \times \Omega_\mu$},
    \end{equation}
 where $\widetilde{F}_r(z)$ is defined on $ \mathbb{C} \times
 \Omega_\mu$ by
     \begin{equation}\label{6.52}
        \widetilde{F}_r(z)=\sum_{s=0}^r \tilde{c}_{r-s}
        \widehat{F}_s(z),
        \qquad \tilde{c}_0=1
     \end{equation}
 with integration constants $\{\tilde{c}_1,\ldots,\tilde{c}_r\}
 \subset \mathbb{C}$ and
    \begin{equation}\label{6.53}
     \widehat{F}_s(z)=(u_x-u) \sum_{\rho=0}^s \hat{c}_\rho(\underline{\widetilde{E}})
     \sum_{l=0}^{s-\rho} \Psi_{s-l-\rho}(\underline{\mu}) z^l.
    \end{equation}
 To prove (\ref{6.51}), let 
     \begin{eqnarray}
      \check{F}_n(z)&=&(u-u_{xx})^{-1}F_n(z), \\
      \check{\widetilde{F}}_r(z)&=&(u-u_{xx})^{-1}\widetilde{F}_r(z),
     \end{eqnarray}
 on $ \mathbb{C} \times \Omega_\mu$. A direct calculation shows that
 (\ref{6.51}) is equivalent to
     \begin{equation}\label{6.56}
        \check{F}_{n,t_r}(z)=2\check{\widetilde{F}}_r(z)\check{F}_{n,x}(z)
         -2\check{F}_n(z)\check{\widetilde{F}}_{r,x}(z),
     \end{equation}
 which is similar to that in the AKNS context \cite{12}. So,
 (\ref{6.56}) holds due to (F.112) \cite{12}. This in turn proves
 (\ref{6.51}). \\
 Next, we denote the polynomial $\widetilde{H}_r$ of degree $r$ by
      \begin{equation}\label{6.57}
        z\widetilde{G}_{r,x}(z)=-\frac{1}{2}(u-u_{xx})\widetilde{H}_r(z)
        -\frac{1}{2}(u-u_{xx})\widetilde{F}_r(z)
        \quad \textrm{on $ \mathbb{C} \times \Omega_\mu$. }
      \end{equation}
 Then, differentiating on both sides of (6.45) with respect to $t_r$ and inserting
 (\ref{6.51}) and (\ref{6.50}), we have
     \begin{eqnarray}\label{6.58}
       F_{n,xt_r}&=&-2(G_n\widetilde{F}_r-F_n\widetilde{G}_r)
       +(u-u_{xx})_{t_r}G_n+(u-u_{xx})G_{n,t_r}
          \\
       &=&
       2F_n\widetilde{G}_r+2\widetilde{F}_{r,x}G_n-2(u-u_{xx})\widetilde{G}_rG_n
       +(u-u_{xx})G_{n,t_r}.
         \nonumber
     \end{eqnarray}
 On the other hand, taking the derivative on both sides of (\ref{6.51}) with respect to
 $x$, and using (6.45), (6.47) and (\ref{6.57}), we
 obtain
     \begin{eqnarray}\label{6.59}
      F_{n,t_rx}&=&+2G_n\widetilde{F}_{r,x}
      +2F_n\widetilde{G}_r-2(u-u_{xx})\widetilde{G}_rG_n
        \nonumber \\
      &&
      +z^{-1}(u-u_{xx})F_n\widetilde{H}_r
      -z^{-1}(u-u_{xx})H_n\widetilde{F}_r.
     \end{eqnarray}
 Consequently, combining (\ref{6.58}) with (\ref{6.59}) we conclude
    \begin{equation}\label{6.60}
        zG_{n,t_r}(z)=F_n(z)\widetilde{H}_r(z)-H_n(z)\widetilde{F}_r(z)
           \quad \textrm{on $ \mathbb{C} \times \Omega_\mu$. }
    \end{equation}

 Next, differentiating both sides of (6.49) with respect to $t_r$, and inserting the
 expressions (\ref{6.51}) and (\ref{6.60}) for $F_{n,t_r}$ and
 $G_{n,t_r}$,  we have
    \begin{equation}\label{6.61}
        H_{n,t_r}(z)=2(H_n(z)\widetilde{G}_r(z)-G_n(z)\widetilde{H}_r(z))
        \quad \textrm{on $ \mathbb{C} \times \Omega_\mu$. }
    \end{equation}
 Finally, taking the derivative on both sides of (\ref{6.60}) with respect to $x$,
 and inserting (6.45), (6.48) and (\ref{6.50}) for $F_{n,x}$,
 $H_{n,x}$ and $\widetilde{F}_{r,x}$, we arrive at 
    \begin{eqnarray}\label{6.62}
     zG_{n,t_rx}&=& (u-u_{xx})G_n\widetilde{H}_r-F_n\widetilde{H}_r
      +F_n\widetilde{H}_{r,x}-(u-u_{xx})G_n\widetilde{F}_r
         \nonumber \\
      &&
      -\frac{1}{2}(u-u_{xx})_{t_r}H_n-(u-u_{xx})H_n\widetilde{G}_r.
    \end{eqnarray}
 On the other hand, differentiating both sides of (6.47) with respect to $t_r$,
 and using (\ref{6.51}) and (\ref{6.61}) for $F_{n,t_r}$ and
 $H_{n,t_r}$, we have
   \begin{eqnarray}\label{6.63}
    zG_{n,xt_r}&=&-\frac{1}{2}(u-u_{xx})_{t_r}H_n
     -\frac{1}{2}(u-u_{xx})(2H_n\widetilde{G}_r-2G_n\widetilde{H}_r)
        \nonumber \\
     &&
     -\frac{1}{2}(u-u_{xx})_{t_r}F_n
     -\frac{1}{2}(u-u_{xx})(2G_n\widetilde{F}_r-2F_n\widetilde{G}_r).
   \end{eqnarray}
Therefore, combining (\ref{6.62}) with (\ref{6.63}) yields
       \begin{equation}\label{6.64}
        -F_n\widetilde{H}_r+F_n\widetilde{H}_{r,x}=
        -\frac{1}{2}(u-u_{xx})_{t_r}F_n
        +(u-u_{xx})F_n\widetilde{G}_r,
       \end{equation}
 which implies
       \begin{equation}\label{6.65}
        -\frac{1}{2}(u-u_{xx})_{t_r}=\widetilde{H}_{r,x}(z)-\widetilde{H}_r(z)
        -(u-u_{xx})\widetilde{G}_r(z)
          \quad \textrm{on $ \mathbb{C} \times \Omega_\mu$. }
       \end{equation}
 So, we proved (5.12)-(5.17) and (5.38)-(5.40) on $ \mathbb{C} \times \Omega_\mu$
 and thus conclude that $u$ satisfies the $r$th MCH equation
 (\ref{5.1}) with initial values satisfying the $n$th stationary MCH
 equation (5.2)  on $ \mathbb{C} \times \Omega_\mu$.

 \newtheorem{rem6.10}[lem6.1]{Remark}
     \begin{rem6.10}
       The result in Theorem $6.9$ is presented in terms of
       $u$ and $\{\mu_j\}_{j=1,\cdots,n}$, but of course
        one can provide the analogous result
       in terms of $u$ and $\{\nu_j\}_{j=1,\cdots,n}$.
     \end{rem6.10}

  The analog of Remark 4.13 directly extends to the current
  time-dependent MCH hierarchy.

\section*{Acknowledgments}
    The work of Fan and Hou was supported by grants from
    the National Science Foundation of China (Project No. 10971031), and
    the Shanghai Shuguang Tracking Project (Project No. 08GG01), and Qiao was partially supported 
by the U. S. Army Research Office (Contract/Grant No. W911NF-08-1-0511) and the Texas Norman Hackerman Advanced Research Program (Grant No. 003599-0001-2009).

\end{document}